\documentclass{article}
\usepackage[dvips]{rotating}

\begin{document}

\title{Isentropic and Isocurvature Axion Perturbations in Inflationary Cosmology}
\author{Stephen D. Burns\thanks{e-mail address:  burns@physics.ucla.edu}
 \\ \begin{em}Department of Physics and Astronomy, University of California\end{em} \\ \begin{em}Los Angeles, California 90025\end{em}}

\maketitle

\begin{abstract}
Isentropic and isocurvature axion inflationary perturbations are studied in detail.  If isocurvature perturbations are present,  COBE requires the inflationary Hubble constant to be rather small, $H_{I} \stackrel{<}{\sim} 10^{12}$GeV, regardless of the inflation model.  Microwave background anisotropies are calculated for CDM models with a mixture of axion and supersymmetric CDM.  The angular power spectrum of the microwave background depends not only on the amplitude of the isocurvature power spectrum but also on $\Omega_{a}$.  The Saskatoon data and $\sigma_{8}$ are used to constrain the contribution of axion isocurvature perturbations.  Consistency with the Saskatoon data requires the square of the amplitude of the isocurvature perturbations to be $\stackrel{<}{\sim} 0.1$ times that of the adiabatic perturbations at $\ell=87$, and $\stackrel{<}{\sim} 10^{-3}-10^{-2}$ that of the adiabatic perturbations at $\ell=237$.  Although not dominant, isocurvature perturbations are not yet ruled out.  The consequences of isocurvature perturbations for axions and inflation are discussed. 
\end{abstract}

\section{Introduction}

There is a considerable body of dynamical evidence which suggests that a substantial part of the universe is composed of dark, non-luminous matter.\footnote{For a review see \cite{KT} or \cite{PeeblesLSS}.}  Mass to light ratios of galaxies, for example, imply that the luminous matter density is $\Omega_{lum} \stackrel{<}{\sim} 0.01$, i.e., just 1\% of the critical density.  Rotation curves of spiral galaxies suggest that $\Omega_{halo} \stackrel{>}{\sim} 0.1 \approx 10\Omega_{lum}$.  And observations of galaxy clusters indicate that $\Omega_{cluster}\stackrel{>}{\sim}0.3$.  Not only is the dark matter non-luminous, but it is apparently non-baryonic as well.  This is because Big Bang nucleosynthesis, which does amazingly well in predicting the abundances of the light elements, constrains $\Omega_{b}h^{2}$, the baryon density, to be in the range $0.008h^{-2} \stackrel{<}{\sim} \Omega_{b} \stackrel{<}{\sim} 0.024h^{-2}$ \cite{Turnerbaryon}.  For a Hubble constant of $h=0.5$, this means $0.032 \stackrel{<}{\sim} \Omega_{b} \stackrel{<}{\sim} 0.096$:  the dark matter can't be baryons.  Fortunately, particle physics provides a plethora of candidates (and then some) for dark matter.  Here we will focus on axions, one of the best motivated extensions of the standard $SU(3)_{c}\times SU(2)_{L}\times U(1)_{em}$ model of particle physics.  Later we will consider the possibility that the universe also contains some other kind of dark matter as well.

Axions have their origin in QCD, the theory of strong interactions.  In QCD, the Lagrangian has both a perturbative and non-perturbative part:
\begin{eqnarray}
  {\cal L}_{QCD} &= &{\cal L}_{pert} + \bar{\Theta} \frac{g^{2}}{32\pi^{2}}G^{a\mu \nu}\tilde{G}_{a\mu \nu}\\
  \bar{\Theta} &= &\bar{\Theta}_{QCD} + \mbox{Arg det }{\cal M},
\end{eqnarray}
where $\bar{\Theta}_{QCD}$ is an arbitrary parameter of the theory to be measured, $G^{a\mu \nu}$ is the gluon field strength tensor and $\tilde{G}_{a\mu \nu}$ is the dual tensor of $G$.  ${\cal M}$ is the quark mass matrix.  This non-perturbative term violates $CP$, and if it were present would lead to a neutron electric dipole moment $d_{n}\sim 10^{-16}\bar{\Theta}$e-cm.  Since the observed electric dipole moment of the neutron is $\stackrel{<}{\sim}10^{-25}$e-cm, $\bar{\Theta}$ must be less than $10^{-9}$.  Because the $G\tilde{G}$ term is a total derivative and has no effect on the equations of motion, one might think that the term could be thrown away.  But t'Hooft \cite{tHooft1,tHooft2} showed that this non-perturbative term is necessary to solve the $U(1)_{A}$ problem.  The question then becomes:  Why is $\bar{\Theta}$ so small?  Or given that $\bar{\Theta}$ depends on the strong ($\bar{\Theta}_{QCD}$) and weak (Arg det ${\cal M}$) sectors of the theory, why should the two contributions nearly cancel each other out?

Peccei and Quinn \cite{PQ1,PQ2} suggested that this question could be answered dynamically.  By adding to the QCD Lagrangian a global $U(1)$ symmetry, now known as a PQ symmetry, which is spontaneously broken at some energy scale $f_{PQ}$, the CP-violating terms $G\tilde{G}$ in the QCD Lagrangian would vanish.  Weinberg \cite{Weinberg} and Wilczek \cite{Wilczek} then noted that there must be a Nambu-Goldstone boson, the axion, associated with the global $U(1)_{PQ}$ symmetry breaking.  At temperatures below the PQ symmetry breaking scale, but above the QCD phase transition, the potential for the complex field $\vec{\phi}$ that carries the PQ charge is \cite{KT}
\begin{equation} 
  V(\phi)=\lambda(\mid \vec{\phi}\mid ^{2}-f^{2}_{PQ}/2)^{2},
\end{equation}
which has a minimum at $f_{PQ}/\sqrt{2}$.  The massless $\bar{\Theta}$ degree of freedom, namely the argument of $\langle \vec{\phi} \rangle$, is the axion:  $a=(f_{PQ}/N)\bar{\Theta}$, where $N$ is the number of flavors felt by the anomaly of the PQ symmetry.  Alternatively, $N$ is the number of degenerate and distinct CP-conserving minima of the axion potential.  $\bar{\Theta}$ is the initial misalignment angle.  The fact that there are $N$  discrete minima of the axion potential means that domain walls will be generated when the PQ symmetry is broken.  This problem can be solved if we assume that the PQ symmetry is broken prior to the end of inflation.  

At the time of the QCD phase transition, instanton effects tilt the potential,  creating a new potential minimum.   A commonly used metaphor is that of a wine bottle.  Suppose that a wine bottle is sitting flat on a table with a little leftover wine at a certain point in the bottom of the bottle.  Now tilt the bottle away from the vertical.  This creates a new minimum (where the bottle makes contact with the table).  Unless the leftover wine was sitting exactly at the point where the bottle makes contact with the table, it will slosh around the bottom of the wine bottle, oscillating about the new minimum.  In the same way, if the axion field is initially displaced by an angle $\bar{\Theta}$, the misalignment angle, the axion field will roll toward the new minimum of the potential, overshoot, and begin to oscillate about the potential minimum.  The mass of the axion corresponds to the inertia of the oscillations.  Since these oscillations correspond to a zero-momentum condensate of axions \cite{KT}, the axions behave like cold dark matter.  Notice that the axion mass is generated in a manner distinct from that of other particles, whose mass arises as a result of their coupling to a scalar field which acquires a vacuum expectation value. 

This unique mass generation mechanism has some interesting cosmological consequences.  If the universe inflated, axion isocurvature perturbations can be produced.  When the axion gets a mass at the QCD phase transition, the isocurvature perturbations in the number of axions, $n_{a}/s$, turn  into energy density perturbations.  Since energy must be locally conserved, this mass-energy must come at the expense of the other fields and the spatial curvature must remain zero.  Therefore the creation of isocurvature axion perturbations at the QCD phase transition is accompanied by the creation of density perturbations in the other fields that balance the axion perturbation to preserve $\delta\rho$=0.  Because causal mechanisms such as pressure and particle interactions are ineffective on super-horizon sized scales, energy cannot be moved around while the isocurvature perturbations are outside the horizon and $\delta \rho$ remains zero.  But once the perturbation crosses inside the horizon, fluctuations in the local pressure, $\delta p = \frac{1}{3}\delta \rho_{r}$, will ``push" around the various components of the isocurvature perturbation, creating an energy density perturbation that can then act as a seed for structure formation.  Depending upon the amplitude of these isocurvature perturbations relative to that of the isentropic perturbations, axions can produce interesting, observable effects in the anisotropy of the microwave background as well as on large scale structure.  The question then is:  Can axions, specifically the inclusion of isocurvature perturbations associated with axions, do any better than standard CDM (sCDM) models with respect to large scale structure?  

In the rest of this paper, we will consider the effects of these isocurvature perturbations on the microwave background and large scale structure.  After a brief review of inflationary perturbations, adiabatic and isocurvature fluctuations are discussed in some detail.  If axions make a sizable contribution to the matter density, we show that the COBE data places a rather strong constraint on the Hubble constant during inflation.  Next, we study the observational consequences of isocurvature fluctuations on the microwave background and large scale structure and use the Saskatoon data \cite{Page} to constrain the amplitude of the isocurvature perturbations.  Finally, we speculate about the consequences for axion isocurvature fluctuations depending upon whether or not such fluctuations are actually observed.

\section{Inflationary Perturbations}

During inflation, quantum fluctuations in massless fields are produced with a nearly scale invariant power spectrum \cite{Davies},
\begin{equation}
  k^{3}|\delta_{k}|^{2}/2\pi^{2} = \left(\frac{H_{I}}{2\pi}\right)^{2}, \label{eq:deSitter}
\end{equation}
where $H_{I}$ is the Hubble constant during inflation, and the fluctuations in the field have been written in terms of their Fourier components:
\begin{equation}
  \delta_{k} = \frac{1}{\sqrt{V_{norm}}}\int d^{3}xe^{i\vec{k}\cdot \vec{x}}\delta(\vec{x})
\end{equation}
Here $V_{norm}$ is a box normalization volume.  By the equivalence principle, these fluctuations are generated in all of the fields \cite{KT}.  Notice that while $H_{I}$ is constant, the amplitude of the fluctuations is independent of the scale $k$:  the fluctuations are scale-invariant.

Fluctuations in the inflaton field lead to ``adiabatic" energy density perturbations.  During the slow roll phase of inflation, the energy density is dominated by the inflaton potential $V(\varphi)$:
\begin{equation}
  \rho = \rho_{\varphi} = \frac{1}{2}\dot{\varphi}^{2} + V(\varphi) \simeq V(\varphi)
\end{equation}
Therefore, fluctuations in $\varphi$ produce fluctuations in the energy density
\begin{equation}
  \delta \rho_{\varphi} = \left(\frac{\partial V}{\partial \varphi}\right)\delta \varphi \label{eq:drhovarphi}
\end{equation}
Through gravity, these fluctuations in the energy density of the inflaton field act as sources for energy fluctuations in the other particle species such as baryons, photons, axions, neutralinos and neutrinos.  As the universe evolves, these energy fluctuations will act as seeds for gravitational instability, leading to the formation of structure.

Since quantum fluctuations are produced in all massless fields during inflation.  this means that if PQ symmetry breaking occurs prior to, or during, inflation,\footnote{Recall that the axion is the Goldstone boson associated with the breaking of PQ symmetry.  Therefore, the axion does not exist when the PQ symmetry is unbroken.} quantum fluctuations will also be generated in the axion field.  These fluctuations correspond to fluctuations in the local number of axions, i.e., the local axion number density to entropy density ratio $n_{a}/s$;  they also correspond to fluctuations in the initial misalignment angle of the axion field.  Since the axion is effectively massless prior to the QCD phase transition, the perturbation to the spatial curvature from these fluctuations is zero.  For this reason these fluctuations are known as isocurvature fluctuations.

\section{Isentropic and Isocurvature Perturbations}

Following Bertschinger \cite{Bertschinger}, let us begin by clearly defining the two types of perturbations.  In the literature, perturbations are typically described as isocurvature or ``adiabatic.''  What distinguishes the two is whether or not there is an initial entropy perturbation, or equivalently, whether or not there is a fluctuation in the equation of state (see below).  A more precise definition, at least from the point of view of thermodynamics, is entropic or isentropic.  Then an entropy fluctuation is one for which $\vec{\nabla} S \neq 0$, whereas an isentropic fluctuation has $\vec{\nabla} S = 0$, i.e., the entropy is the same everywhere.  In general, it is not possible for entropy fluctuations to survive inflation.  This is because the thermal equilibrium produced by reheating will wipe out any primordial entropy fluctuatioons.  In the context of inflation, then, then, entropy fluctuations should be produced after inflation is over.  As described in the Introduction, this is exactly what happens for isocurvature axion perturbations produced at the QCD phase transition.

The two types of perturbations also correspond to fluctuations in the particle number, $n/s$, where $n$ is the number density and $s$ is the entropy density.  Isentropic perturbations are fluctuations in the energy density, not the particle number, so they have $\delta(n_{i}/s) = 0$.  Entropy perturbations, however, are perturbations in the number of particles, so they have $\delta(n_{a}/s) \neq 0$.  Why does $\delta(n_{a}/s) \neq 0$ correspond to an entropy gradient?  If $\delta(n_{a}/s) \neq 0$, that means that there are spatial fluctuations in the equation of state, since the local pressure depends not just on the density $\rho$ but also on the composition of the fluctuation.  As an example, consider an ideal gas with constant density.  Euler's equation for such a comoving fluid element is \cite{Bertschinger}
\begin{equation}
  \frac{1}{\bar{\rho}}\vec{\nabla}p = \frac{2}{3}T\vec{\nabla}S
\end{equation}
So if $\delta(n_{a}/s) \neq 0$, we have pressure or entropy gradients:  $\vec{\nabla}S\neq 0$.  When the perturbations become sub-horizon sized and microphysics can kick back in, these pressure gradients can push particles in the perturbation around, eventually leading to a density fluctuation.

Suppose the universe is made up of axions (subscript $a$), baryons (subscript $b$), some other form of CDM such as neutralinos (subscript $x$), photons and (massless) neutrinos.  Consider isocurvature axion fluctuations, which are fluctuations in the number of axions:  $\delta(n_{a}/s) \neq 0$.  In all of the other species, neutralinos, baryons, photons and neutrinos, there is no such fluctuation and so
\begin{equation}
  \delta\left(\frac{n_{i}}{s}\right) = 0 \Rightarrow \frac{\delta n_{i}}{n_{i}} = \frac{\delta s}{s}
\end{equation}
Except for the axions, we see that the fluctuation in the local number density of any species relative to the fluctuation in the entropy density vanishes, i.e., the fluctuations in the local number density of a given species grows adiabatically with $\frac{\delta s}{s}$.  Because the entropy density $s\propto T^{3}$,
\begin{equation}
  \frac{\delta n_{x}}{n_{x}} = \frac{\delta n_{b}}{n_{b}} = \frac{\delta n_{\gamma}}{n_{\gamma}} = \frac{\delta n_{\nu}}{n_{\nu}} = \frac{\delta s}{s} = 3\frac{\delta T}{T}
\end{equation}
The energy density for the non-relativistic matter is $\rho_{m}=mn_{m}$, so the matter fluctuation is
\begin{equation}
  \delta_{m} \equiv \frac{\delta \rho_{m}}{\rho_{m}} = \frac{\delta n_{m}}{n_{m}}
\end{equation}
For the radiation, $\rho_{r} \propto T^{4}$, which means the fluctuation in the radiation is
\begin{equation}
  \delta_{r} \equiv \frac{\delta \rho_{r}}{\rho_{r}} = 4\frac{\delta T}{T}
\end{equation}
So for the non-axionic components,  $\delta_{x}=\delta_{b}=\frac{3}{4}\delta_{\gamma}=\frac{3}{4}\delta_{\nu}$.  Of course, for isentropic axion perturbations $\delta(n_{a}/s) = 0$ and $\delta_{a}=\delta_{x}=\delta_{b}=\frac{3}{4}\delta_{\gamma}=\frac{3}{4}\delta_{\nu}$.  These relations between $\delta_{x}, \delta_{b}, \delta_{\gamma}$, etc., hold as long as the perturbations are super-horizon-sized, i.e., as long as microphysical processes (such as pressure) are unimportant.  Once the perturbations have been laid down, and while they remain outside the horizon, both isentropic and isocurvature perturbations evolve adiabatically, i.e., $\dot{S}=0$.

\subsection{Super-horizon Sized Perturbations}

Recall that during inflation, density perturbations are produced by quantum fluctuations in the inflaton field.  These perturbations are then stretched out to super-horizon sized scales by the exponential expansion.  Although physical observables, such as the temperature fluctuations in the microwave background, are gauge-invariant, we are still confronted with the following question:  Since the perturbations depend on the choice of gauge, how are they to be described while they are outside the horizon?  This is an especially important question if we wish to relate the fluctuations produced during inflation to the perturbations that are just crossing back inside the horizon today, as is the case if we wish to know the contribution of the perturbations to the quadrupole in the microwave background anisotropy.

As an example of this gauge ambiguity, consider isentropic fluctuations in the synchronous gauge.  In this gauge, the metric has the form \cite{MaBertschinger}
\begin{equation}
  ds^{2}=a^{2}(\tau)\{-d\tau^{2}+(\delta_{ij}+h_{ij})dx^{i}dx^{j}\},
\end{equation}
Here $a$ is the scale factor, $\tau$ is the conformal time, which is related to the proper time $t$ by $d\tau = dt/a(\tau)$, and $h_{ij}$ is the metric perturbation.  In this gauge, the proper time is just the coordinate time since $g_{00}=1$.  Note also that the time coordinates are orthogonal to the space coordinates, $g_{0i}=0$, which means that this gauge corresponds to freely falling observers.  Physically, then, this gauge is the natural choice for cold dark matter, since the cold dark matter has zero mean velocity with respect to the expansion.  Ma and Bertschinger \cite{MaBertschinger} showed that deep in the radiation-dominated era, the growing, isentropic, super-horizon sized perturbations behave as 
\begin{eqnarray}
  h &= &C(k\tau)^{2} \nonumber \\
  \delta_{\gamma} &= &-\frac{2}{3}C(k\tau)^{2} \nonumber \\
  \delta_{c} &= &\delta_{b}=\frac{3}{4}\delta_{\nu}=\frac{3}{4}\delta_{\gamma}
\end{eqnarray}
Here $h=tr(h_{ii})$ and $C$ is an arbitrary constant.  Note that in the synchronous gauge, the perturbations grow while they are outside of the horizon. 

Another common choice is the conformal Newtonian gauge, in which the metric has the form
\begin{equation}
  ds^{2} = a^{2}(\tau)[-(1+2\psi)d\tau^{2} + (1-2\phi)dx^{i}dx_{i}],
\end{equation}
In this gauge, the metric perturbations are simply the Newtonian potential $\psi$ and the fractional perturbation to the spatial curvature $\phi$.  This gauge is then a natural choice for understanding effects such as gravitational infall and redshift.  Deep in the radiation-dominated era, the solutions for the growing, isentropic, super-horizon sized perturbations are \cite{MaBertschinger}:
\begin{eqnarray}
  \psi &= &\frac{20C}{15+4R_{\nu}} \nonumber \\
  \delta_{\gamma} &= &-2\psi \nonumber \\
  \delta_{c} &= &\delta_{b}=\frac{3}{4}\delta_{\nu}=\frac{3}{4}\delta_{\gamma},
\end{eqnarray}
where $R_{\nu}\equiv \bar{\rho}_{\nu}/(\bar{\rho}_{\nu}+\bar{\rho}_{\gamma})$.  Note that unlike their counterparts in the synchronous gauge, the conformal Newtonian gauge perturbations are constant while they are outside the horizon.  

This gauge ambiguity complicates attempts to relate fluctuations that are produced during inflation to those that are just crossing back inside the horizon today.  Fortunately, however,
\begin{equation}
  \zeta \equiv \frac{\delta \rho}{\rho + p} = (1+w)^{-1}\frac{\delta \rho}{\rho},
\end{equation}
where $w=p/\rho$, is gauge invariant, and in fact is a constant outside the horizon \cite{Bardeen}.  This provides a simple solution to the problem of matching the amplitude of the fluctuations on present day horizon-sized scales to the amplitude produced during inflation.

\subsection{Isentropic Perturbations}
As an example of the usefulness of $\zeta$, let's calculate the present day value of $\delta \rho/\rho$ in terms of that produced during inflation.  Using Equations~\ref{eq:deSitter} and~\ref{eq:drhovarphi}, we find that during inflation,
\begin{equation}
  \delta \rho(k) = \frac{1}{\sqrt{2}}H_{I}V'(\varphi)k^{-3/2},
\end{equation}
where $V'(\varphi) = \frac{\partial V}{\partial \varphi}$.  To calculate $\rho +p$, we can use the energy-momentum tensor of the scalar field which is responsible for inflation.  The inflaton has a Lagrangian density
\begin{equation}
  {\cal L}  = \frac{1}{2}\dot{\varphi}^{2} - V(\varphi)
\end{equation}
and an energy-momentum tensor 
\begin{equation}
  T^{\mu\nu} = \partial^{\mu}\varphi\partial^{\nu}\varphi - g^{\mu \nu}\cal L
\end{equation}
If $\varphi$ is spatially homogeneous, which is a requirement for inflation to occur,\footnote{If gradients in $\varphi$ dominated the stress energy, $\varphi$ would behave like a fluid with pressure $p=-\rho/3$, which would lead to $a\propto t$, not inflation \cite{KT}.} the energy density $\rho$ and the pressure $p$ of the inflaton field are
\begin{eqnarray}
  \rho &= &T^{00} = \frac{1}{2}\dot{\varphi}^{2} + V(\varphi) \nonumber \\
  p &= &T^{ii} = \frac{1}{2}\dot{\varphi}^{2} - V(\varphi)
\end{eqnarray}
Since $\rho + p = \dot{\varphi}^{2}$, the gauge-invariant variable $\zeta_{isen}$ for the isentropic perturbations is 
\begin{equation}
  \zeta_{isen}(k) = \frac{\delta\rho}{\dot{\varphi}^{2}} = \frac{1}{\sqrt{2}}\frac{H_{I}V'(\varphi)}{\dot{\varphi}^{2}}k^{-3/2}
\end{equation}

During inflation, the equation of motion for $\varphi$ is
\begin{equation}
  \ddot{\varphi} + 3H_{I}\dot{\varphi} + \Gamma_{\varphi}\dot{\varphi} + V'(\varphi) = 0,
\end{equation}
where $\Gamma_{\varphi}$ is the decay width of the inflaton field into other particles.  Since during slow roll $\ddot{\varphi}$ is negligible and the inflaton field is not decaying, the equation of motion is simply
\begin{equation}
  \dot{\varphi} = -\frac{V'(\varphi)}{3H_{I}}
\end{equation}
For isentropic perturbations then, 
\begin{equation}
  \zeta_{isen}(k) = \frac{9}{\sqrt{2}}\frac{H_{I}^{3}}{V'(\varphi)}k^{-3/2} \label{eq:zetaad}
\end{equation}
This expression can be used to match the adiabatic perturbations generated during inflation to the fluctuation in $\delta \rho/\rho$ that is just crossing back inside the horizon today.

\subsection{Isocurvature Perturbations}
The isocurvature evolution for super-horizon sized modes is simple:  $\delta \rho =0$, which means that $\zeta_{iso}=0$.  But this is rather deceptive, because the perturbations in the individual species of the isocurvature perturbation arrange themselves to maintain $\delta \rho =0$ as the universe evolves.

As a function of temperature, the mean photon energy density is $\rho_{\gamma} = a_{\gamma}T^{4}$ and the mean neutrino energy density is $\rho_{\nu} = a_{\nu}T^{4}$.  Let $R_{\nu \gamma} = \rho_{\nu}/\rho_{\gamma}$ (for three massless neutrinos, $R_{\nu \gamma} =0.68$).  For non-relativistic particles with mass $m_{i}$ and number density $n_{i}$, $\rho_{i}=mn_{i}$.  In a universe consisting of axions, baryons, some other form of CDM such as neutralinos, photons and (massless) neutrinos, $\delta \rho=0$ implies that
\begin{equation}
  m_{a}\delta n_{a} + m_{b}\delta n_{b} + m_{x}\delta n_{x} + 4\rho_{\gamma}(1+R_{\nu \gamma})\frac{\delta T}{T} = 0
\end{equation}
Using the fact that $\delta_{b}=\delta_{x}=\frac{3}{4}\delta_{\gamma}$, $\delta \rho =0$ becomes
\begin{equation}
  \rho_{a}\frac{\delta n_{a}}{n_{a}} + [3(\rho_{b}+\rho_{x}) + 4\rho_{\gamma}(1+R_{\nu \gamma})]\frac{\delta T}{T} = 0,
\end{equation}
which can be rearranged to find the temperature perturbation:
\begin{equation}
  \frac{\delta T}{T} = -\frac{\rho_{a}}{[3(\rho_{b}+\rho_{x}) + 4\rho_{\gamma}(1+R_{\nu \gamma})]}\frac{\delta n_{a}}{n_{a}}
\end{equation}
Since the number of particles per comoving volume is $n/s$, the initial fluctuation in the number of axions per comoving volume is
\begin{equation}
  (\delta_{a})_{i} = \frac{\delta(n_{a}/s)}{(n_{a}/s)}=\frac{\delta n_{a}}{n_{a}} - 3\frac{\delta T}{T}
\end{equation}
It is straightforward to relate $(\delta_{a})_{i}$ to the fluctuations produced during inflation.  Assuming that inflation occurs well before the QCD phase transition, the axion is massless during inflation, and so quantum fluctuations in the axion field have the same power spectrum as those of the inflaton field:
\begin{equation}
  k^{3}|\delta_{a}(k)|^{2}/2\pi^{2} = \left(\frac{H_{I}}{2\pi}\right)^{2}
\end{equation}
At the time of the QCD phase transition, the axion field is $a=(f_{PQ}/N)\bar{\Theta}$.  Therefore the fluctuation in the axion field is simply $\delta a = (f_{PQ}/N)\delta \bar{\Theta}$.  Since the number density of the axions is proportional to $\bar{\Theta}^{2}$, $\delta n_{a}/n_{a} = 2\delta a/a$.

Now it may be that the PQ field $|\vec{\phi}|$ is still evolving toward its SSB value of $f_{PQ}/\sqrt{2}$ during inflation.  In this case, let $f_{a}$ be the value of $|\vec{\phi}|$ when the interesting fluctuations (such as the quadrupole) are pushed outside of the inflationary horizon.  Then the initial axion perturbation is
\begin{equation}
  (\delta_{a})_{i} = \frac{\delta n_{a}}{n_{a}} = \sqrt{2}\frac{H_{I}}{f_{a}\bar{\Theta}}k^{-3/2}  \label{eq:dai}
\end{equation}

If we define $R_{am}\equiv (\rho_{b}+\rho_{x})/\rho_{a}$, then $\delta n_{a}/n_{a}$ and $\delta T/T$ at some later time are related to the initial axion perturbation by
\begin{eqnarray}
  \frac{\delta n_{a}}{n_{a}} &= &(\delta_{a})_{i} + 3\frac{\delta T}{T} \nonumber \\
  \frac{\delta T}{T} &= &- \frac{\rho_{a}/\rho_{\gamma}}{[3(1+R_{am})(\rho_{a}/\rho_{\gamma})+4(1+R_{\nu \gamma})]}(\delta_{a})_{i}
\end{eqnarray}
There are two interesting limits:  $\rho_{\gamma} \gg \rho_{a}$ and $\rho_{a} \gg \rho_{\gamma}$.  In the first case,
\begin{equation}
  \frac{\delta T}{T} = -\frac{1}{4(1+R_{\nu \gamma})}\frac{\rho_{a}}{\rho_{\gamma}}(\delta_{a})_{i} \ll (\delta_{a})_{i}
\end{equation}
Since $\delta T/T$ is so small, these fluctuations are sometimes referred to as isothermal fluctuations.  This is sensible:  Since $\rho_{\gamma} \gg \rho_{a}$, it doesn't take much of a fluctuation in the photon energy density (or temperature) to compensate for the axion perturbation.  In this limit we also have
\begin{equation}
  \frac{\delta n_{a}}{n_{a}} = (\delta_{a})_{i}
\end{equation}
In the other limit, $\rho_{a} \gg \rho_{\gamma}$.  Since $1/(1+R_{am})=\Omega_{a}/\Omega_{0}$, the temperature fluctuation becomes
\begin{equation}
  \frac{\delta T}{T}\rightarrow-\frac{1}{3}\frac{\Omega_{a}}{\Omega_{0}} (\delta_{a})_{i}\label{eq:isoT}
\end{equation}
In terms of the photon perturbation,
\begin{equation}
  (\delta_{\gamma})_{iso}=-\frac{4}{3}\frac{\Omega_{a}}{\Omega_{0}} (\delta_{a})_{i}\label{eq:isophoton}
\end{equation}
The axion perturbation decreases to
\begin{equation}
  \frac{\delta n_{a}}{n_{a}} \rightarrow \left(\frac{\Omega_{0}-\Omega_{a}}{\Omega_{0}}\right)(\delta_{a})_{i} < (\delta_{a})_{i}\label{eq:isoaxion}
\end{equation}
This is also reasonable:  As $\rho_{a}$ becomes greater than $\rho_{\gamma}$, the only way to maintain $\delta \rho =0$ is for $\delta n_{a}/n_{a}$ to decrease. 

\subsection{Relative Contribution to the Power Spectrum}
We are now in a position to compute the relative contributions to the power spectrum or quadrupole.  For the purpose of CMB anisotropies, we are interested in the photon perturbations.  For the isocurvature perturbations,
\begin{equation}
  (\delta_{\gamma})_{iso} \equiv \left(\frac{\delta \rho_{\gamma}}{\rho_{\gamma}}\right)_{iso} = -\frac{4}{3}\frac{\Omega_{a}}{\Omega_{0}}(\delta_{a})_{i},
\end{equation}
while for the isentropic perturbations, 
\begin{equation}
  (\delta_{\gamma})_{isen} \equiv \left(\frac{\delta \rho_{\gamma}}{\rho_{\gamma}}\right)_{isen} = \frac{4}{3}\zeta_{isen}
\end{equation}
When the perturbations cross back inside the horizon, the ratio of the isocurvature to isentropic power spectra for the photons is
\begin{equation}
  \beta \equiv \left[\frac{(\delta_{\gamma})_{iso}}{(\delta_{\gamma})_{isen}}\right]^{2} = \left(\frac{\Omega_{a}}{\Omega_{0}}\right)^{2}\left[\frac{(\delta_{a})_{i}}{\zeta_{isen}} \right]^{2}
\end{equation}
As might be expected, there are no isocurvature perturbations if $\Omega_{a}=0$, since in that case there are no axions.

Provided that $(\delta_{a})_{i} \ll 1$, which corresponds to $\frac{H_{I}}{f\bar{\Theta}} \ll 1$, and which should be true in any realistic model of inflation, we can rewrite the equation for $\beta$ by using Equations~\ref{eq:zetaad} and~\ref{eq:dai}:
\begin{equation}
  \beta = \frac{4}{81}\left(\frac{\Omega_{a}}{\Omega_{0}}\right)^{2}\left[\frac{V'(\varphi)}{H_{I}^{2}}\right]^{2} \left(\frac{1}{f_{a}\bar{\Theta}}\right)^{2}
\end{equation}
In that case, since the energy density is dominated by the potential during inflation, the Hubble constant is
\begin{equation}
  H_{I}^{2} \approx \frac{8\pi}{3m_{Pl}^{2}}V(\varphi),
\end{equation}
and so the expression for $\beta$ is
\begin{equation}
  \beta = \frac{1}{144\pi^{2}}\left(\frac{\Omega_{a}}{\Omega_{0}}\right)^{2} \left[\frac{m_{Pl}V'(\varphi)}{V(\varphi)}\right]^{2} \left(\frac{m_{Pl}}{f_{a}\bar{\Theta}}\right)^{2}
\end{equation}
Now consider the term $m_{Pl}V'/V$, which measures the deviation from scale invariance.  Following Turner\cite{Turnertilt}, we can define $x_{50}$ to be the value of $m_{Pl}V'/V$ fifty $e$-foldings before the end of inflation:
\begin{equation}
  x_{50} \equiv \frac{m_{P}V'(\varphi_{50})}{V(\varphi_{50})},
\end{equation}
where $\varphi_{50}$ is the value of the inflaton field fifty $e$-foldings before the end of inflation.  $x_{50}$ is related to $n_{I}$, the spectral index of the isocurvature power spectrum ($P_{I}(k)\propto k^{n_{I}}$), by
\begin{equation}
  n_{I} = -\frac{x_{50}^{2}}{8\pi}
\end{equation}
Notice that the isocurvature power spectrum is the same as the tensor power spectrum, $n_{I}=n_{T}$.  This is not surprising, since both correspond to excitations of massless fields in de Sitter space.   

Suppose, however, that the tensor spectral index $n_{T}$ is known.  This is, of course, a very challenging quantity to measure.  It was recently suggested \cite{Seljakpol,SZpol,Kamionpol1,Kamionpol2} that if polarization maps can be made of the CMB by the MAP and PLANCK satellites, there may be a chance of directly observing the effects of gravitational waves and measuring the gravitational wave power spectrum.  Even so, however, it is estimated that $n_{T}$ can only be determined to within $\Delta n_{T} = 0.2$ \cite{SZpol}.  Since it may at least be possible to constrain $n_{T}$, or equivalently $m_{Pl}V'/V$, we will henceforth write $\beta$ in terms of $n_{T}$ rather than $n_{I}$.  Then the ratio of the isocurvature to isentropic power spectrum is
\begin{equation}
  \beta = \frac{1}{18\pi}\left(\frac{\Omega_{a}}{\Omega_{0}}\right)^{2} \left(\frac{m_{Pl}}{f_{a}\bar{\Theta}}\right)^{2}(-n_{T}) \label{eq:beta1}
\end{equation}
Notice that although $\beta \propto n_{T} \propto m_{Pl}V'/V$, the term $m_{Pl}V'/V$ really comes from the isentropic power spectrum, $P_{isen} \propto 1/V'$, not from the isocurvature power spectrum.  In future, then, when we mention the dependence of the isocurvature power spectrum (or rather $\beta$) on $n_{T}$, what is really being considered is how large the isentropic perturbations can be relative to the isocurvature perturbations for the isocurvature perturbations to make an interesting contribution to the power spectrum.

If the axions are produced by misalignment (the scenario corresponding to isocurvature perturbations), then for $f_{PQ}\stackrel{<}{\sim}1.6\times 10^{18}$GeV, the mass density in axions is \cite{axTurner}
\begin{equation}
  \Omega_{a}h^{2} = 2.86 \times 10^{-2\pm 0.4} \Lambda_{200}^{-0.7}\left ( \frac{f_{PQ}}{10^{12}\mbox{GeV}}\right )^{1.18}\bar{\Theta}^{2}f(\bar{\Theta}), 
\end{equation}
while if $f_{PQ}\stackrel{>}{\sim}1.6\times 10^{18}$GeV,
\begin{equation}
  \Omega_{a}h^{2} = 10^{5\pm 0.1} \left(\frac{f_{PQ}}{10^{18}\mbox{GeV}}\right)^{1.5}\bar{\Theta}^{2}f(\bar{\Theta}) 
\end{equation}
Here we have assumed that $N=6$ and we have used the notation of  \cite{KT}, in which $N\bar{\Theta}\rightarrow \bar{\Theta}\Rightarrow -\pi\leq \bar{\Theta} \leq \pi$.  $\Lambda_{200}=\Lambda_{QCD}/200$MeV, where $\Lambda_{QCD}$ is the QCD energy scale and $f(\bar{\Theta})$ is a monotonically increasing function with $f(0)=1$.  Provided that $\bar{\Theta} \stackrel{<}{\sim}2.5$, $f(\bar{\Theta}) \stackrel{<}{\sim} 2$ and the correction term is of order unity.  Here we are also assuming that there has been no significant entropy production since the QCD phase transition; if this were not the case, and the entropy increased by a factor $\gamma$, then $\Omega_{a}h^{2}$ would decrease by a factor $\gamma$.  For $f_{PQ}\stackrel{<}{\sim}1.6\times 10^{18}$GeV, the misalignment angle is
\begin{equation}
\frac{1}{\bar{\Theta}^{2}} = \frac{6.58\times 10^{6}\times 10^{\pm 0.4}\Lambda_{200}^{-0.7}}{\Omega_{a}h^{2}} \left(\frac{m_{Pl}}{f_{PQ}}\right)^{-1.18}f(\bar{\Theta}),
\end{equation}
while for $f_{PQ}\stackrel{>}{\sim}1.6\times 10^{18}$GeV,
\begin{equation}
\frac{1}{\bar{\Theta}^{2}} = \frac{4.26\times 10^{6\pm 0.1}}{\Omega_{a}h^{2}}\left(\frac{m_{Pl}}{f_{PQ}}\right)^{-1.5}f(\bar{\Theta})
\end{equation}

Two measurements of the strong coupling constant $\alpha_{S}$ at an energy scale $M_{Z}$ correspond to the following values for $\Lambda_{QCD}$:  $\approx$ 250Mev \cite{Altarelli} and 253$^{+130}_{-96}$ \cite{Abe}.  The prefactors for $\beta$ therefore fall within the ranges 
\begin{eqnarray}
  3.0\times 10^{4} \leq 1.7\times 10^{7}\times 10^{\pm 0.4}\Lambda_{200}^{-0.7} \leq 3.5\times 10^{5}, f_{PQ}\stackrel{<}{\sim}1.6\times 10^{18}\mbox{GeV} \nonumber \\
  6.0\times 10^{4} \leq 7.5\times 10^{3\pm 0.1}\leq 9.6\times 10^{4}, f_{PQ}\stackrel{>}{\sim}1.6\times 10^{18}\mbox{GeV} \nonumber
\end{eqnarray}
So if $ f_{PQ}\stackrel{<}{\sim}1.6\times 10^{18}$, the ratio of the power spectra is approximately
\begin{equation}
  \beta \approx 3 \times (10^{4}-10^{5})\left(\frac{m_{Pl}}{f_{PQ}}\right)^{0.82}\left[\frac{\Omega_{a}}{(\Omega_{0}h)^{2}}\right] \left[\left(\frac{f_{PQ}}{f_{a}}\right)^{2}f(\bar{\Theta})\right](-n_{T}) \label{eq:powerl},
\end{equation}
while for $ f_{PQ}\stackrel{>}{\sim}1.6\times 10^{18}$GeV,
\begin{equation}
\beta \approx (0.6-1.0)\times 10^{5}\sqrt{\frac{m_{Pl}}{f_{PQ}}}\left[\frac{\Omega_{a}}{(\Omega_{0}h)^{2}}\right] \left[\left(\frac{f_{PQ}}{f_{a}}\right)^{2}f(\bar{\Theta})\right](-n_{T}) \label{eq:powerg}
\end{equation}
As we shall see, $\beta \stackrel{<}{\sim} 1$, which implies that $\Omega_{a}$, $f_{PQ}/f_{a}$ or $-n_{T}$ must combine to reduce $\beta$ by at least a factor of $10^{-5}$.

\section{Implications for $f_{a}$}
Great care must be taken with these equations for $\beta$, however, since they have been derived under the assumption that the isocurvature perturbations are indeed small, i.e.,
\begin{equation}
  (\delta_{a})_{i} \sim \frac{H_{I}}{f_{a}\bar{\Theta}} \ll 1,
\end{equation}
We know this to be experimentally true, since COBE tells us that $\Delta T/T \sim 10^{-5}$.  In some models of inflation, however, $(\delta_{a})_{i}$ is not always much less than one.  If PQ symmetry is broken before inflation, for example, both chaotic and exponential inflation predict $(\delta_{a})_{i} \stackrel{>}{\sim} 0.1$ (see below).  Therefore, before using Equation~\ref{eq:powerl} or~\ref{eq:powerg} to constrain the axion or inflationary sector, it is first necessary to make sure that $(\delta_{a})_{i} \sim \frac{H_{I}}{f_{a}\bar{\Theta}} \ll 1$.  First, consider the term $f_{a}\bar{\Theta}$.  For simplicity, assume that $\Omega_{a}$ is ${\cal O}(1)$ and that PQ symmetry breaking occurs before inflation begins, in which case $f_{a}=f_{PQ}$.  Table~\ref{tab:ftheta} shows some representative values of $f\bar{\Theta}$ for various values of $f_{PQ}$.  Note that $\bar{\Theta}$ decreases from about 1 to about $10^{-3}$ as $f_{PQ}$ increases from $10^{12}$GeV to $m_{Pl}$.

\begin{table}
\begin{center}
\begin{tabular}{|c|c|c|} \hline
 $f_{PQ}$		&$\bar{\Theta}$		&$f\bar{\Theta}$	\\ \hline
 $10^{12}$GeV		&$\sim 1$		&$10^{12}$GeV		\\ \hline
 $10^{18}$GeV		&$\sim 10^{-3}$		&$10^{15}$GeV		\\ \hline 
 $m_{Pl}$		&$\sim 10^{-3}$		&$10^{15}$GeV		\\ \hline
\end{tabular}  

\end{center}
\caption{Some representative values of $f_{PQ}$, the initial misalignment angle $\bar{\Theta}$ and $f\bar{\Theta}$.}\label{tab:ftheta}
\end{table}

\begin{figure}
\begin{center}
\begin{picture}(0,0)%
\includegraphics{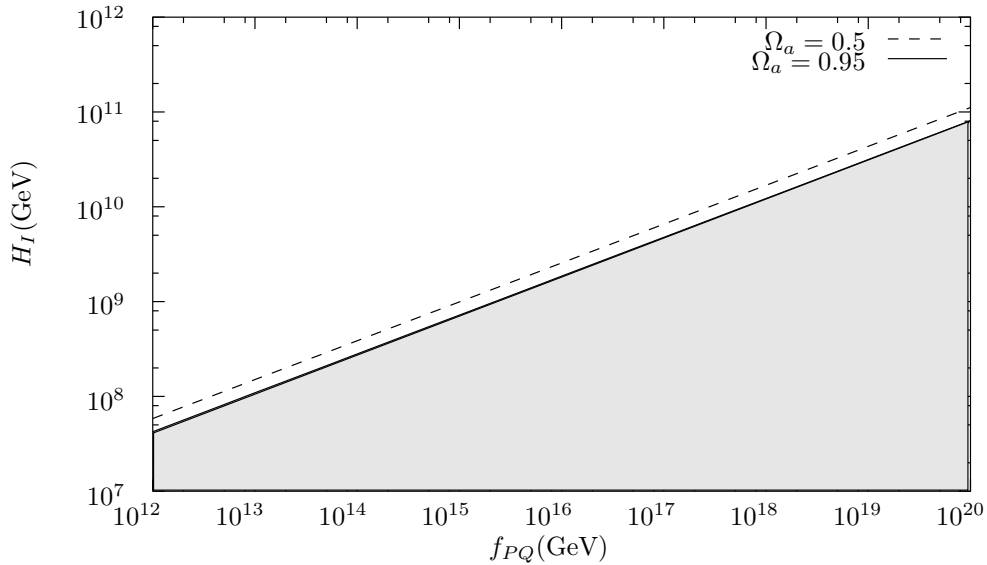}%
\end{picture}%
\setlength{\unitlength}{0.00083300in}%
\begingroup\makeatletter\ifx\SetFigFont\undefined%
\gdef\SetFigFont#1#2#3#4#5{%
  \reset@font\fontsize{#1}{#2pt}%
  \fontfamily{#3}\fontseries{#4}\fontshape{#5}%
  \selectfont}%
\fi\endgroup%
\begin{picture}(6269,3440)(304,-2967)
\put(425,-1137){\makebox(0,0)[b]{\smash{\SetFigFont{10}{12.0}{\familydefault}{\mddefault}{\updefault}\begin{rotate}{90}$H_{I}$(GeV)\end{rotate}}}}
\put(3659,-2937){\makebox(0,0)[b]{\smash{\SetFigFont{10}{12.0}{\familydefault}{\mddefault}{\updefault}$f_{PQ}$(GeV)}}}
\put(5634,216){\makebox(0,0)[rb]{\smash{\SetFigFont{10}{12.0}{\familydefault}{\mddefault}{\updefault}$\Omega_{a}=0.5$}}}
\put(5634, 92){\makebox(0,0)[rb]{\smash{\SetFigFont{10}{12.0}{\familydefault}{\mddefault}{\updefault}$\Omega_{a}=0.95$}}}
\put(1017,-2627){\makebox(0,0)[rb]{\smash{\SetFigFont{10}{12.0}{\familydefault}{\mddefault}{\updefault}$10^{7}$}}}
\put(1017,-2031){\makebox(0,0)[rb]{\smash{\SetFigFont{10}{12.0}{\familydefault}{\mddefault}{\updefault}$10^{8}$}}}
\put(1017,-1435){\makebox(0,0)[rb]{\smash{\SetFigFont{10}{12.0}{\familydefault}{\mddefault}{\updefault}$10^{9}$}}}
\put(1017,-839){\makebox(0,0)[rb]{\smash{\SetFigFont{10}{12.0}{\familydefault}{\mddefault}{\updefault}$10^{10}$}}}
\put(1017,-243){\makebox(0,0)[rb]{\smash{\SetFigFont{10}{12.0}{\familydefault}{\mddefault}{\updefault}$10^{11}$}}}
\put(1017,353){\makebox(0,0)[rb]{\smash{\SetFigFont{10}{12.0}{\familydefault}{\mddefault}{\updefault}$10^{12}$}}}
\put(1091,-2751){\makebox(0,0)[b]{\smash{\SetFigFont{10}{12.0}{\familydefault}{\mddefault}{\updefault}$10^{12}$}}}
\put(1733,-2751){\makebox(0,0)[b]{\smash{\SetFigFont{10}{12.0}{\familydefault}{\mddefault}{\updefault}$10^{13}$}}}
\put(2375,-2751){\makebox(0,0)[b]{\smash{\SetFigFont{10}{12.0}{\familydefault}{\mddefault}{\updefault}$10^{14}$}}}
\put(3017,-2751){\makebox(0,0)[b]{\smash{\SetFigFont{10}{12.0}{\familydefault}{\mddefault}{\updefault}$10^{15}$}}}
\put(3659,-2751){\makebox(0,0)[b]{\smash{\SetFigFont{10}{12.0}{\familydefault}{\mddefault}{\updefault}$10^{16}$}}}
\put(4301,-2751){\makebox(0,0)[b]{\smash{\SetFigFont{10}{12.0}{\familydefault}{\mddefault}{\updefault}$10^{17}$}}}
\put(4943,-2751){\makebox(0,0)[b]{\smash{\SetFigFont{10}{12.0}{\familydefault}{\mddefault}{\updefault}$10^{18}$}}}
\put(5585,-2751){\makebox(0,0)[b]{\smash{\SetFigFont{10}{12.0}{\familydefault}{\mddefault}{\updefault}$10^{19}$}}}
\put(6227,-2751){\makebox(0,0)[b]{\smash{\SetFigFont{10}{12.0}{\familydefault}{\mddefault}{\updefault}$10^{20}$}}}
\end{picture}
\caption[Allowed regions of $H_{I}-f_{PQ}$ parameter space]{The allowed values of $H_{I}$ fifty $e$-foldings before the end of inflation for various values of $f_{PQ}$, assuming PQ symmetry breaking occurs before inflation.  The top line corresponds to $\Omega_{a}=0.5$ and the bottom line to $\Omega_{a}=0.95$.  Here $\Omega_{b}=0.05$ and $h=0.5$.} \label{fig:Hf}
\end{center}
\end{figure}

Now consider $H_{I}$, the Hubble constant during inflation.  Since the power spectrum is normalized to present day horizon-sized scales, i.e., the quadrupole, the relevant value of $H_{I}$ is that when the present-day horizon-sized scales exited the inflationary horizon, which occurred roughly fifty to sixty $e$-foldings before the end of inflation.  The value of the Hubble constant at that time is constrained by the COBE quadrupole anisotropy, which, if fit to an $n=1$ scale invariant power spectrum, is  $Q = 18\mu$K \cite{Bennett, Gorski}.  This is because the temperature perturbation associated with the isocurvature perturbations is (cf. Equations~\ref{eq:dai} and ~\ref{eq:isoT})
\begin{equation}
  \frac{\Delta T}{T} = \frac{\sqrt{2}}{3}\frac{\Omega_{a}}{\Omega_{0}}\ \frac{H_{I}}{f_{a}\bar{\Theta}}
\end{equation}
Assuming PQ symmetry is broken before inflation, Figure~\ref{fig:Hf} shows the allowed values of $H_{I}$ as a function of $f_{PQ}$ for two values of $\Omega_{a}$.  Provided that $\Omega_{a}$ doesn't vary too much, neither does the allowed region of parameter space.  Note that when PQ symmetry breaking occurs prior to inflation, the Hubble constant must be less than about $10^{11}$GeV, even if $f_{PQ}$ ranges all the way up to the Planck mass.

The assumption that axions exist and that PQ symmetry is broken before inflation places quite a strong constraint on inflationary models.  This is because the COBE quadrupole can be used to normalize $V_{50}$, the inflaton potential fifty $e$-foldings before the end of inflation.  Turner \cite{Turnererice} finds that $V_{50}$ is
\begin{equation}
  V_{50} = 1.8\times 10^{-11} m_{Pl}^{4}x_{50}^{2},
\end{equation}
Since both $H_{I}$ and $\beta$ depend on $x_{50}^{2}$ (or $-n_{T}$), it is instructive to consider some typical inflationary models.  Table~\ref{tab:modelnT} shows the expected values of $x_{50}$ and $n_{T}$ for some standard models.  Following Turner \cite{Turnertilt}, we can use $x_{50}$ and $V_{50}$ to calculate $H_{I}$ and $(\delta_{a})_{i}$ for several inflation models, including chaotic inflation, extended inflation, new inflation and natural inflation.   

\begin{table}
\begin{center}
\begin{tabular}{|c|c|c|c|} \hline
 Inflation Model				&$x_{50}$			&$-n_{T}$	\\ \hline
 Chaotic (b=2)					&0.71				&0.02		\\ \hline
 Chaotic (b=4)					&1				&0.04		\\ \hline 
 Extended					&$\sqrt{\frac{64\pi}{2\omega+3}}$			&$\frac{8}{2\omega+3}$	\\ \hline 
 New						&$\ll 1$				&$\ll 1$	\\ \hline 
 Natural ($f \sim m_{Pl}/2$)			&0.06				&$\sim 10^{-4}$	\\ \hline 
 Natural ($f \sim m_{Pl}$)			&0.6				&0.01		\\ \hline 
\end{tabular}  
\end{center}
\caption{Values of the tensor spectral index for some typical inflationary models.}\label{tab:modelnT}
\end{table}

\subsection{Chaotic Inflation}
Chaotic inflation has a potential $V(\varphi) = A\varphi^{B}$, where $B=2$ corresponds to a model based upon a massive scalar field with $m^{2}=2A$ \cite{Belinsky,Jensen}, and $B=4$ corresponds to Linde's original model of chaotic inflation \cite{Lindechaotic}.  In these simple models, $\varphi$ is initially displaced from the minimum of the potential, and inflation occurs as the scalar field slowly rolls back to the minimum.  Fifty $e$-foldings before the end of inflation, $\varphi_{50}\approx 4m_{Pl}$, which leads to the following expressions for the $V_{50}$, $x_{50}$ and $n_{T}$:
\begin{eqnarray}
  V_{50} &= &Am_{Pl}^{B}\left[\frac{50B}{4\pi}\right]^{B/2}\\
  x_{50} &= &\left[\frac{4\pi B}{50}\right]^{1/2}\\
  n_{T}  &= &-\frac{B}{100}
\end{eqnarray}
If the potential is normalized to COBE,
\begin{equation}
  A = 1.6\times 10^{-11}B^{(1-B/2)}\left[\frac{4\pi}{50}\right]^{(\frac{B}{2}+1)}m_{Pl}^{(4-B)}
\end{equation}
For chaotic inflation, then, the Hubble constant fifty $e$-foldings before the end of inflation is
\begin{equation}
  H_{I} \simeq 1.2\times 10^{-5}m_{Pl} \sim 10^{14}\mbox{GeV}
\end{equation}
If PQ symmetry is broken before chaotic inflation begins, we can see from Table~\ref{tab:ftheta} that at a minimum, $(\delta_{a})_{i}$ is at least $10^{-1}$, which is far too large, especially considering that the quadrupole anisotropy $\Delta T = 18\mu$K.

\begin{figure}
\begin{center}
\begin{picture}(0,0)%
\includegraphics{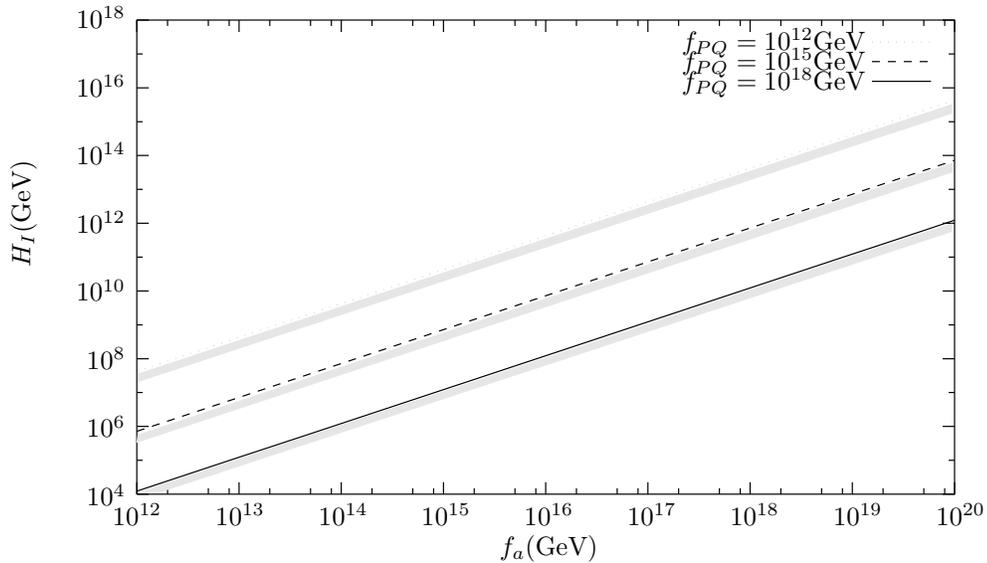}%
\end{picture}%
\setlength{\unitlength}{0.00083300in}%
\begingroup\makeatletter\ifx\SetFigFont\undefined%
\gdef\SetFigFont#1#2#3#4#5{%
  \reset@font\fontsize{#1}{#2pt}%
  \fontfamily{#3}\fontseries{#4}\fontshape{#5}%
  \selectfont}%
\fi\endgroup%
\begin{picture}(6115,3404)(304,-2967)
\put(1017,-2627){\makebox(0,0)[rb]{\smash{\SetFigFont{10}{12.0}{\familydefault}{\mddefault}{\updefault}$10^{4}$}}}
\put(1017,-2201){\makebox(0,0)[rb]{\smash{\SetFigFont{10}{12.0}{\familydefault}{\mddefault}{\updefault}$10^{6}$}}}
\put(1017,-1776){\makebox(0,0)[rb]{\smash{\SetFigFont{10}{12.0}{\familydefault}{\mddefault}{\updefault}$10^{8}$}}}
\put(1017,-1350){\makebox(0,0)[rb]{\smash{\SetFigFont{10}{12.0}{\familydefault}{\mddefault}{\updefault}$10^{10}$}}}
\put(1017,-924){\makebox(0,0)[rb]{\smash{\SetFigFont{10}{12.0}{\familydefault}{\mddefault}{\updefault}$10^{12}$}}}
\put(1017,-498){\makebox(0,0)[rb]{\smash{\SetFigFont{10}{12.0}{\familydefault}{\mddefault}{\updefault}$10^{14}$}}}
\put(1017,-73){\makebox(0,0)[rb]{\smash{\SetFigFont{10}{12.0}{\familydefault}{\mddefault}{\updefault}$10^{16}$}}}
\put(1017,353){\makebox(0,0)[rb]{\smash{\SetFigFont{10}{12.0}{\familydefault}{\mddefault}{\updefault}$10^{18}$}}}
\put(1091,-2751){\makebox(0,0)[b]{\smash{\SetFigFont{10}{12.0}{\familydefault}{\mddefault}{\updefault}$10^{12}$}}}
\put(1733,-2751){\makebox(0,0)[b]{\smash{\SetFigFont{10}{12.0}{\familydefault}{\mddefault}{\updefault}$10^{13}$}}}
\put(2375,-2751){\makebox(0,0)[b]{\smash{\SetFigFont{10}{12.0}{\familydefault}{\mddefault}{\updefault}$10^{14}$}}}
\put(3017,-2751){\makebox(0,0)[b]{\smash{\SetFigFont{10}{12.0}{\familydefault}{\mddefault}{\updefault}$10^{15}$}}}
\put(3659,-2751){\makebox(0,0)[b]{\smash{\SetFigFont{10}{12.0}{\familydefault}{\mddefault}{\updefault}$10^{16}$}}}
\put(4301,-2751){\makebox(0,0)[b]{\smash{\SetFigFont{10}{12.0}{\familydefault}{\mddefault}{\updefault}$10^{17}$}}}
\put(4943,-2751){\makebox(0,0)[b]{\smash{\SetFigFont{10}{12.0}{\familydefault}{\mddefault}{\updefault}$10^{18}$}}}
\put(5585,-2751){\makebox(0,0)[b]{\smash{\SetFigFont{10}{12.0}{\familydefault}{\mddefault}{\updefault}$10^{19}$}}}
\put(6227,-2751){\makebox(0,0)[b]{\smash{\SetFigFont{10}{12.0}{\familydefault}{\mddefault}{\updefault}$10^{20}$}}}
\put(5634,216){\makebox(0,0)[rb]{\smash{\SetFigFont{10}{12.0}{\familydefault}{\mddefault}{\updefault}$f_{PQ}=10^{12}$GeV}}}
\put(3659,-2937){\makebox(0,0)[b]{\smash{\SetFigFont{10}{12.0}{\familydefault}{\mddefault}{\updefault}$f_{a}$(GeV)}}}
\put(425,-1137){\makebox(0,0)[b]{\smash{\SetFigFont{10}{12.0}{\familydefault}{\mddefault}{\updefault}\begin{rotate}{90}$H_{I}$(GeV)\end{rotate}}}}
\put(5634, 92){\makebox(0,0)[rb]{\smash{\SetFigFont{10}{12.0}{\familydefault}{\mddefault}{\updefault}$f_{PQ}=10^{15}$GeV}}}
\put(5634,-32){\makebox(0,0)[rb]{\smash{\SetFigFont{10}{12.0}{\familydefault}{\mddefault}{\updefault}$f_{PQ}=10^{18}$GeV}}}
\end{picture}
\caption[Allowed regions of $H-f_{a}$ parameter space]{The allowed values of $H$ 50 $e$-foldings before the end of inflation for various values of $f_{a}$.  The top line corresponds to $f_{PQ}=10^{12}$GeV, the middle line to $f_{PQ}=10^{15}$GeV and the bottom line to $f_{PQ}=10^{18}$GeV.  Here $\Omega_{b}=0.05$ and $h=0.5$.} \label{fig:varyfa}
\end{center}
\end{figure}

Linde \cite{Lindeaxion}, however, has suggested that if $f_{a}$ varies with time during inflation, it is possible to produce acceptable temperature fluctuations.  In the simplest case, consider the possibility that the radial component of the PQ field is the inflaton field.  Then, in the context of chaotic inflation, fifty $e$-foldings before the end of inflation, $f_{a}=\varphi_{50}=4m_{Pl}$.  If $\bar{\Theta} \approx 1$, corresponding to $f_{PQ}=10^{12}$GeV, the isocurvature perturbation $(\delta_{a})_{i} \approx 10^{-6}$, which now easily satisfies the COBE constraints.  Figure~\ref{fig:varyfa} shows the values of $H_{I}$ allowed by the COBE quadrupole if $f_{a}$ varies with time.  In this scenario, if $f_{a} \stackrel{>}{\sim} 10^{18}$GeV, $H_{I}$ may easily take on the chaotic inflation value of $10^{14}$GeV.  Notice, however, that this scenario still generally requires a small value of the Hubble constant.  This is because a large Hubble constant correlates to a small value of $f_{PQ}$ ($\stackrel{<}{\sim} 10^{15}$GeV) but a rather large value of $f_{a}$ ($\stackrel{>}{\sim} 10^{18}$GeV).  If $f_{PQ}=10^{12}$GeV, for example, $H_{I}$ can only take on the chaotic inflation value of about $10^{14}$GeV if $f_{a}\stackrel{>}{\sim}10^{18}$GeV.  Physically, this means that the PQ symmetry breaking phase transition would have to span an energy range of about six orders of magnitude, a very long phase transition indeed!  Still working within the context of the chaotic inflation scenario, a more reasonable choice of $f_{PQ}=10^{18}$GeV requires a very small Hubble constant, $H_{I} \stackrel{<}{\sim} 10^{10}$GeV, as can be seen in Figure~\ref{fig:varyfa}.  So although making $f_{a}$ a dynamical variable can produce isocurvature fluctuations of an acceptable magnitude, it also seems somewhat unnatural.

\subsection{Exponential Inflation}
Exponential inflation has a potential $V(\varphi) = V_{0}\exp(-b\varphi/m_{Pl})$.  This type of potential was first invoked in power-law inflation \cite{Abbott,Lucchin,Fabbri} and later in extended inflation \cite{extended}, which is based upon a Brans-Dicke-Jordan theory of gravity with $b^{2}=64\pi /(2\omega+3)$.  Normalizing to COBE, $V_{50}$, $x_{50}$ and $n_{T}$ for extended inflation are
\begin{eqnarray}
  V_{50} &= &\frac{1.6 \times 10^{-11}m_{Pl}^{4}b^{2}}{1+0.28b^{2}}\\
  x_{50} &= &-b\\
  n_{T}  &= &-\frac{8}{(2\omega+3)}
\end{eqnarray}
Since large scale structure formation requires $\omega < 20$,\footnote{Note that this is in strong disagreement with observational measurements, which give $\omega > 500$!} these models have a significant amount of tilt, $-n_{T}\stackrel{>}{\sim} 0.3$.  Fifty $e$-foldings before the end of inflation, the Hubble constant is
\begin{equation}
  H_{I} \simeq 2\times 10^{-5}m_{Pl} \sim 10^{14}\mbox{GeV}
\end{equation}
Notice that $H_{I}$ for exponential inflation is roughly the same as that for chaotic inflation.  This means that if PQ symmetry breaking occurs before inflation takes place, exponential inflation suffers from the same problem as chaotic inflation, i.e., $(\delta_{a})_{i} \stackrel{>}{\sim} 10^{-1}$.

Although it seems unlikely that the PQ field could play the role of the Brans-Dicke-Jordan scalar field, one can imagine that the PQ field is chaotically distributed, like the inflaton in the chaotic inflation model.  In this case, it could be that the PQ field just happens to be rolling back down to its SSB value of $f_{PQ}$ as exponential inflation is occurring.  In that case, it is not inconceivable that isocurvature perturbations of an acceptable magnitude could be produced as described in the previous section on chaotic inflation.  But since exponential inflation also predicts a Hubble constant $H_{I} \approx 10^{14}$GeV, such a solution seems as unnatural as that of chaotic inflation.

\subsection{New Inflation}
\begin{figure}
\begin{center}
\begin{picture}(0,0)%
\includegraphics{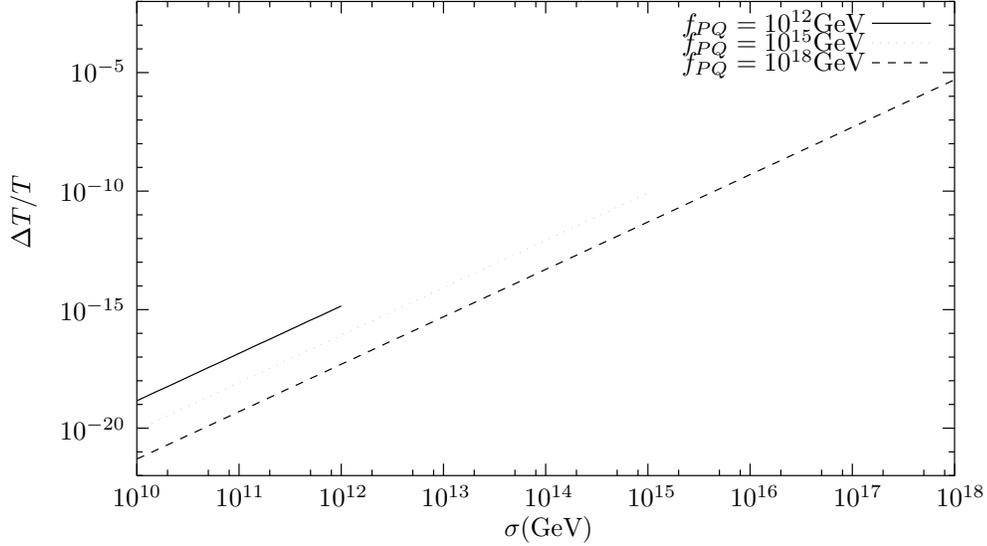}%
\end{picture}%
\setlength{\unitlength}{0.00083300in}%
\begingroup\makeatletter\ifx\SetFigFont\undefined%
\gdef\SetFigFont#1#2#3#4#5{%
  \reset@font\fontsize{#1}{#2pt}%
  \fontfamily{#3}\fontseries{#4}\fontshape{#5}%
  \selectfont}%
\fi\endgroup%
\begin{picture}(6115,3394)(304,-2967)
\put(1017,-2329){\makebox(0,0)[rb]{\smash{\SetFigFont{10}{12.0}{\familydefault}{\mddefault}{\updefault}$10^{-20}$}}}
\put(1017,-1584){\makebox(0,0)[rb]{\smash{\SetFigFont{10}{12.0}{\familydefault}{\mddefault}{\updefault}$10^{-15}$}}}
\put(1017,-839){\makebox(0,0)[rb]{\smash{\SetFigFont{10}{12.0}{\familydefault}{\mddefault}{\updefault}$10^{-10}$}}}
\put(1017,-94){\makebox(0,0)[rb]{\smash{\SetFigFont{10}{12.0}{\familydefault}{\mddefault}{\updefault}$10^{-5}$}}}
\put(1091,-2751){\makebox(0,0)[b]{\smash{\SetFigFont{10}{12.0}{\familydefault}{\mddefault}{\updefault}$10^{10}$}}}
\put(1733,-2751){\makebox(0,0)[b]{\smash{\SetFigFont{10}{12.0}{\familydefault}{\mddefault}{\updefault}$10^{11}$}}}
\put(2375,-2751){\makebox(0,0)[b]{\smash{\SetFigFont{10}{12.0}{\familydefault}{\mddefault}{\updefault}$10^{12}$}}}
\put(3017,-2751){\makebox(0,0)[b]{\smash{\SetFigFont{10}{12.0}{\familydefault}{\mddefault}{\updefault}$10^{13}$}}}
\put(3659,-2751){\makebox(0,0)[b]{\smash{\SetFigFont{10}{12.0}{\familydefault}{\mddefault}{\updefault}$10^{14}$}}}
\put(4301,-2751){\makebox(0,0)[b]{\smash{\SetFigFont{10}{12.0}{\familydefault}{\mddefault}{\updefault}$10^{15}$}}}
\put(4943,-2751){\makebox(0,0)[b]{\smash{\SetFigFont{10}{12.0}{\familydefault}{\mddefault}{\updefault}$10^{16}$}}}
\put(5585,-2751){\makebox(0,0)[b]{\smash{\SetFigFont{10}{12.0}{\familydefault}{\mddefault}{\updefault}$10^{17}$}}}
\put(6227,-2751){\makebox(0,0)[b]{\smash{\SetFigFont{10}{12.0}{\familydefault}{\mddefault}{\updefault}$10^{18}$}}}
\put(425,-1137){\makebox(0,0)[b]{\smash{\SetFigFont{10}{12.0}{\familydefault}{\mddefault}{\updefault}\begin{rotate}{90}$\Delta T/T$\end{rotate}}}}
\put(3659,-2937){\makebox(0,0)[b]{\smash{\SetFigFont{10}{12.0}{\familydefault}{\mddefault}{\updefault}$\sigma$(GeV)}}}
\put(5634,216){\makebox(0,0)[rb]{\smash{\SetFigFont{10}{12.0}{\familydefault}{\mddefault}{\updefault}$f_{PQ}=10^{12}$GeV}}}
\put(5634, 92){\makebox(0,0)[rb]{\smash{\SetFigFont{10}{12.0}{\familydefault}{\mddefault}{\updefault}$f_{PQ}=10^{15}$GeV}}}
\put(5634,-32){\makebox(0,0)[rb]{\smash{\SetFigFont{10}{12.0}{\familydefault}{\mddefault}{\updefault}$f_{PQ}=10^{18}$GeV}}}
\end{picture}
\caption[$\Delta T/T$ for isocurvature fluctuations in the new inflationary scenario]{Possible values of $\Delta T/T$ for isocurvature perturbations as a function of $\sigma$, the expectation value of the inflaton field in the new inflation scenario.  The cutoffs are the result of requiring $f_{PQ} \geq \sigma$.  Here $\Omega_{b}=0.05$, $\Omega_{a}=0.95$ and $h=0.5$.} \label{fig:new}
\end{center}
\end{figure}

New inflation \cite{Lindenew,Albrecht} has a very flat potential that is based upon the Coleman-Weinberg potential:
\begin{equation}
 V(\varphi) = \frac{C\sigma^{4}}{2} + C\varphi^{4}\left [\ln\left(\frac{\varphi^{2}}{\sigma^{2}}\right)-\frac{1}{2}\right],
\end{equation}
where $C$ is typically of the order of $10^{-15}$.  During inflation, the scalar field rolls from $\varphi\approx 0$ to the spontaneous symmetry breaking value of $\varphi = \sigma$.  50 $e$-foldings before the end of inflation, $\varphi_{50} \sim 5\times 10^{-5}\sigma$ and thus
\begin{eqnarray}
  V_{50} &\simeq &\frac{C\sigma^{4}}{2}\\
  x_{50} &\simeq &-\frac{(\pi/25)^{3/2}}{\sqrt{|\ln(\varphi_{50}^{2}/\sigma^{2})|}}\left(\frac{\sigma}{m_{Pl}}\right)^{2} \approx 0.01\left(\frac{\sigma}{m_{Pl}}\right)^{2}\ll 1\\
  n_{T}  &\approx &-4 \times 10^{-6} \left(\frac{\sigma}{m_{Pl}}\right)^{4}\ll 1
\end{eqnarray}
At that time, the Hubble constant is
\begin{equation}
  H_{I} \simeq 10^{12}\left(\frac{\sigma}{m_{Pl}}\right)^{2}\mbox{GeV}
\end{equation}
If we assume that PQ symmetry breaking occurs before inflation starts, then it is reasonable to assume that $f_{PQ} \geq \sigma$.  Figure~\ref{fig:new} shows $\Delta T/T$ as a function of $\sigma$ for various values of $f_{PQ}$.  As can be seen from this plot, only if $f_{PQ}$ and $\sigma$ are near the Planck mass will the isocurvature perturbations make a significant contribution.  This is not surprising, since $H \propto (\sigma/m_{Pl})^{2}$.  Typically, however, new inflation is associated with a GUT phase transition, in which case $\sigma \sim 10^{16}$GeV, and so isocurvature perturbations are highly suppressed.  

Suppose, however, that PQ symmetry is broken during inflation, as in the model of Pi \cite{Pi}.  Then $f_{a}<f_{PQ}/\sqrt{2}$.  This will increase the amplitude of the isocurvature perturbations, since $\Delta T/T \propto 1/f_{a}$.  In Pi's model, for example, $f_{PQ} \simeq 10^{18}$GeV and $f_{a}\simeq f_{PQ}/300$.  This increases $\Delta T/T$ by a factor of $\approx 10^{3}$.  If $\sigma \sim 10^{16}$GeV, isocurvature perturbations with an amplitude $\sim 10^{-6}$ are easily produced.  As we will see, isocurvature perturbations of this amplitude are entirely compatible with the observed microwave background anisotropy.  

\subsection{Natural Inflation}
Finally, consider natural inflation \cite{natural}, which has the potential
\begin{equation}
 V(\varphi) = \Lambda^{4}[1+\cos(\varphi/f)]
\end{equation}
Here $f$ is the energy scale of spontaneous symmetry breakdown and $\Lambda$ is the scale of explicit symmetry breaking.  In this case the flatness of the potential is due to the fact that $\varphi$ is a Nambu-Goldstone boson \cite{natural,Turnertilt}.  There are two simplifying regimes:  $f\gg m_{Pl}$ and $f\stackrel{<}{\sim}m_{Pl}$.  In the first regime, inflation occurs near the minimum of the potential ($\sigma=\pi f$), and the potential can be expanded about $\varphi=\sigma$.  Defining $\varphi_{\sigma}=\varphi-\sigma$,  the potential is $V(\varphi_{\sigma})=m^{2}\varphi_{\sigma}^{2}/2$.  This is simply the same potential as chaotic inflation with $B=2$, and the same results apply, i.e., $H_{I} \sim 10^{14}\mbox{GeV}$, and, if PQ symmetry is broken before the start of inflation, $(\delta_{a})_{i} \stackrel{>}{\sim} 10^{-1}$.

\begin{figure}
\begin{center}
\begin{picture}(0,0)%
\includegraphics{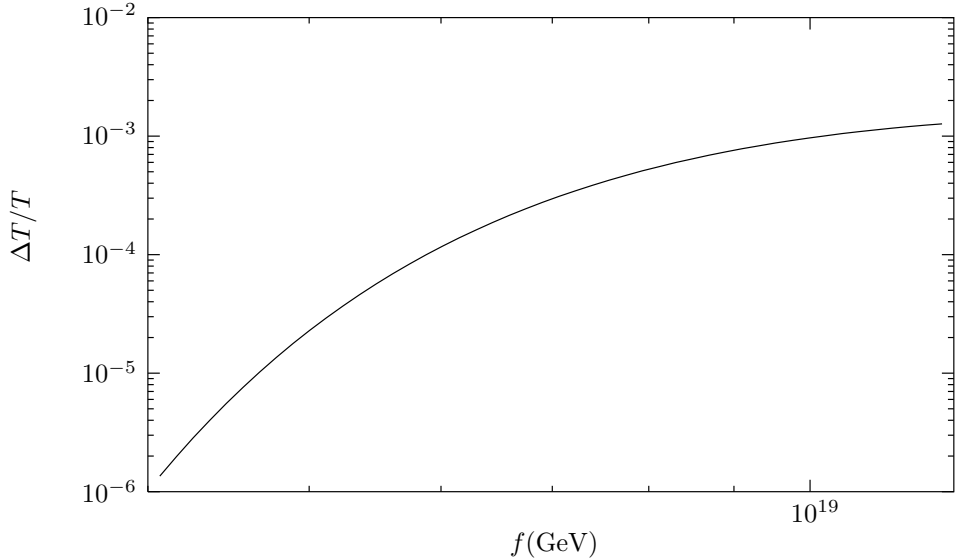}%
\end{picture}%
\setlength{\unitlength}{0.00083300in}%
\begingroup\makeatletter\ifx\SetFigFont\undefined%
\gdef\SetFigFont#1#2#3#4#5{%
  \reset@font\fontsize{#1}{#2pt}%
  \fontfamily{#3}\fontseries{#4}\fontshape{#5}%
  \selectfont}%
\fi\endgroup%
\begin{picture}(6019,3512)(220,-2991)
\put(1091,-2627){\makebox(0,0)[rb]{\smash{\SetFigFont{10}{12.0}{\familydefault}{\mddefault}{\updefault}$10^{-6}$}}}
\put(1091,-1882){\makebox(0,0)[rb]{\smash{\SetFigFont{10}{12.0}{\familydefault}{\mddefault}{\updefault}$10^{-5}$}}}
\put(1091,-1137){\makebox(0,0)[rb]{\smash{\SetFigFont{10}{12.0}{\familydefault}{\mddefault}{\updefault}$10^{-4}$}}}
\put(1091,-392){\makebox(0,0)[rb]{\smash{\SetFigFont{10}{12.0}{\familydefault}{\mddefault}{\updefault}$10^{-3}$}}}
\put(1091,353){\makebox(0,0)[rb]{\smash{\SetFigFont{10}{12.0}{\familydefault}{\mddefault}{\updefault}$10^{-2}$}}}
\put(5324,-2751){\makebox(0,0)[b]{\smash{\SetFigFont{10}{12.0}{\familydefault}{\mddefault}{\updefault}$10^{19}$}}}
\put(425,-1137){\makebox(0,0)[b]{\smash{\SetFigFont{10}{12.0}{\familydefault}{\mddefault}{\updefault}\begin{rotate}{90}$\Delta T/T$\end{rotate}}}}
\put(3696,-2937){\makebox(0,0)[b]{\smash{\SetFigFont{10}{12.0}{\familydefault}{\mddefault}{\updefault}$f$(GeV)}}}
\end{picture}
\caption[$\Delta T/T$ for isocurvature fluctuations in the natural inflationary scenario with $f \stackrel{<}{\sim} m_{Pl}$]{Possible values of $\Delta T/T$ for isocurvature perturbations as a function of $f$, the expectation value of the inflaton field in the natural inflation scenario.  Here $\Omega_{b}=0.05$, $\Omega_{a}=0.95$ and $h=0.5$.} \label{fig:natural}
\end{center}
\end{figure}

In the other regime, $f \stackrel{<}{\sim} m_{Pl}$.  In this case, sufficient inflation requires $f$ to be greater than about $m_{Pl}/3$, which some consider unnatural, since $ m_{Pl}/3 \stackrel{<}{\sim} f \stackrel{<}{\sim} m_{Pl}$ is a somewhat constricted region of parameter space.  On the other hand, it doesn't really seem sensible to talk about physics above the Planck mass either.  For this scenario, $\varphi_{50} \stackrel{<}{\sim} 0.1(\pi f)$, and $V_{50} \approx 2\Lambda^{4}$, where
\begin{equation}
  \Lambda = 6.7\times 10^{-4}\sqrt{\frac{m_{Pl}}{f}}\exp \left(-\frac{25 m_{Pl}^{2}}{16\pi f^{2}}\right)
\end{equation}
As before, suppose that PQ symmetry breaking happens before inflation.  In that case, we would expect that $f_{PQ} \sim m_{Pl}$.  To illustrate, consider two examples, $f\sim m_{Pl}$ and $f\sim \frac{1}{2}m_{Pl}$.  In the first case, $x_{50} \sim 0.6$, $n_{T} \sim 0.01$, $H_{I}\sim 10^{13}$GeV, $(\delta_{a})_{i} \stackrel{>}{\sim} 10^{-3}$ and $\Delta T/T \sim 10^{-4}$.  Clearly, the microwave background anisotropies are too large for $f \sim m_{Pl}$.  In the second case, $x_{50} \sim 0.06$, $n_{T} \sim 10^{-4}$, $H_{I}\sim 10^{12}$GeV, $(\delta_{a})_{i} \stackrel{>}{\sim} 10^{-5}$ and $\Delta T/T \sim 10^{-6}$, which is much more reasonable.  So if PQ symmetry breaking occurs before natural inflation, it is entirely possible to produce isocurvature perturbations that give sensible values of $\Delta T/T$.  Figure~\ref{fig:natural} shows $\Delta T/T$ as a function of $f$ for the ``unnatural" regime of natural inflation.  Notice that for a rather limited range of $f$ values, $m_{Pl}/3 \stackrel{<}{\sim} f \stackrel{<}{\sim} 5 \times 10^{18}$GeV, it is quite possible to have isocurvature perturbations which are consistent with the COBE data.

\section{Observational Consequences of Axions} 

Several authors have considered the question of isocurvature perturbations.  Efstathiou and Bond \cite{EB} considered the possibility that axions make up all of the cold dark matter in the case that isocurvature perturbations have the same amplitude as the adiabatic perturbations.  Stompor et al. \cite{Stompor} and Kawasaki et al. \cite{Kawasaki} considered arbitrary mixtures of isentropic and isocurvature perturbations in the case that axions make up all of the cold dark matter.  Here we will consider the effects of varying both the axion content and the mix of fluctuations.  Then we will describe how this may be used to probe axion physics and inflation.

\subsection{Axions and CMB Anisotropies}
Efstathiou and Bond \cite{EB} found that the most significant effect of isocurvature fluctuations is a boost in the anisotropy on large angular scales.  Recall from Equations~\ref{eq:isophoton} and~\ref{eq:isoaxion} that in the matter-dominated era,
\begin{eqnarray}
  \frac{\delta T}{T} &\rightarrow &-\frac{1}{3}\frac{\Omega_{a}}{\Omega_{0}} (\delta_{a})_{i}\nonumber \\
  \frac{\delta n_{a}}{n_{a}} &\rightarrow &\left(\frac{\Omega_{0}-\Omega_{a}}{\Omega_{0}}\right)(\delta_{a})_{i} < (\delta_{a})_{i}\nonumber
\end{eqnarray}
In order not to affect the curvature of the universe, the isocurvature axion perturbation must be transferred to the radiation as the universe becomes matter dominated.  This leads to the above boost in the photon anisotropy on large angular scales.

To get a sense of the effect of isocurvature perturbations on the CMB, we will ignore any possible tensor contribution to the quadrupole, i.e., we will assume that $-n_{T} \ll 1$.  (Later we will relax this assumption.)  In this case, the temperature perturbation is the sum of the isocurvature boost and the ordinary Sachs-Wolfe effect.  Suppose that $\beta=1$.  Then, the contribution to the isocurvature quadrupole is
\begin{equation}
  C_{2}^{iso} \propto \left(\frac{4}{3}\frac{\Omega_{a}}{\Omega_{0}}+\gamma\right)^{2}, \label{eq:C2iso}
\end{equation}
where $\gamma=4/15$ describes the ordinary Sachs-Wolfe effect.  By comparison, isentropic perturbations only get a contribution from the ordinary Sachs-Wolfe effect, and so $C_{2}^{isen} \propto \gamma^{2}.$

Provided that $\Omega_{a}\simeq\Omega_{0}$, the ratio of the two quadrupoles is $C_{2}^{iso}/C_{2}^{isen}=36$.  Since $C_{l} \approx (\Delta T/T)^{2}$, this means that on large angular scales, $\Delta T/T$ is approximately six times larger for isocurvature perturbations than it is for adiabatic perturbations.  Efstathiou and Bond used this fact to argue against a purely isocurvature CDM model, since when the fluctuation spectrum is normalized to $J_{3}$(10$h^{-1}$Mpc), the second moment of the two-point galaxy correlation function, $\Delta T/T$ is approximately six times too large.

Stompor et al. \cite{Stompor} and Kawasaki et al. \cite{Kawasaki} investigated the possibility that the amplitude of the isocurvature perturbations is not equal to the amplitude of the adiabatic perturbations.  Stompor et al. took a model-independent approach, considering the possibility that there might be an isocurvature perturbation component of the power spectrum without speculating about its origin.  To describe this they introduced a mixing parameter $\alpha$, which interpolates between a purely isocurvature power spectrum $P_{iso}(k)$ and a purely isentropic power spectrum $P_{isen}(k)$:
\begin{equation}
  P(k) = \alpha P_{isen}(k) + (1-\alpha)P_{iso}(k),
\end{equation}
where $0\leq\alpha\leq1$.  Kawasaki et al. worked within the context of chaotic inflation and they parametrized their models by the ratio of the power spectra, $P_{iso}(k)/P_{isen}(k)$.  We shall follow the philosophy of Stompor et al., but define the power spectrum in the same way as Kawasaki et al.:
\begin{equation}
  C_{2} = C_{2}^{isen}\left(1 + \beta\frac{C_{2}^{iso}}{C_{2}^{isen}}\right) \label{eq:C2SI}
\end{equation}
Since $C_{2}^{iso} \propto A_{iso}^{2}(k)(\frac{4}{3}\frac{\Omega_{a}}{\Omega_{0}}+\gamma)^{2}$ and $C_{2}^{isen} \propto A^{2}_{isen}\gamma^{2}$, $\beta$ is simply the ratio of the square of the amplitudes of the isocurvature to adiabatic perturbations, as defined in Section 3:
\begin{equation}
  \beta = \left[\frac{A_{iso}(k)}{A_{isen}(k)}\right]^{2}
\end{equation}

\begin{figure}
\begin{center}
\begin{picture}(0,0)%
\includegraphics{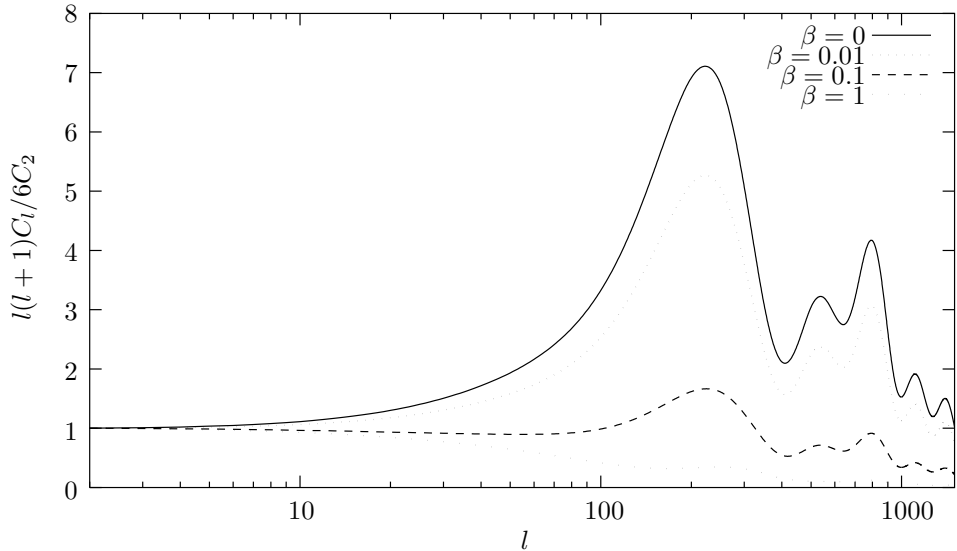}%
\end{picture}%
\setlength{\unitlength}{0.00083300in}%
\begingroup\makeatletter\ifx\SetFigFont\undefined%
\gdef\SetFigFont#1#2#3#4#5{%
  \reset@font\fontsize{#1}{#2pt}%
  \fontfamily{#3}\fontseries{#4}\fontshape{#5}%
  \selectfont}%
\fi\endgroup%
\begin{picture}(5935,3401)(304,-2964)
\put(721,-2627){\makebox(0,0)[rb]{\smash{\SetFigFont{10}{12.0}{\familydefault}{\mddefault}{\updefault}0}}}
\put(721,-2254){\makebox(0,0)[rb]{\smash{\SetFigFont{10}{12.0}{\familydefault}{\mddefault}{\updefault}1}}}
\put(721,-1882){\makebox(0,0)[rb]{\smash{\SetFigFont{10}{12.0}{\familydefault}{\mddefault}{\updefault}2}}}
\put(721,-1509){\makebox(0,0)[rb]{\smash{\SetFigFont{10}{12.0}{\familydefault}{\mddefault}{\updefault}3}}}
\put(721,-1137){\makebox(0,0)[rb]{\smash{\SetFigFont{10}{12.0}{\familydefault}{\mddefault}{\updefault}4}}}
\put(721,-764){\makebox(0,0)[rb]{\smash{\SetFigFont{10}{12.0}{\familydefault}{\mddefault}{\updefault}5}}}
\put(721,-392){\makebox(0,0)[rb]{\smash{\SetFigFont{10}{12.0}{\familydefault}{\mddefault}{\updefault}6}}}
\put(721,-19){\makebox(0,0)[rb]{\smash{\SetFigFont{10}{12.0}{\familydefault}{\mddefault}{\updefault}7}}}
\put(721,353){\makebox(0,0)[rb]{\smash{\SetFigFont{10}{12.0}{\familydefault}{\mddefault}{\updefault}8}}}
\put(2116,-2751){\makebox(0,0)[b]{\smash{\SetFigFont{10}{12.0}{\familydefault}{\mddefault}{\updefault}10}}}
\put(4005,-2751){\makebox(0,0)[b]{\smash{\SetFigFont{10}{12.0}{\familydefault}{\mddefault}{\updefault}100}}}
\put(5894,-2751){\makebox(0,0)[b]{\smash{\SetFigFont{10}{12.0}{\familydefault}{\mddefault}{\updefault}1000}}}
\put(425,-1509){\makebox(0,0)[b]{\smash{\SetFigFont{10}{12.0}{\familydefault}{\mddefault}{\updefault}\begin{rotate}{90}$l(l+1)C_{l}/6C_{2}$\end{rotate}}}}
\put(3511,-2937){\makebox(0,0)[b]{\smash{\SetFigFont{10}{12.0}{\familydefault}{\mddefault}{\updefault}$l$}}}
\put(5634,216){\makebox(0,0)[rb]{\smash{\SetFigFont{10}{12.0}{\familydefault}{\mddefault}{\updefault}$\beta=0$}}}
\put(5634, 92){\makebox(0,0)[rb]{\smash{\SetFigFont{10}{12.0}{\familydefault}{\mddefault}{\updefault}$\beta=0.01$}}}
\put(5634,-32){\makebox(0,0)[rb]{\smash{\SetFigFont{10}{12.0}{\familydefault}{\mddefault}{\updefault}$\beta=0.1$}}}
\put(5634,-156){\makebox(0,0)[rb]{\smash{\SetFigFont{10}{12.0}{\familydefault}{\mddefault}{\updefault}$\beta=1$}}}
\end{picture}
\caption[$l(l+1)C_{l}$:  Varying $\beta$]{The angular power spectrum for various values of $\beta$.  From top to bottom, $\beta$=0 (standard CDM with only adiabatic perturbations), 0.01, 0.1 and 1.  In this plot $\Omega_{b}=0.096, h=0.5$ and $\Omega_{a}=0.904$.} \label{fig:varyalpha}
\end{center}
\end{figure}

Figure~\ref{fig:varyalpha} shows the effect of varying the mixing parameter $\beta$ from zero to one.  From the plots of the anisotropy, we can see that two generic features of adding an isocurvature component are:
\begin{enumerate}
\item  A flattening of the angular power spectrum on large angular scales and a decrease in the heights of the Doppler peaks relative to the standard CDM scenario.  This flattening of the angular power spectrum is due to the isocurvature boost on large angular scales, as described above.  Since the smaller angular scales do not experience this effect, the Doppler peaks are suppressed relative to the quadrupole.

\item  The Doppler peaks for isocurvature perturbations are shifted with respect to the Doppler peaks for isentropic perturbations:  the peaks and troughs are switched.  This is because the isocurvature oscillations start out with $\delta \rho=0$ and therefore oscillate like a sine function, whereas the isentropic oscillations start out with $\delta \rho\neq 0$ and thus oscillate like a cosine function.
\end{enumerate}
These are features which should be resolvable by the next generation of satellite experiments.\footnote{If the universe were somehow reionized after recombination, the new Thomson scattering would tend to wipe out the anisotropy generated at recombination, lowering the Doppler peaks.  It is worth mentioning that the lowering of the peaks in the isocurvature case appears to be distinguishable from that produced by reionization \cite{Stompor}.}

\begin{figure}
\begin{center}
\begin{picture}(0,0)%
\includegraphics{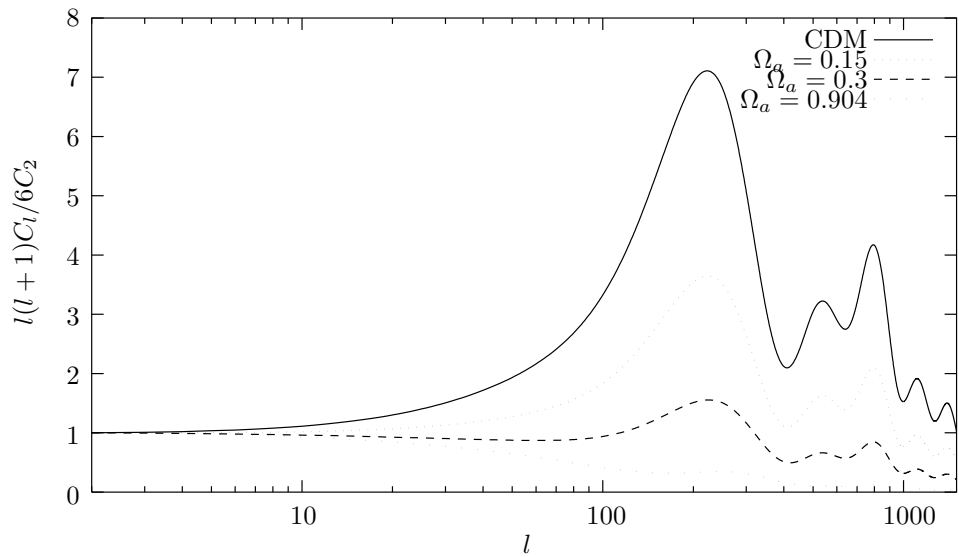}%
\end{picture}%
\setlength{\unitlength}{0.00083300in}%
\begingroup\makeatletter\ifx\SetFigFont\undefined%
\gdef\SetFigFont#1#2#3#4#5{%
  \reset@font\fontsize{#1}{#2pt}%
  \fontfamily{#3}\fontseries{#4}\fontshape{#5}%
  \selectfont}%
\fi\endgroup%
\begin{picture}(5935,3401)(304,-2964)
\put(721,-2627){\makebox(0,0)[rb]{\smash{\SetFigFont{10}{12.0}{\familydefault}{\mddefault}{\updefault}0}}}
\put(721,-2254){\makebox(0,0)[rb]{\smash{\SetFigFont{10}{12.0}{\familydefault}{\mddefault}{\updefault}1}}}
\put(721,-1882){\makebox(0,0)[rb]{\smash{\SetFigFont{10}{12.0}{\familydefault}{\mddefault}{\updefault}2}}}
\put(721,-1509){\makebox(0,0)[rb]{\smash{\SetFigFont{10}{12.0}{\familydefault}{\mddefault}{\updefault}3}}}
\put(721,-1137){\makebox(0,0)[rb]{\smash{\SetFigFont{10}{12.0}{\familydefault}{\mddefault}{\updefault}4}}}
\put(721,-764){\makebox(0,0)[rb]{\smash{\SetFigFont{10}{12.0}{\familydefault}{\mddefault}{\updefault}5}}}
\put(721,-392){\makebox(0,0)[rb]{\smash{\SetFigFont{10}{12.0}{\familydefault}{\mddefault}{\updefault}6}}}
\put(721,-19){\makebox(0,0)[rb]{\smash{\SetFigFont{10}{12.0}{\familydefault}{\mddefault}{\updefault}7}}}
\put(721,353){\makebox(0,0)[rb]{\smash{\SetFigFont{10}{12.0}{\familydefault}{\mddefault}{\updefault}8}}}
\put(2116,-2751){\makebox(0,0)[b]{\smash{\SetFigFont{10}{12.0}{\familydefault}{\mddefault}{\updefault}10}}}
\put(4005,-2751){\makebox(0,0)[b]{\smash{\SetFigFont{10}{12.0}{\familydefault}{\mddefault}{\updefault}100}}}
\put(5894,-2751){\makebox(0,0)[b]{\smash{\SetFigFont{10}{12.0}{\familydefault}{\mddefault}{\updefault}1000}}}
\put(425,-1509){\makebox(0,0)[b]{\smash{\SetFigFont{10}{12.0}{\familydefault}{\mddefault}{\updefault}\begin{rotate}{90}$l(l+1)C_{l}/6C_{2}$\end{rotate}}}}
\put(3511,-2937){\makebox(0,0)[b]{\smash{\SetFigFont{10}{12.0}{\familydefault}{\mddefault}{\updefault}$l$}}}
\put(5634,216){\makebox(0,0)[rb]{\smash{\SetFigFont{10}{12.0}{\familydefault}{\mddefault}{\updefault}CDM}}}
\put(5634, 92){\makebox(0,0)[rb]{\smash{\SetFigFont{10}{12.0}{\familydefault}{\mddefault}{\updefault}$\Omega_{a}=0.15$}}}
\put(5634,-32){\makebox(0,0)[rb]{\smash{\SetFigFont{10}{12.0}{\familydefault}{\mddefault}{\updefault}$\Omega_{a}=0.3$}}}
\put(5634,-156){\makebox(0,0)[rb]{\smash{\SetFigFont{10}{12.0}{\familydefault}{\mddefault}{\updefault}$\Omega_{a}=0.904$}}}
\end{picture}
\caption[$l(l+1)C_{l}$:  Varying $\Omega_{a}$]{The angular power spectrum for various values of $\Omega_{a}$ in the case that the adiabatic and isocurvature perturbations have equal amplitudes ($\beta$=1, which is not the same as equal contributions to the quadrupole).  The top line is standard CDM with adiabatic perturbations.  Going down, $\Omega_{a}$=0.15, 0.3 and 0.904.  In this plot $\Omega_{b}=0.096$ and $h=0.5$.} \label{fig:cdmmix}
\end{center}
\end{figure}

Now suppose that some of the dark matter is SUSY CDM as well as axions.  First, consider the case when the amplitude of the adiabatic fluctuations is the same as that of the isocurvature perturbations, i.e., $\beta=1$.  In this case, Figure~\ref{fig:cdmmix} shows the CMB anisotropy for various values of $\Omega_{a}$.  Qualitatively, these results show that the large scale isocurvature contribution decreases as the axion density decreases, which would be expected since the isocurvature contribution has its source in the axion field.  This is just the result predicted by Equations~\ref{eq:powerl} (or~\ref{eq:powerg}) and Equation~\ref{eq:C2iso}.  So adding a SUSY CDM component to the matter density will decrease the amplitude of the isocurvature perturbations relative to the case in which all of the cold dark matter is in the form of axions.

\begin{figure}
\begin{center}
\begin{picture}(0,0)%
\includegraphics{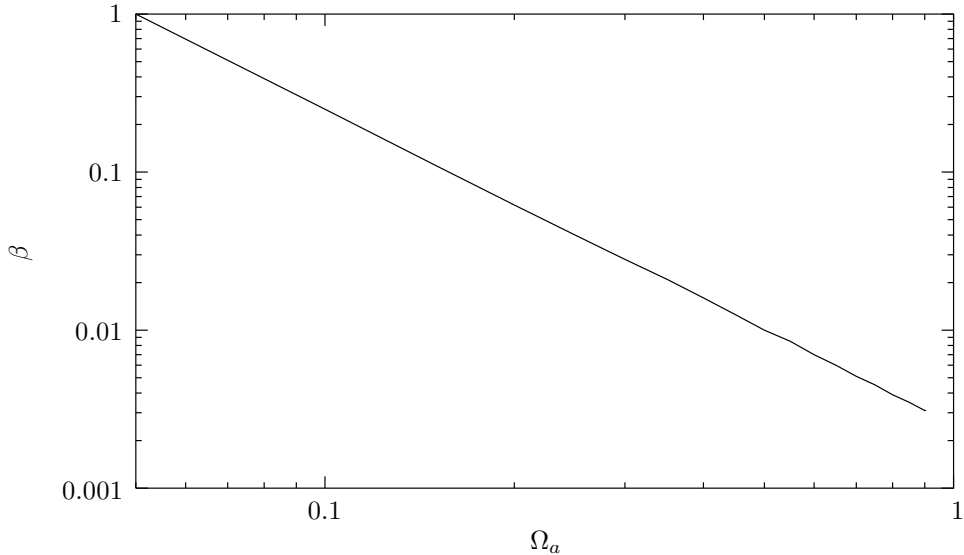}%
\end{picture}%
\setlength{\unitlength}{0.00083300in}%
\begingroup\makeatletter\ifx\SetFigFont\undefined%
\gdef\SetFigFont#1#2#3#4#5{%
  \reset@font\fontsize{#1}{#2pt}%
  \fontfamily{#3}\fontseries{#4}\fontshape{#5}%
  \selectfont}%
\fi\endgroup%
\begin{picture}(5949,3404)(316,-2967)
\put(1017,-2627){\makebox(0,0)[rb]{\smash{\SetFigFont{10}{12.0}{\familydefault}{\mddefault}{\updefault}0.001}}}
\put(1017,-1634){\makebox(0,0)[rb]{\smash{\SetFigFont{10}{12.0}{\familydefault}{\mddefault}{\updefault}0.01}}}
\put(1017,-640){\makebox(0,0)[rb]{\smash{\SetFigFont{10}{12.0}{\familydefault}{\mddefault}{\updefault}0.1}}}
\put(1017,353){\makebox(0,0)[rb]{\smash{\SetFigFont{10}{12.0}{\familydefault}{\mddefault}{\updefault}1}}}
\put(2279,-2751){\makebox(0,0)[b]{\smash{\SetFigFont{10}{12.0}{\familydefault}{\mddefault}{\updefault}0.1}}}
\put(6227,-2751){\makebox(0,0)[b]{\smash{\SetFigFont{10}{12.0}{\familydefault}{\mddefault}{\updefault}1}}}
\put(425,-1137){\makebox(0,0)[b]{\smash{\SetFigFont{10}{12.0}{\familydefault}{\mddefault}{\updefault}\begin{rotate}{90}$\beta$\end{rotate}}}}
\put(3659,-2937){\makebox(0,0)[b]{\smash{\SetFigFont{10}{12.0}{\familydefault}{\mddefault}{\updefault}$\Omega_{a}$}}}
\end{picture}
\caption[Degeneracy in $\beta$ and $\Omega_{a}$]{In this figure, values of $\Omega_{a}$ and $\beta$ which lie on the solid line give identical angular power spectra ($l(l+1)C_{l}$), matter power spectra ($P(k)$) and rms mass fluctuations on $8h^{-1}$Mpc scales ($\sigma_{8}$).  Here $h=0.5$ and $\Omega_{b}=0.096$.} \label{fig:degen}
\end{center}
\end{figure}

One might hope that the CMB and/or large scale structure could be used together to determine $\beta$ and $\Omega_{a}$.  Unfortunately, however, there is a lot of degeneracy in these parameters.  In figure~\ref{fig:degen}, we show a combination of the parameters $\beta$ and $\Omega_{a}$, which produce identical angular power spectra, matter power spectra $P(k)$ and rms mass-fluctuations on $8h^{-1}$Mpc scales.  Since the parameters $\beta$ and $\Omega_{a}$ affect the large scale anisotropy in a similar way (isocurvature boost), and since $P(k)$ and $\sigma_{8}$ are normalized to the COBE data, this is not surprising.  At best, CMB measurements would only be able to identify a curve in the $\beta-\Omega_{a}$ parameter space.

\subsection{CMB Parameter Space} \label{sec-CMBparam}
Here we consider the region of parameter space allowed by the CMB.  Although it is still early and the data are scattered, it would appear that there is a rise in the angular power spectrum as one goes to smaller angular scales.  This rise is inconsistent with a purely isocurvature angular power spectrum, which instead predicts a lot of large scale power but not much small scale power.  This information can be used to probe the axion parameter space.  

\begin{figure}
\begin{center}
\begin{picture}(0,0)%
\includegraphics{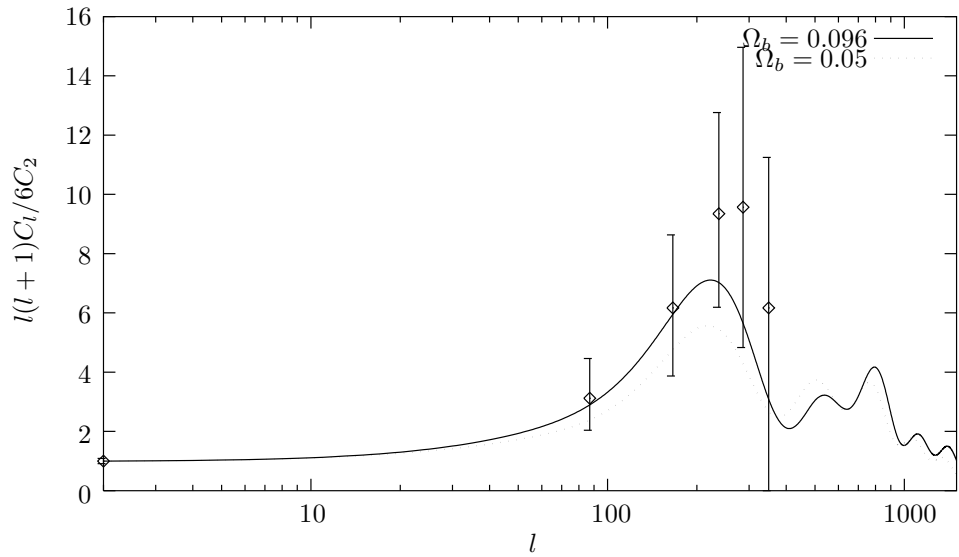}%
\end{picture}%
\setlength{\unitlength}{0.00083300in}%
\begingroup\makeatletter\ifx\SetFigFont\undefined%
\gdef\SetFigFont#1#2#3#4#5{%
  \reset@font\fontsize{#1}{#2pt}%
  \fontfamily{#3}\fontseries{#4}\fontshape{#5}%
  \selectfont}%
\fi\endgroup%
\begin{picture}(5935,3401)(304,-2964)
\put(795,-2627){\makebox(0,0)[rb]{\smash{\SetFigFont{10}{12.0}{\familydefault}{\mddefault}{\updefault}0}}}
\put(795,-2254){\makebox(0,0)[rb]{\smash{\SetFigFont{10}{12.0}{\familydefault}{\mddefault}{\updefault}2}}}
\put(795,-1882){\makebox(0,0)[rb]{\smash{\SetFigFont{10}{12.0}{\familydefault}{\mddefault}{\updefault}4}}}
\put(795,-1509){\makebox(0,0)[rb]{\smash{\SetFigFont{10}{12.0}{\familydefault}{\mddefault}{\updefault}6}}}
\put(795,-1137){\makebox(0,0)[rb]{\smash{\SetFigFont{10}{12.0}{\familydefault}{\mddefault}{\updefault}8}}}
\put(795,-764){\makebox(0,0)[rb]{\smash{\SetFigFont{10}{12.0}{\familydefault}{\mddefault}{\updefault}10}}}
\put(795,-392){\makebox(0,0)[rb]{\smash{\SetFigFont{10}{12.0}{\familydefault}{\mddefault}{\updefault}12}}}
\put(795,-19){\makebox(0,0)[rb]{\smash{\SetFigFont{10}{12.0}{\familydefault}{\mddefault}{\updefault}14}}}
\put(795,353){\makebox(0,0)[rb]{\smash{\SetFigFont{10}{12.0}{\familydefault}{\mddefault}{\updefault}16}}}
\put(2172,-2751){\makebox(0,0)[b]{\smash{\SetFigFont{10}{12.0}{\familydefault}{\mddefault}{\updefault}10}}}
\put(4035,-2751){\makebox(0,0)[b]{\smash{\SetFigFont{10}{12.0}{\familydefault}{\mddefault}{\updefault}100}}}
\put(5899,-2751){\makebox(0,0)[b]{\smash{\SetFigFont{10}{12.0}{\familydefault}{\mddefault}{\updefault}1000}}}
\put(425,-1509){\makebox(0,0)[b]{\smash{\SetFigFont{10}{12.0}{\familydefault}{\mddefault}{\updefault}\begin{rotate}{90}$l(l+1)C_{l}/6C_{2}$\end{rotate}}}}
\put(3548,-2937){\makebox(0,0)[b]{\smash{\SetFigFont{10}{12.0}{\familydefault}{\mddefault}{\updefault}$l$}}}
\put(5634,216){\makebox(0,0)[rb]{\smash{\SetFigFont{10}{12.0}{\familydefault}{\mddefault}{\updefault}$\Omega_{b}=0.096$}}}
\put(5634, 92){\makebox(0,0)[rb]{\smash{\SetFigFont{10}{12.0}{\familydefault}{\mddefault}{\updefault}$\Omega_{b}=0.05$}}}
\end{picture}
\caption[$l(l+1)C_{l}$:  Saskatoon]{The angular power spectrum, $l (l+1)C_{l}$, for two CDM models, one with $\Omega_{b}=0.096$ and the other with $\Omega_{b}=0.05$.  In both cases, $h=0.5$.  Also shown are the error bars from the Saskatoon experiment, including a 14\% calibration uncertainty, and the COBE quadrupole.} \label{fig:sask}
\end{center}
\end{figure}

The consensus seems to be that the Saskatoon experiment \cite{Page} is the best present day microwave background anisotropy experiment at intermediate to small angular scales.  For that reason, we will use the recent Saskatoon data to constrain the axion parameter space.  Not including a 14\% calibration uncertainty, Netterfield et al. \cite{Page} found that $\delta T=49^{+8}_{-3}\mu$K for $l=87$ and $\delta T=85^{+10}_{-8}\mu$K for $l=237$.  The angular power spectrum $C_{l}$ is related to $\delta T_{l}$ by
\begin{equation}
  C_{l} = \frac{4\pi}{l (2l+1)}(\delta T_{l})^{2}
\end{equation}
The Saskatoon data are shown in Figure~\ref{fig:sask}, along with theoretical predictions for a standard CDM model with $\Omega_{b}=0.05$ and a CDM model with $\Omega_{b}=0.096$, the maximum value of $\Omega_{b}$ allowed by nucleosynthesis \cite{Turnerbaryon}.  It is interesting to note that standard cold dark matter models with low values of $\Omega_{b}$ appear to have difficulty fitting the observed anisotropy at $l=237$.  From Figure~\ref{fig:sask}, it would qualitatively seem that in the context of pure cold dark matter, a higher value of $\Omega_{b}$ appears more consistent with this data point.  

Other models that are consistent with the Saskatoon data point at $l=237$ are cosmological constant models and MDM models.  Tilting the power spectrum index, however, is not consistent with this data point in the context of sCDM with $\Omega_{0}=1$, since for $n<1$, the height of the Doppler peak decreases relative to the quadrupole.  In the context of $\Omega_{0}=1$, the most dramatic effect on the Doppler peak comes from increasing the baryon density $\Omega_{b}h^{2}$.  Increasing $\Omega_{b}h^{2}$ from 0.008 to 0.024, the range allowed by nucleosynthesis \cite{Turnerbaryon}, can roughly double the height of the Doppler peak relative to the quadrupole \cite{Jungman}.\footnote{Note that increasing $\Omega_{\Lambda}$ from 0 to 0.7 has a similar effect.}  Since we are interested in determining how well pure CDM models with an isocurvature component do in reproducing the observed microwave background anisotropies and large scale structure, we will explore the parameter space by varying $\Omega_{b}$.

\begin{figure}
\begin{center}
\begin{picture}(0,0)%
\includegraphics{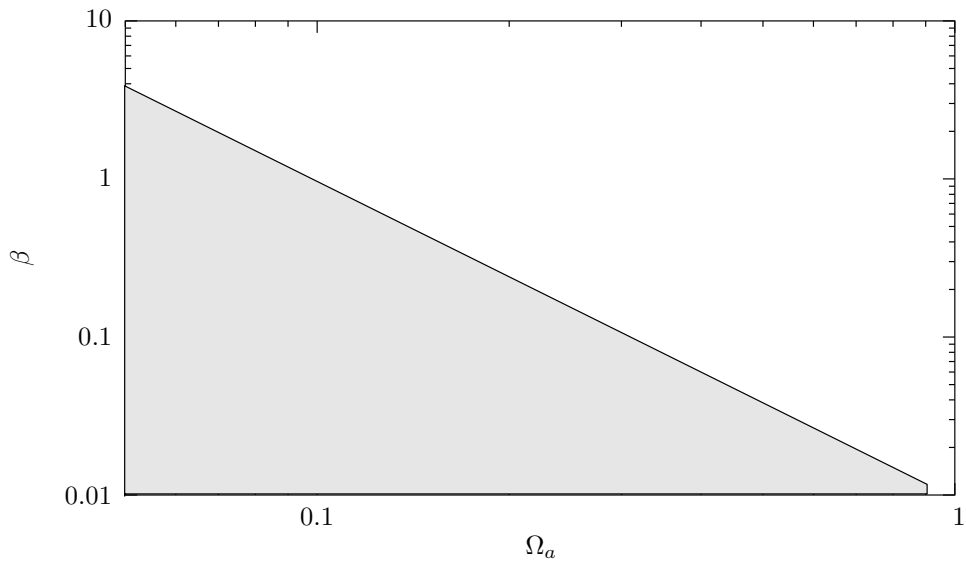}%
\end{picture}%
\setlength{\unitlength}{0.00083300in}%
\begingroup\makeatletter\ifx\SetFigFont\undefined%
\gdef\SetFigFont#1#2#3#4#5{%
  \reset@font\fontsize{#1}{#2pt}%
  \fontfamily{#3}\fontseries{#4}\fontshape{#5}%
  \selectfont}%
\fi\endgroup%
\begin{picture}(6071,3515)(232,-2994)
\put(943,-2627){\makebox(0,0)[rb]{\smash{\SetFigFont{10}{12.0}{\familydefault}{\mddefault}{\updefault}0.01}}}
\put(943,-1634){\makebox(0,0)[rb]{\smash{\SetFigFont{10}{12.0}{\familydefault}{\mddefault}{\updefault}0.1}}}
\put(943,-640){\makebox(0,0)[rb]{\smash{\SetFigFont{10}{12.0}{\familydefault}{\mddefault}{\updefault}1}}}
\put(943,353){\makebox(0,0)[rb]{\smash{\SetFigFont{10}{12.0}{\familydefault}{\mddefault}{\updefault}10}}}
\put(2222,-2751){\makebox(0,0)[b]{\smash{\SetFigFont{10}{12.0}{\familydefault}{\mddefault}{\updefault}0.1}}}
\put(6227,-2751){\makebox(0,0)[b]{\smash{\SetFigFont{10}{12.0}{\familydefault}{\mddefault}{\updefault}1}}}
\put(3622,-2937){\makebox(0,0)[b]{\smash{\SetFigFont{10}{12.0}{\familydefault}{\mddefault}{\updefault}$\Omega_{a}$}}}
\put(425,-1137){\makebox(0,0)[b]{\smash{\SetFigFont{10}{12.0}{\familydefault}{\mddefault}{\updefault}\begin{rotate}{90}$\beta$\end{rotate}}}}
\end{picture}
\caption[$\beta-\Omega_{a}$ Parameter Space for Saskatoon Data Point $l=87$]{Acceptable regions of the axion parameter space for $\l=87$.  In this plot $\Omega_{b}=0.096$ and $h=0.5$.} \label{fig:par87}
\end{center}
\end{figure}

\begin{figure}
\begin{center}
\begin{picture}(0,0)%
\includegraphics{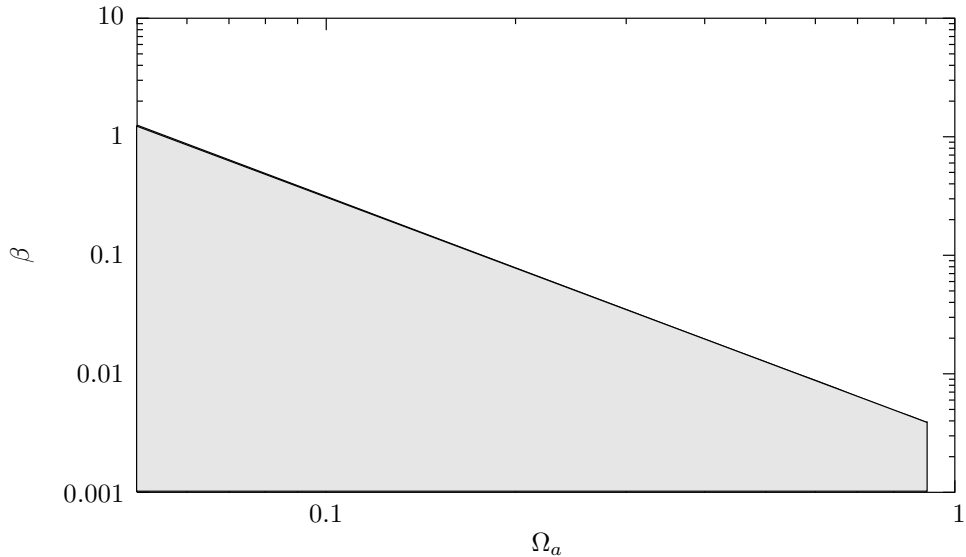}%
\end{picture}%
\setlength{\unitlength}{0.00083300in}%
\begingroup\makeatletter\ifx\SetFigFont\undefined%
\gdef\SetFigFont#1#2#3#4#5{%
  \reset@font\fontsize{#1}{#2pt}%
  \fontfamily{#3}\fontseries{#4}\fontshape{#5}%
  \selectfont}%
\fi\endgroup%
\begin{picture}(6071,3515)(232,-2994)
\put(1017,-2627){\makebox(0,0)[rb]{\smash{\SetFigFont{10}{12.0}{\familydefault}{\mddefault}{\updefault}0.001}}}
\put(1017,-1882){\makebox(0,0)[rb]{\smash{\SetFigFont{10}{12.0}{\familydefault}{\mddefault}{\updefault}0.01}}}
\put(1017,-1137){\makebox(0,0)[rb]{\smash{\SetFigFont{10}{12.0}{\familydefault}{\mddefault}{\updefault}0.1}}}
\put(1017,-392){\makebox(0,0)[rb]{\smash{\SetFigFont{10}{12.0}{\familydefault}{\mddefault}{\updefault}1}}}
\put(1017,353){\makebox(0,0)[rb]{\smash{\SetFigFont{10}{12.0}{\familydefault}{\mddefault}{\updefault}10}}}
\put(2279,-2751){\makebox(0,0)[b]{\smash{\SetFigFont{10}{12.0}{\familydefault}{\mddefault}{\updefault}0.1}}}
\put(6227,-2751){\makebox(0,0)[b]{\smash{\SetFigFont{10}{12.0}{\familydefault}{\mddefault}{\updefault}1}}}
\put(425,-1137){\makebox(0,0)[b]{\smash{\SetFigFont{10}{12.0}{\familydefault}{\mddefault}{\updefault}\begin{rotate}{90}$\beta$\end{rotate}}}}
\put(3659,-2937){\makebox(0,0)[b]{\smash{\SetFigFont{10}{12.0}{\familydefault}{\mddefault}{\updefault}$\Omega_{a}$}}}
\end{picture}
\caption[$\beta-\Omega_{a}$ Parameter Space for Saskatoon Data Point $l=237$]{Acceptable regions of the axion parameter space for $\l=237$.  In this plot $\Omega_{b}=0.096$ and $h=0.5$.} \label{fig:par237}
\end{center}
\end{figure}

In mapping out the parameter space, we would like to know what is the largest isocurvature contribution we can have that is consistent with the data.  In the context of cold dark matter models, this corresponds to a large value of $\Omega_{b}$.  This is because the effect of the isocurvature perturbations is to decrease the height of the peaks relative to the quadrupole.  Therefore if we increase the peak height in the standard adiabatic case by increasing $\Omega_{b}$, this allows a larger isocurvature component to produce the observed peak height.  Figures~\ref{fig:par87} and~\ref{fig:par237} show plots of the allowed regions of parameter space for $\l=87$ and $\l=237$ with $\Omega_{b}=0.096$ and $h=0.5$.  Note that any variations in the parameters which lower the first Doppler peak (such as $n<1$, lowering $\Omega_{b}$) will constrict the parameter space even further. 

Qualitatively, these results suggest that unless $\Omega_{a}$ makes a small contribution to the matter density, the ratio of the isocurvature to the isentropic power spectrum, $\beta$, should be relatively small.  For the Saskatoon data point at $\l=87$, $\beta$ should be less than about 0.1 for most values of $\Omega_{a}$, and for the data point at $\l=237$, $\beta$ should be at most $\approx 10^{-3}-10^{-2}$ for most values of $\Omega_{a}$.  The parameter space is more restricted for $\l=237$ because the anisotropy at that point is greater compared to the standard CDM model than the anisotropy at $\l=87$.  Therefore, the data at $\l=237$ admit a smaller isocurvature contribution than the data at $\l=87$.

\subsection{Tensor vs. Isocurvature Perturbations}
\begin{figure}
\begin{center}
\begin{picture}(0,0)%
\includegraphics{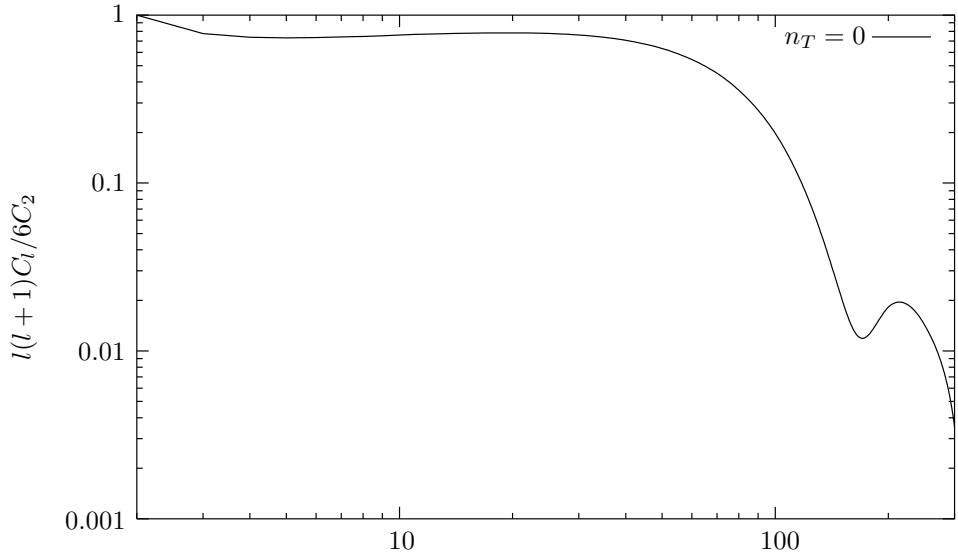}%
\end{picture}%
\setlength{\unitlength}{0.00083300in}%
\begingroup\makeatletter\ifx\SetFigFont\undefined%
\gdef\SetFigFont#1#2#3#4#5{%
  \reset@font\fontsize{#1}{#2pt}%
  \fontfamily{#3}\fontseries{#4}\fontshape{#5}%
  \selectfont}%
\fi\endgroup%
\begin{picture}(5935,3395)(304,-2958)
\put(1017,-2813){\makebox(0,0)[rb]{\smash{\SetFigFont{10}{12.0}{\familydefault}{\mddefault}{\updefault}0.001}}}
\put(1017,-1758){\makebox(0,0)[rb]{\smash{\SetFigFont{10}{12.0}{\familydefault}{\mddefault}{\updefault}0.01}}}
\put(1017,-702){\makebox(0,0)[rb]{\smash{\SetFigFont{10}{12.0}{\familydefault}{\mddefault}{\updefault}0.1}}}
\put(1017,353){\makebox(0,0)[rb]{\smash{\SetFigFont{10}{12.0}{\familydefault}{\mddefault}{\updefault}1}}}
\put(2741,-2937){\makebox(0,0)[b]{\smash{\SetFigFont{10}{12.0}{\familydefault}{\mddefault}{\updefault}10}}}
\put(5101,-2937){\makebox(0,0)[b]{\smash{\SetFigFont{10}{12.0}{\familydefault}{\mddefault}{\updefault}100}}}
\put(425,-1758){\makebox(0,0)[b]{\smash{\SetFigFont{10}{12.0}{\familydefault}{\mddefault}{\updefault}\begin{rotate}{90}$l(l+1)C_{l}/6C_{2}$\end{rotate}}}}
\put(5634,216){\makebox(0,0)[rb]{\smash{\SetFigFont{10}{12.0}{\familydefault}{\mddefault}{\updefault}$n_{T}=0$}}}
\end{picture}
\caption[$l(l+1)C_{l}$:  Tensor Perturbations with $n_{T}=0$]{The scale invariant ($n_{T}=0$) angular power spectrum of tensor perturbations.} \label{fig:tensorn0}
\end{center}
\end{figure}

\begin{figure}
\begin{center}
\begin{picture}(0,0)%
\includegraphics{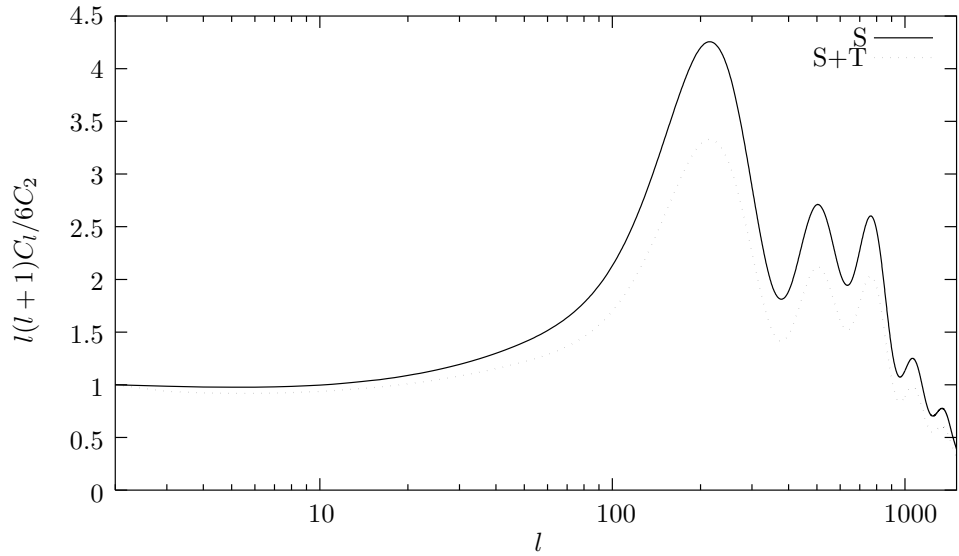}%
\end{picture}%
\setlength{\unitlength}{0.00083300in}%
\begingroup\makeatletter\ifx\SetFigFont\undefined%
\gdef\SetFigFont#1#2#3#4#5{%
  \reset@font\fontsize{#1}{#2pt}%
  \fontfamily{#3}\fontseries{#4}\fontshape{#5}%
  \selectfont}%
\fi\endgroup%
\begin{picture}(5935,3401)(304,-2964)
\put(869,-2627){\makebox(0,0)[rb]{\smash{\SetFigFont{10}{12.0}{\familydefault}{\mddefault}{\updefault}0}}}
\put(869,-2296){\makebox(0,0)[rb]{\smash{\SetFigFont{10}{12.0}{\familydefault}{\mddefault}{\updefault}0.5}}}
\put(869,-1965){\makebox(0,0)[rb]{\smash{\SetFigFont{10}{12.0}{\familydefault}{\mddefault}{\updefault}1}}}
\put(869,-1634){\makebox(0,0)[rb]{\smash{\SetFigFont{10}{12.0}{\familydefault}{\mddefault}{\updefault}1.5}}}
\put(869,-1303){\makebox(0,0)[rb]{\smash{\SetFigFont{10}{12.0}{\familydefault}{\mddefault}{\updefault}2}}}
\put(869,-971){\makebox(0,0)[rb]{\smash{\SetFigFont{10}{12.0}{\familydefault}{\mddefault}{\updefault}2.5}}}
\put(869,-640){\makebox(0,0)[rb]{\smash{\SetFigFont{10}{12.0}{\familydefault}{\mddefault}{\updefault}3}}}
\put(869,-309){\makebox(0,0)[rb]{\smash{\SetFigFont{10}{12.0}{\familydefault}{\mddefault}{\updefault}3.5}}}
\put(869, 22){\makebox(0,0)[rb]{\smash{\SetFigFont{10}{12.0}{\familydefault}{\mddefault}{\updefault}4}}}
\put(869,353){\makebox(0,0)[rb]{\smash{\SetFigFont{10}{12.0}{\familydefault}{\mddefault}{\updefault}4.5}}}
\put(2228,-2751){\makebox(0,0)[b]{\smash{\SetFigFont{10}{12.0}{\familydefault}{\mddefault}{\updefault}10}}}
\put(4065,-2751){\makebox(0,0)[b]{\smash{\SetFigFont{10}{12.0}{\familydefault}{\mddefault}{\updefault}100}}}
\put(5903,-2751){\makebox(0,0)[b]{\smash{\SetFigFont{10}{12.0}{\familydefault}{\mddefault}{\updefault}1000}}}
\put(425,-1634){\makebox(0,0)[b]{\smash{\SetFigFont{10}{12.0}{\familydefault}{\mddefault}{\updefault}\begin{rotate}{90}$l(l+1)C_{l}/6C_{2}$\end{rotate}}}}
\put(3585,-2937){\makebox(0,0)[b]{\smash{\SetFigFont{10}{12.0}{\familydefault}{\mddefault}{\updefault}$l$}}}
\put(5634,216){\makebox(0,0)[rb]{\smash{\SetFigFont{10}{12.0}{\familydefault}{\mddefault}{\updefault}S}}}
\put(5634, 92){\makebox(0,0)[rb]{\smash{\SetFigFont{10}{12.0}{\familydefault}{\mddefault}{\updefault}S+T}}}
\end{picture}
\caption[$l(l+1)C_{l}$:  Scalar + Tensor Perturbations with $n_{S}=0.94$ and $n_{T}=-0.04$]{The angular power spectrum for purely isentropic perturbations ($S$) and isentropic + tensor ($S+T$) perturbations in a chaotic inflation scenario with $n_{S}=0.94$ and $n_{T}=-0.04$.} \label{fig:ST}
\end{center}
\end{figure}

\begin{figure}
\begin{center}
\begin{picture}(0,0)%
\includegraphics{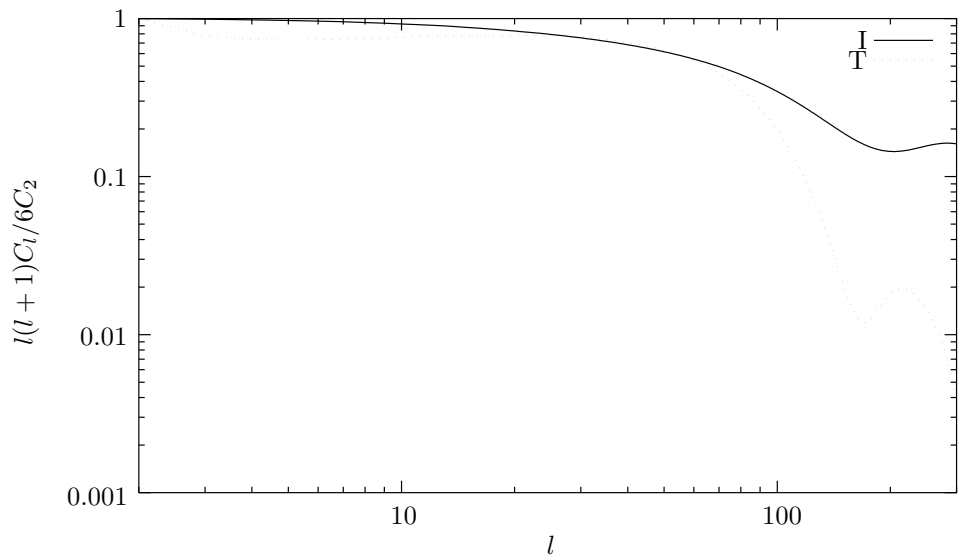}%
\end{picture}%
\setlength{\unitlength}{0.00083300in}%
\begingroup\makeatletter\ifx\SetFigFont\undefined%
\gdef\SetFigFont#1#2#3#4#5{%
  \reset@font\fontsize{#1}{#2pt}%
  \fontfamily{#3}\fontseries{#4}\fontshape{#5}%
  \selectfont}%
\fi\endgroup%
\begin{picture}(5935,3401)(304,-2964)
\put(1017,-2627){\makebox(0,0)[rb]{\smash{\SetFigFont{10}{12.0}{\familydefault}{\mddefault}{\updefault}0.001}}}
\put(1017,-1634){\makebox(0,0)[rb]{\smash{\SetFigFont{10}{12.0}{\familydefault}{\mddefault}{\updefault}0.01}}}
\put(1017,-640){\makebox(0,0)[rb]{\smash{\SetFigFont{10}{12.0}{\familydefault}{\mddefault}{\updefault}0.1}}}
\put(1017,353){\makebox(0,0)[rb]{\smash{\SetFigFont{10}{12.0}{\familydefault}{\mddefault}{\updefault}1}}}
\put(2741,-2751){\makebox(0,0)[b]{\smash{\SetFigFont{10}{12.0}{\familydefault}{\mddefault}{\updefault}10}}}
\put(5101,-2751){\makebox(0,0)[b]{\smash{\SetFigFont{10}{12.0}{\familydefault}{\mddefault}{\updefault}100}}}
\put(425,-1634){\makebox(0,0)[b]{\smash{\SetFigFont{10}{12.0}{\familydefault}{\mddefault}{\updefault}\begin{rotate}{90}$l(l+1)C_{l}/6C_{2}$\end{rotate}}}}
\put(3659,-2937){\makebox(0,0)[b]{\smash{\SetFigFont{10}{12.0}{\familydefault}{\mddefault}{\updefault}$l$}}}
\put(5634,216){\makebox(0,0)[rb]{\smash{\SetFigFont{10}{12.0}{\familydefault}{\mddefault}{\updefault}I}}}
\put(5634, 92){\makebox(0,0)[rb]{\smash{\SetFigFont{10}{12.0}{\familydefault}{\mddefault}{\updefault}T}}}
\end{picture}
\caption[$l(l+1)C_{l}$:  Tensor and Isocurvature Perturbations]{The angular power spectra for scale invariant tensor ($T$) and isocurvature ($I$)  perturbations.} \label{fig:TandI}
\end{center}
\end{figure}

Until now we have ignored the tensor contribution to the microwave background anisotropy.  This is not unreasonable, in that for $-n_{T} \stackrel{<}{\sim} 10^{-4}$, it is still possible to have isocurvature perturbations with $\beta \stackrel{<}{\sim} 1$ (cf. Equations~\ref{eq:powerl} and~\ref{eq:powerg}).  If, however, $n_{T} \sim 0.01-0.04$, the tensor contributions can make a significant contribution to the quadrupole.  Since the tensor perturbations are important only on large angular scales (see Figure~\ref{fig:tensorn0}), the primary effect of the tensor perturbations is on the normalization of the power spectrum.  Inflation predicts that $r$, the ratio of the tensor to isentropic contribution to the quadrupole, is \cite{Steinhardt}
\begin{equation}
  r \approx -7n_{T}
\end{equation}
Figure~\ref{fig:ST} shows the anisotropy for a typical chaotic inflation scenario with $n_{S}=0.94$ and $n_{T}=-0.04$.  Note that the addition of the tensor perturbations lowers the Doppler peaks in a way that is reminiscent of the isocurvature perturbations.  An important question is then whether or not the isocurvature perturbations are distinguishable from the tensor perturbations.  As Figure~\ref{fig:TandI} shows, the two types of perturbations leave different signatures on the microwave background anisotropy.  Although the large scale power is roughly the same for the two perturbations, the small scale tensor perturbations are considerably smaller than the isocurvature perturbations.  This is because, unlike the scalar perturbations, the gravitational waves, which are the source of the tensor perturbations, redshift away rather quickly.   Therefore, it should be possible to distinguish between an isentropic + tensor ($S+T$) and an isentropic + isocurvature ($S+I$) scenario at high $l$.

\begin{figure}
\begin{center}
\begin{picture}(0,0)%
\includegraphics{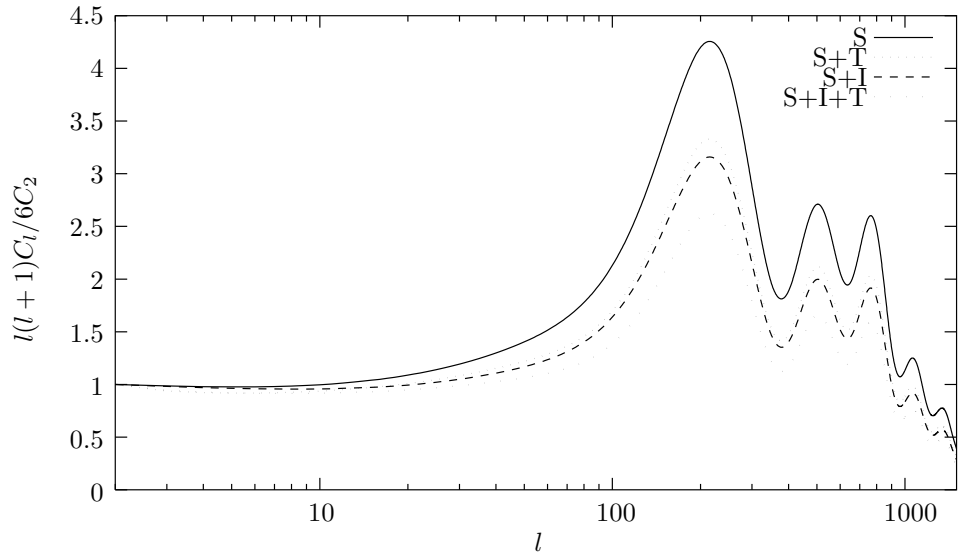}%
\end{picture}%
\setlength{\unitlength}{0.00083300in}%
\begingroup\makeatletter\ifx\SetFigFont\undefined%
\gdef\SetFigFont#1#2#3#4#5{%
  \reset@font\fontsize{#1}{#2pt}%
  \fontfamily{#3}\fontseries{#4}\fontshape{#5}%
  \selectfont}%
\fi\endgroup%
\begin{picture}(5935,3401)(304,-2964)
\put(869,-2627){\makebox(0,0)[rb]{\smash{\SetFigFont{10}{12.0}{\familydefault}{\mddefault}{\updefault}0}}}
\put(869,-2296){\makebox(0,0)[rb]{\smash{\SetFigFont{10}{12.0}{\familydefault}{\mddefault}{\updefault}0.5}}}
\put(869,-1965){\makebox(0,0)[rb]{\smash{\SetFigFont{10}{12.0}{\familydefault}{\mddefault}{\updefault}1}}}
\put(869,-1634){\makebox(0,0)[rb]{\smash{\SetFigFont{10}{12.0}{\familydefault}{\mddefault}{\updefault}1.5}}}
\put(869,-1303){\makebox(0,0)[rb]{\smash{\SetFigFont{10}{12.0}{\familydefault}{\mddefault}{\updefault}2}}}
\put(869,-971){\makebox(0,0)[rb]{\smash{\SetFigFont{10}{12.0}{\familydefault}{\mddefault}{\updefault}2.5}}}
\put(869,-640){\makebox(0,0)[rb]{\smash{\SetFigFont{10}{12.0}{\familydefault}{\mddefault}{\updefault}3}}}
\put(869,-309){\makebox(0,0)[rb]{\smash{\SetFigFont{10}{12.0}{\familydefault}{\mddefault}{\updefault}3.5}}}
\put(869, 22){\makebox(0,0)[rb]{\smash{\SetFigFont{10}{12.0}{\familydefault}{\mddefault}{\updefault}4}}}
\put(869,353){\makebox(0,0)[rb]{\smash{\SetFigFont{10}{12.0}{\familydefault}{\mddefault}{\updefault}4.5}}}
\put(2228,-2751){\makebox(0,0)[b]{\smash{\SetFigFont{10}{12.0}{\familydefault}{\mddefault}{\updefault}10}}}
\put(4065,-2751){\makebox(0,0)[b]{\smash{\SetFigFont{10}{12.0}{\familydefault}{\mddefault}{\updefault}100}}}
\put(5903,-2751){\makebox(0,0)[b]{\smash{\SetFigFont{10}{12.0}{\familydefault}{\mddefault}{\updefault}1000}}}
\put(425,-1634){\makebox(0,0)[b]{\smash{\SetFigFont{10}{12.0}{\familydefault}{\mddefault}{\updefault}\begin{rotate}{90}$l(l+1)C_{l}/6C_{2}$\end{rotate}}}}
\put(3585,-2937){\makebox(0,0)[b]{\smash{\SetFigFont{10}{12.0}{\familydefault}{\mddefault}{\updefault}$l$}}}
\put(5634,216){\makebox(0,0)[rb]{\smash{\SetFigFont{10}{12.0}{\familydefault}{\mddefault}{\updefault}S}}}
\put(5634, 92){\makebox(0,0)[rb]{\smash{\SetFigFont{10}{12.0}{\familydefault}{\mddefault}{\updefault}S+T}}}
\put(5634,-32){\makebox(0,0)[rb]{\smash{\SetFigFont{10}{12.0}{\familydefault}{\mddefault}{\updefault}S+I}}}
\put(5634,-156){\makebox(0,0)[rb]{\smash{\SetFigFont{10}{12.0}{\familydefault}{\mddefault}{\updefault}S+I+T}}}
\end{picture}
\caption[$l(l+1)C_{l}$:  Isentropic, Isocurvature and Tensor Perturbations]{The angular power spectra for isentropic ($S$), isentropic + tensor ($S+T$), isentropic + isocurvature ($S+I$) and isentropic + isocurvature + tensor ($S+I+T$) perturbations.  In this example $n=0.94$, $n_{T}=-0.04$, $h=0.5$ and $\Omega_{a}=0.95$.} \label{fig:sitbeta01}
\end{center}
\end{figure}

\begin{figure}
\begin{center}
\begin{picture}(0,0)%
\includegraphics{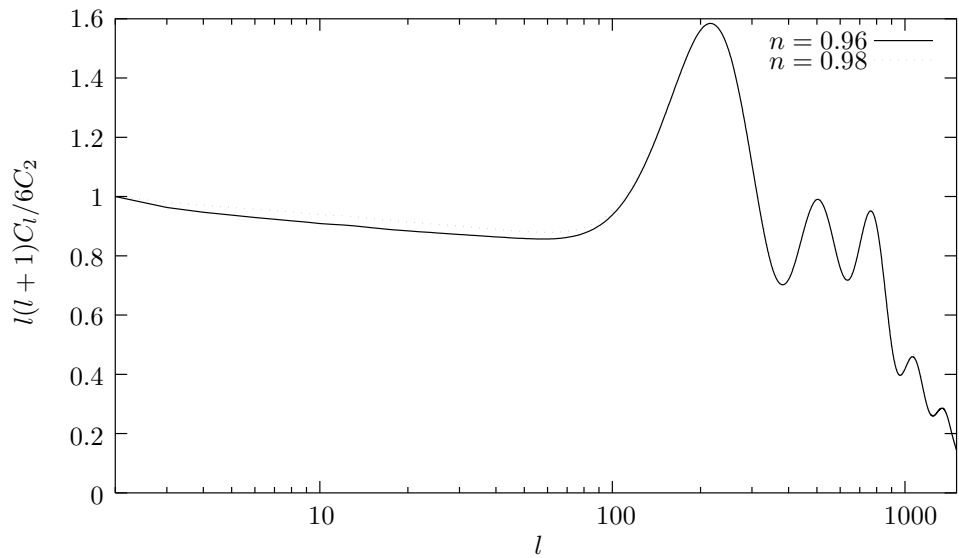}%
\end{picture}%
\setlength{\unitlength}{0.00083300in}%
\begingroup\makeatletter\ifx\SetFigFont\undefined%
\gdef\SetFigFont#1#2#3#4#5{%
  \reset@font\fontsize{#1}{#2pt}%
  \fontfamily{#3}\fontseries{#4}\fontshape{#5}%
  \selectfont}%
\fi\endgroup%
\begin{picture}(5935,3401)(304,-2964)
\put(869,-2627){\makebox(0,0)[rb]{\smash{\SetFigFont{10}{12.0}{\familydefault}{\mddefault}{\updefault}0}}}
\put(869,-2254){\makebox(0,0)[rb]{\smash{\SetFigFont{10}{12.0}{\familydefault}{\mddefault}{\updefault}0.2}}}
\put(869,-1882){\makebox(0,0)[rb]{\smash{\SetFigFont{10}{12.0}{\familydefault}{\mddefault}{\updefault}0.4}}}
\put(869,-1509){\makebox(0,0)[rb]{\smash{\SetFigFont{10}{12.0}{\familydefault}{\mddefault}{\updefault}0.6}}}
\put(869,-1137){\makebox(0,0)[rb]{\smash{\SetFigFont{10}{12.0}{\familydefault}{\mddefault}{\updefault}0.8}}}
\put(869,-764){\makebox(0,0)[rb]{\smash{\SetFigFont{10}{12.0}{\familydefault}{\mddefault}{\updefault}1}}}
\put(869,-392){\makebox(0,0)[rb]{\smash{\SetFigFont{10}{12.0}{\familydefault}{\mddefault}{\updefault}1.2}}}
\put(869,-19){\makebox(0,0)[rb]{\smash{\SetFigFont{10}{12.0}{\familydefault}{\mddefault}{\updefault}1.4}}}
\put(869,353){\makebox(0,0)[rb]{\smash{\SetFigFont{10}{12.0}{\familydefault}{\mddefault}{\updefault}1.6}}}
\put(2228,-2751){\makebox(0,0)[b]{\smash{\SetFigFont{10}{12.0}{\familydefault}{\mddefault}{\updefault}10}}}
\put(4065,-2751){\makebox(0,0)[b]{\smash{\SetFigFont{10}{12.0}{\familydefault}{\mddefault}{\updefault}100}}}
\put(5903,-2751){\makebox(0,0)[b]{\smash{\SetFigFont{10}{12.0}{\familydefault}{\mddefault}{\updefault}1000}}}
\put(425,-1509){\makebox(0,0)[b]{\smash{\SetFigFont{10}{12.0}{\familydefault}{\mddefault}{\updefault}\begin{rotate}{90}$l(l+1)C_{l}/6C_{2}$\end{rotate}}}}
\put(3585,-2937){\makebox(0,0)[b]{\smash{\SetFigFont{10}{12.0}{\familydefault}{\mddefault}{\updefault}$l$}}}
\put(5634,216){\makebox(0,0)[rb]{\smash{\SetFigFont{10}{12.0}{\familydefault}{\mddefault}{\updefault}$n=0.96$}}}
\put(5634, 92){\makebox(0,0)[rb]{\smash{\SetFigFont{10}{12.0}{\familydefault}{\mddefault}{\updefault}$n=0.98$}}}
\end{picture}
\caption[$l(l+1)C_{l}$:  Isentropic, Isocurvature and Tensor Perturbations]{Two nearly degenerate angular power spectra for two combinations of $n$, $n_{T}$ and $\beta$.} \label{fig:Clnbeta}
\end{center}
\end{figure}

If, however, both isocurvature and tensor perturbations make a significant contribution to the quadrupole, then it will be considerably more difficult to separate the two using the microwave background.  To see this, recall that the quadrupole is the sum of each perturbation.  If we include a tensor component, the equation for the quadrupole (Equation~\ref{eq:C2SI}) becomes
\begin{eqnarray}
  C_{2} &= &C_{2}^{isen}\left(1 + \frac{C_{2}^{T}}{C_{2}^{isen}} + \beta\frac{C_{2}^{iso}}{C_{2}^{isen}}\right)\\ \nonumber
  &= &C_{2}^{isen}\left(1 - 7n_{T} + \beta\frac{C_{2}^{iso}}{C_{2}^{isen}}\right)
\end{eqnarray}
Using Equation~\ref{eq:beta1} for $\beta$, this can be rewritten as
\begin{eqnarray}
  C_{2} &= &C_{2}^{isen}\left\{1 + \left[7 + \frac{1}{18\pi}\left(\frac{\Omega_{a}}{\Omega_{0}}\right)^{2} \left(\frac{m_{Pl}}{f_{a}\bar{\Theta}}\right)^{2}\frac{C_{2}^{iso}}{C_{2}^{isen}}\right](-n_{T})\right\}\\ \nonumber
	&\approx &C_{2}^{isen}\left\{1 + \left[7 + \frac{2}{\pi}\left(\frac{m_{Pl}}{f_{a}\bar{\Theta}}\right)^{2}\right](-n_{T})\right\}, 
\end{eqnarray}
where we have assumed that $\Omega_{a} \approx \Omega_{0} = 1$, in which case $C_{2}^{iso}/C_{2}^{isen} \approx 36$.  It is then easy to see that $-n_{T}$ and $(m_{Pl}/f_{a}\bar{\Theta})^{2}$ can be varied in such a way as to keep the second term constant, which suggests that there is a degeneracy between the isocurvature and tensor perturbations.  Figure~\ref{fig:Clnbeta} depicts the degeneracy in the angular power spectra for two models, one with $n_{T} = -0.02$, $\beta = 0.0633$ and $n = 0.98$, and another with $n_{T} = -0.04$, $\beta = 0.05$ and $n=0.96$.  In both of these models $\Omega_{b}=0.05$, $\Omega_{a}=0.95$ and $h=0.5$.  Despite this degeneracy, large scale structure should be able to distinguish between these two models.  This is because the tensor perturbations make no contribution to large scale structure, and so the degeneracy is broken by the variation of $n=1+n_{T}$.  Consider $\sigma_{8}$, the rms mass fluctuation on $8h^{-1}$Mpc scales.  The model with $n_{T} = -0.02$, $\beta = 0.0633$ and $n = 0.98$, for example, has $\sigma_{8}=0.66$, while the model with $n_{T} = -0.04$, $\beta = 0.05$ and $n=0.96$ has $\sigma_{8}=0.71$.  Therefore, even if isocurvature and tensor perturbations are present, large scale structure measurements should be able to distinguish between the effects of the two.

\subsection{Large Scale Structure}
In this section, we study the ($\beta, \Omega_{a}$) parameter space for the large scale structure statistics of these models.  Fortunately, it is not necessary to perform complicated non-linear simulations to get important information about large scale structure that can then be compared to observations.  Instead, many quantities that describe large scale structure, such as the power spectrum, $\sigma_{8}$ and the excess power (EP) can be studied using linear perturbation theory.  A discussion of large scale structure statistics usually begins with the contribution to the fractional density variance per bin of $\ln k$, which is defined as \cite{PeeblesLSS,PD}
\begin{equation}
  \Delta^{2}(k) = \frac{d\sigma^{2}}{d\ln k} = 4\pi k^{3}|\delta_{k}|^{2} = 4\pi k^{3}P(k),
\end{equation}
where $\delta_{k}$ is the amplitude of the matter contribution and $P(k)=|\delta_{k}|^{2}$ is known as the power spectrum.  Since the universe contains several matter components, we use the density weighted power spectrum to calculate the total power spectrum of all of the components.  For a universe consisting of baryons (subscript $b$), neutralinos or other CDM (subscript $x$),  axions (subscript $a$), massive neutrinos (subscript $\nu$) and warm dark matter (subscript $w$), the total density fluctuation is
\begin{eqnarray}
  \delta \rho &= &\delta \rho_{b} + \delta \rho_{x} + \delta \rho_{a} + \delta \rho_{\nu} + \delta \rho_{w}\nonumber \\
   &= &\bar{\rho}_{b}\delta_{b} + \bar{\rho}_{x}\delta_{x} + \bar{\rho}_{a}\delta_{a} + \bar{\rho}_{\nu}\delta_{\nu} + \bar{\rho}_{w}\delta_{w}
\end{eqnarray}
The density perturbation $\delta$ is then
\begin{equation}
  \delta_{k} = \frac{\delta \rho}{\rho} = \frac{\bar{\rho}_{b}}{\rho}\delta_{b}(k) + \frac{\bar{\rho}_{x}}{\rho}\delta_{x}(k) + \frac{\bar{\rho}_{a}}{\rho}\delta_{a}(k) + \frac{\bar{\rho}_{\nu}}{\rho}\delta_{\nu}(k) + \frac{\bar{\rho}_{w}}{\rho}\delta_{w}(k)
\end{equation}
Since the density fluctuations occur only in the baryons, CDM, HDM and WDM, $\delta_{k}$ is just
\begin{equation}
  \delta_{k} = \frac{\Omega_{b}\delta_{b}(k)+\Omega_{x}\delta_{x}(k)+\Omega_{a}\delta_{a}(k)+\Omega_{\nu}\delta_{\nu}(k)+\Omega_{w}\delta_{w}(k)}{\Omega_{0} + \Omega_{\nu} + \Omega_{w}}
\end{equation}
It used to be that the power spectrum was normalized on $8h^{-1}$Mpc scales.  But the COBE satellite has given a much cleaner measurement of the power spectrum, one that has been unaffected by any microphysical processing and is therefore primordial.  Thus the power spectrum is now normalized to the COBE data \cite{Turnererice,Bunn}.   

Once we know $P(k)$, or equivalently $\Delta^{2}(k)$, we can calculate the density contrast on a scale $R$:
\begin{equation}
  \sigma_{R}^{2} = \int \frac{dk}{k}\Delta^{2}(k)W^{2}(kR),
\end{equation}
where $W=3\jmath_{1}(x)/x$ is the top hat window function \cite{KT}.  The standard scale in cosmology is $8h^{-1}$Mpc, which is close to the scales that are just beginning to go non-linear.  Using different methods, several groups \cite{BondMyers1,Carlberg,BondMyers2,WEF,Viana} find $\sigma_{8}$ in the range 0.5-0.8 for $\Omega_{0}=1$.\footnote{Note that these values of $\sigma_{8}$ come from linear perturbation theory.  Non-linear effects actually push $\sigma_{8}$ up to its observed value of 1.}  Viana and Liddle \cite{Viana}, for example, find $\sigma_{8} \approx (0.6\pm 0.1)\Omega_{0}^{-\alpha}$, with $\alpha\approx 0.4$ for CDM and 0.45 for $\Lambda$CDM.  These values appear consistent with those inferred from large-scale flows \cite{Dekel,Strauss}, as well as direct observations of galaxies \cite{Loveday}.  Other fittings for $\sigma_{8}$ are those of Peacock and Dodds \cite{PD},
\begin{equation}
 \sigma_{8} = 0.75\Omega_{0}^{-0.15},
\end{equation}
and White, Efstathiou and Frenk \cite{WEF}:
\begin{equation}
 \sigma_{8} = 0.57\Omega_{0}^{-0.56}
\end{equation}
We shall assume that for $\Omega_{0}=1$, $\sigma_{8}$ falls in the range 0.5-0.8, and that if $\Omega_{0}\neq 1$, $\sigma_{8}$ scales as $\Omega_{0}^{-0.56}$.

\begin{figure}
\begin{center}
\begin{picture}(0,0)%
\includegraphics{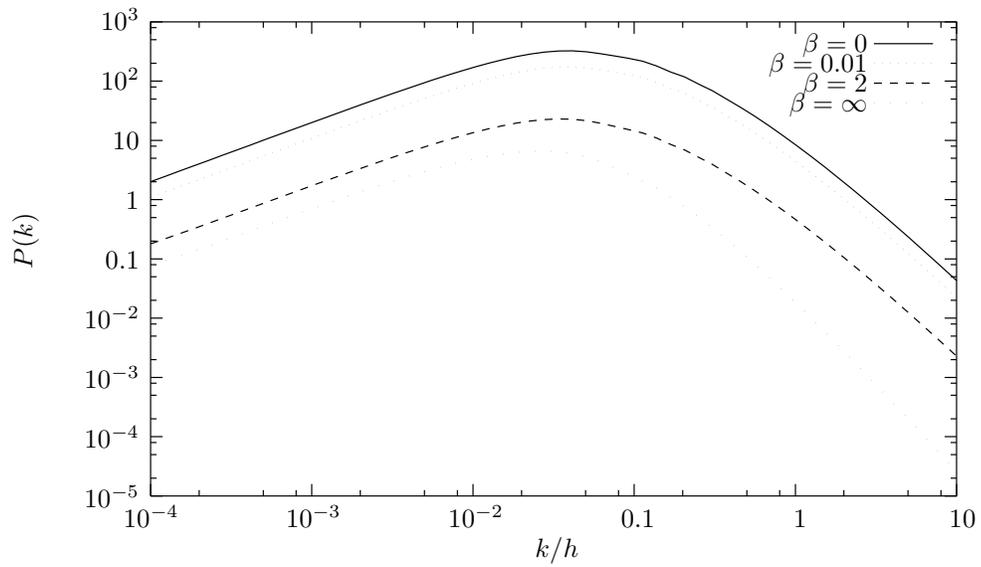}%
\end{picture}%
\setlength{\unitlength}{0.00083300in}%
\begingroup\makeatletter\ifx\SetFigFont\undefined%
\gdef\SetFigFont#1#2#3#4#5{%
  \reset@font\fontsize{#1}{#2pt}%
  \fontfamily{#3}\fontseries{#4}\fontshape{#5}%
  \selectfont}%
\fi\endgroup%
\begin{picture}(5999,3404)(304,-2967)
\put(1091,-2627){\makebox(0,0)[rb]{\smash{\SetFigFont{10}{12.0}{\familydefault}{\mddefault}{\updefault}$10^{-5}$}}}
\put(1091,-2254){\makebox(0,0)[rb]{\smash{\SetFigFont{10}{12.0}{\familydefault}{\mddefault}{\updefault}$10^{-4}$}}}
\put(1091,-1882){\makebox(0,0)[rb]{\smash{\SetFigFont{10}{12.0}{\familydefault}{\mddefault}{\updefault}$10^{-3}$}}}
\put(1091,-1509){\makebox(0,0)[rb]{\smash{\SetFigFont{10}{12.0}{\familydefault}{\mddefault}{\updefault}$10^{-2}$}}}
\put(1091,-1137){\makebox(0,0)[rb]{\smash{\SetFigFont{10}{12.0}{\familydefault}{\mddefault}{\updefault}0.1}}}
\put(1091,-764){\makebox(0,0)[rb]{\smash{\SetFigFont{10}{12.0}{\familydefault}{\mddefault}{\updefault}1}}}
\put(1091,-392){\makebox(0,0)[rb]{\smash{\SetFigFont{10}{12.0}{\familydefault}{\mddefault}{\updefault}10}}}
\put(1091,-19){\makebox(0,0)[rb]{\smash{\SetFigFont{10}{12.0}{\familydefault}{\mddefault}{\updefault}$10^{2}$}}}
\put(1091,353){\makebox(0,0)[rb]{\smash{\SetFigFont{10}{12.0}{\familydefault}{\mddefault}{\updefault}$10^{3}$}}}
\put(1165,-2751){\makebox(0,0)[b]{\smash{\SetFigFont{10}{12.0}{\familydefault}{\mddefault}{\updefault}$10^{-4}$}}}
\put(2177,-2751){\makebox(0,0)[b]{\smash{\SetFigFont{10}{12.0}{\familydefault}{\mddefault}{\updefault}$10^{-3}$}}}
\put(3190,-2751){\makebox(0,0)[b]{\smash{\SetFigFont{10}{12.0}{\familydefault}{\mddefault}{\updefault}$10^{-2}$}}}
\put(4202,-2751){\makebox(0,0)[b]{\smash{\SetFigFont{10}{12.0}{\familydefault}{\mddefault}{\updefault}0.1}}}
\put(5215,-2751){\makebox(0,0)[b]{\smash{\SetFigFont{10}{12.0}{\familydefault}{\mddefault}{\updefault}1}}}
\put(6227,-2751){\makebox(0,0)[b]{\smash{\SetFigFont{10}{12.0}{\familydefault}{\mddefault}{\updefault}10}}}
\put(425,-1137){\makebox(0,0)[b]{\smash{\SetFigFont{10}{12.0}{\familydefault}{\mddefault}{\updefault}\begin{rotate}{90}$P(k)$\end{rotate}}}}
\put(3696,-2937){\makebox(0,0)[b]{\smash{\SetFigFont{10}{12.0}{\familydefault}{\mddefault}{\updefault}$k/h$}}}
\put(5634,-156){\makebox(0,0)[rb]{\smash{\SetFigFont{10}{12.0}{\familydefault}{\mddefault}{\updefault}$\beta=\infty$}}}
\put(5634,-32){\makebox(0,0)[rb]{\smash{\SetFigFont{10}{12.0}{\familydefault}{\mddefault}{\updefault}$\beta=2$}}}
\put(5634, 92){\makebox(0,0)[rb]{\smash{\SetFigFont{10}{12.0}{\familydefault}{\mddefault}{\updefault}$\beta=0.01$}}}
\put(5634,216){\makebox(0,0)[rb]{\smash{\SetFigFont{10}{12.0}{\familydefault}{\mddefault}{\updefault}$\beta=0$}}}
\end{picture}
\caption[$P(k)$:  Varying $\beta$]{The matter power spectrum $P(k)$ for various values of $\beta$.  From top to bottom, $\beta$=0 (standard CDM with only adiabatic perturbations), 0.01, 2 and $\beta=\infty$ (pure isocurvature model).  In this plot $\Omega_{b}=0.096, h=0.5$ and $\Omega_{a}=0.5$.} \label{fig:betaPk}
\end{center}
\end{figure}

\begin{figure}
\begin{center}
\begin{picture}(0,0)%
\includegraphics{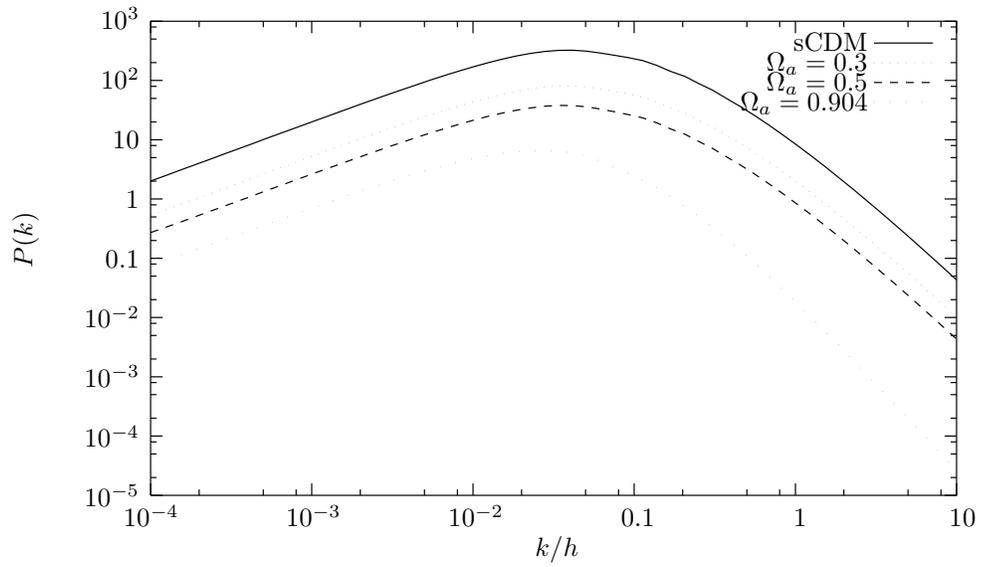}%
\end{picture}%
\setlength{\unitlength}{0.00083300in}%
\begingroup\makeatletter\ifx\SetFigFont\undefined%
\gdef\SetFigFont#1#2#3#4#5{%
  \reset@font\fontsize{#1}{#2pt}%
  \fontfamily{#3}\fontseries{#4}\fontshape{#5}%
  \selectfont}%
\fi\endgroup%
\begin{picture}(5999,3404)(304,-2967)
\put(1091,-2627){\makebox(0,0)[rb]{\smash{\SetFigFont{10}{12.0}{\familydefault}{\mddefault}{\updefault}$10^{-5}$}}}
\put(1091,-2254){\makebox(0,0)[rb]{\smash{\SetFigFont{10}{12.0}{\familydefault}{\mddefault}{\updefault}$10^{-4}$}}}
\put(1091,-1882){\makebox(0,0)[rb]{\smash{\SetFigFont{10}{12.0}{\familydefault}{\mddefault}{\updefault}$10^{-3}$}}}
\put(1091,-1509){\makebox(0,0)[rb]{\smash{\SetFigFont{10}{12.0}{\familydefault}{\mddefault}{\updefault}$10^{-2}$}}}
\put(1091,-1137){\makebox(0,0)[rb]{\smash{\SetFigFont{10}{12.0}{\familydefault}{\mddefault}{\updefault}0.1}}}
\put(1091,-764){\makebox(0,0)[rb]{\smash{\SetFigFont{10}{12.0}{\familydefault}{\mddefault}{\updefault}1}}}
\put(1091,-392){\makebox(0,0)[rb]{\smash{\SetFigFont{10}{12.0}{\familydefault}{\mddefault}{\updefault}10}}}
\put(1091,-19){\makebox(0,0)[rb]{\smash{\SetFigFont{10}{12.0}{\familydefault}{\mddefault}{\updefault}$10^{2}$}}}
\put(1091,353){\makebox(0,0)[rb]{\smash{\SetFigFont{10}{12.0}{\familydefault}{\mddefault}{\updefault}$10^{3}$}}}
\put(1165,-2751){\makebox(0,0)[b]{\smash{\SetFigFont{10}{12.0}{\familydefault}{\mddefault}{\updefault}$10^{-4}$}}}
\put(2177,-2751){\makebox(0,0)[b]{\smash{\SetFigFont{10}{12.0}{\familydefault}{\mddefault}{\updefault}$10^{-3}$}}}
\put(3190,-2751){\makebox(0,0)[b]{\smash{\SetFigFont{10}{12.0}{\familydefault}{\mddefault}{\updefault}$10^{-2}$}}}
\put(4202,-2751){\makebox(0,0)[b]{\smash{\SetFigFont{10}{12.0}{\familydefault}{\mddefault}{\updefault}0.1}}}
\put(5215,-2751){\makebox(0,0)[b]{\smash{\SetFigFont{10}{12.0}{\familydefault}{\mddefault}{\updefault}1}}}
\put(6227,-2751){\makebox(0,0)[b]{\smash{\SetFigFont{10}{12.0}{\familydefault}{\mddefault}{\updefault}10}}}
\put(425,-1137){\makebox(0,0)[b]{\smash{\SetFigFont{10}{12.0}{\familydefault}{\mddefault}{\updefault}\begin{rotate}{90}$P(k)$\end{rotate}}}}
\put(3696,-2937){\makebox(0,0)[b]{\smash{\SetFigFont{10}{12.0}{\familydefault}{\mddefault}{\updefault}$k/h$}}}
\put(5634,216){\makebox(0,0)[rb]{\smash{\SetFigFont{10}{12.0}{\familydefault}{\mddefault}{\updefault}sCDM}}}
\put(5634, 92){\makebox(0,0)[rb]{\smash{\SetFigFont{10}{12.0}{\familydefault}{\mddefault}{\updefault}$\Omega_{a}=0.3$}}}
\put(5634,-32){\makebox(0,0)[rb]{\smash{\SetFigFont{10}{12.0}{\familydefault}{\mddefault}{\updefault}$\Omega_{a}=0.5$}}}
\put(5634,-156){\makebox(0,0)[rb]{\smash{\SetFigFont{10}{12.0}{\familydefault}{\mddefault}{\updefault}$\Omega_{a}=0.904$}}}
\end{picture}
\caption[$P(k)$:  Varying $\Omega_{a}$]{The matter power spectrum $P(k)$ for various values of $\Omega_{a}$ in the case that the adiabatic and isocurvature perturbations have equal amplitudes ($\beta$=1).  The top line is standard CDM with adiabatic perturbations.  Going down, $\Omega_{a}$=0.3, 0.5 and 0.904.  In this plot $\Omega_{b}=0.096$ and $h=0.5$.} \label{fig:omegaPk}
\end{center}
\end{figure}

What does this have to do with isocurvature perturbations?  One hopes that by adding an isocurvature component to the mix that the amount of small scale power can be reduced when the power spectrum is normalized to the large scale COBE data.  This is because in a purely isentropic model, the amplitude of the photon perturbation on large scales (which is what is measured by COBE) is approximately the same as that of the matter fluctuation.   The COBE normalization reveals that matter fluctuations on small scales are too large for standard CDM models.  For isocurvature fluctuations, the fluctuation in the matter on large scales is smaller than the fluctuation in the photon perturbations because of the isocurvature boost in the anisotropy on large angular scales as described in Equation~\ref{eq:C2iso}.  So when the power spectrum is normalized to the COBE data, the amplitude of the matter power spectrum is lower in the isocurvature case than in the isentropic case.  As some examples, Figure~\ref{fig:betaPk} shows the matter power spectrum $P(k)$ for various values of $\beta$, while Figure~\ref{fig:omegaPk} shows the matter power spectrum $P(k)$ for various values of $\Omega_{a}$.  As can be seen from both of these figures, adding an isocurvature component, either by varying $\beta$ or $\Omega_{a}$, decreases the amount of small scale (high $k$) power as compared to the standard isentropic CDM scenarios.

\begin{figure}
\begin{center}
\begin{picture}(0,0)%
\includegraphics{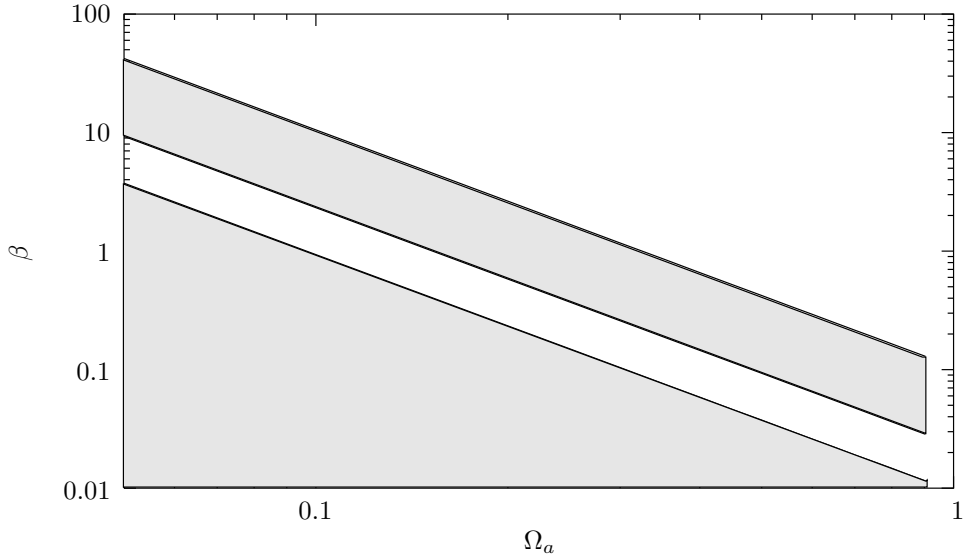}%
\end{picture}%
\setlength{\unitlength}{0.00083300in}%
\begingroup\makeatletter\ifx\SetFigFont\undefined%
\gdef\SetFigFont#1#2#3#4#5{%
  \reset@font\fontsize{#1}{#2pt}%
  \fontfamily{#3}\fontseries{#4}\fontshape{#5}%
  \selectfont}%
\fi\endgroup%
\begin{picture}(5949,3404)(316,-2967)
\put(943,-2627){\makebox(0,0)[rb]{\smash{\SetFigFont{10}{12.0}{\familydefault}{\mddefault}{\updefault}0.01}}}
\put(943,-1882){\makebox(0,0)[rb]{\smash{\SetFigFont{10}{12.0}{\familydefault}{\mddefault}{\updefault}0.1}}}
\put(943,-1137){\makebox(0,0)[rb]{\smash{\SetFigFont{10}{12.0}{\familydefault}{\mddefault}{\updefault}1}}}
\put(943,-392){\makebox(0,0)[rb]{\smash{\SetFigFont{10}{12.0}{\familydefault}{\mddefault}{\updefault}10}}}
\put(943,353){\makebox(0,0)[rb]{\smash{\SetFigFont{10}{12.0}{\familydefault}{\mddefault}{\updefault}100}}}
\put(2222,-2751){\makebox(0,0)[b]{\smash{\SetFigFont{10}{12.0}{\familydefault}{\mddefault}{\updefault}0.1}}}
\put(6227,-2751){\makebox(0,0)[b]{\smash{\SetFigFont{10}{12.0}{\familydefault}{\mddefault}{\updefault}1}}}
\put(425,-1137){\makebox(0,0)[b]{\smash{\SetFigFont{10}{12.0}{\familydefault}{\mddefault}{\updefault}\begin{rotate}{90}$\beta$\end{rotate}}}}
\put(3622,-2937){\makebox(0,0)[b]{\smash{\SetFigFont{10}{12.0}{\familydefault}{\mddefault}{\updefault}$\Omega_{a}$}}}
\end{picture}
\caption[$\beta-\Omega_{a}$ Parameter Space for Saskatoon Data Point $l=87$ and $\sigma_{8}$]{Acceptable regions of the axion parameter space using $\sigma_{8}$ and the Saskatoon experiment.  The top region (between the dotted lines) is the allowed parameter space for $\sigma_{8}$ and the bottom region (beneath the solid line) is the allowed parameter space for the Saskatoon experiment at $l=87$.  In this plot $\Omega_{b}=0.096$ and $h=0.5$.} \label{fig:87sigpar}
\end{center}
\end{figure}

\begin{figure}
\begin{center}
\begin{picture}(0,0)%
\includegraphics{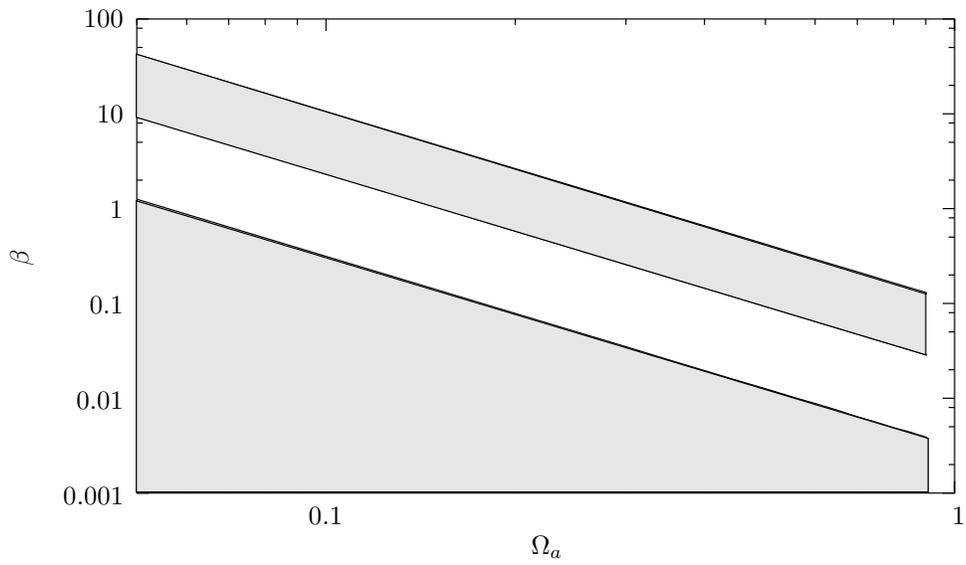}%
\end{picture}%
\setlength{\unitlength}{0.00083300in}%
\begingroup\makeatletter\ifx\SetFigFont\undefined%
\gdef\SetFigFont#1#2#3#4#5{%
  \reset@font\fontsize{#1}{#2pt}%
  \fontfamily{#3}\fontseries{#4}\fontshape{#5}%
  \selectfont}%
\fi\endgroup%
\begin{picture}(5949,3404)(316,-2967)
\put(1017,-2627){\makebox(0,0)[rb]{\smash{\SetFigFont{10}{12.0}{\familydefault}{\mddefault}{\updefault}0.001}}}
\put(1017,-2031){\makebox(0,0)[rb]{\smash{\SetFigFont{10}{12.0}{\familydefault}{\mddefault}{\updefault}0.01}}}
\put(1017,-1435){\makebox(0,0)[rb]{\smash{\SetFigFont{10}{12.0}{\familydefault}{\mddefault}{\updefault}0.1}}}
\put(1017,-839){\makebox(0,0)[rb]{\smash{\SetFigFont{10}{12.0}{\familydefault}{\mddefault}{\updefault}1}}}
\put(1017,-243){\makebox(0,0)[rb]{\smash{\SetFigFont{10}{12.0}{\familydefault}{\mddefault}{\updefault}10}}}
\put(1017,353){\makebox(0,0)[rb]{\smash{\SetFigFont{10}{12.0}{\familydefault}{\mddefault}{\updefault}100}}}
\put(2279,-2751){\makebox(0,0)[b]{\smash{\SetFigFont{10}{12.0}{\familydefault}{\mddefault}{\updefault}0.1}}}
\put(6227,-2751){\makebox(0,0)[b]{\smash{\SetFigFont{10}{12.0}{\familydefault}{\mddefault}{\updefault}1}}}
\put(425,-1137){\makebox(0,0)[b]{\smash{\SetFigFont{10}{12.0}{\familydefault}{\mddefault}{\updefault}\begin{rotate}{90}$\beta$\end{rotate}}}}
\put(3659,-2937){\makebox(0,0)[b]{\smash{\SetFigFont{10}{12.0}{\familydefault}{\mddefault}{\updefault}$\Omega_{a}$}}}
\end{picture}
\caption[$\beta-\Omega_{a}$ Parameter Space for Saskatoon Data Point $l=237$ and $\sigma_{8}$]{Acceptable regions of the axion parameter space using $\sigma_{8}$ and the Saskatoon experiment.  The top region (between the dotted lines) is the allowed parameter space for $\sigma_{8}$ and the bottom region (beneath the solid line) is the allowed parameter space for the Saskatoon experiment at $l=237$.  In this plot $\Omega_{b}=0.096$ and $h=0.5$.} \label{fig:237sigpar}
\end{center}
\end{figure}

Does adding an isocurvature component to CDM really work?  The fact that $\sigma_{8}$ seems to lie in the range $0.5\stackrel{<}{\sim}\sigma_{8}\stackrel{<}{\sim}0.8$ can also be used to constrain the parameter space of $\beta$ and $\Omega_{a}$.  Figure~\ref{fig:87sigpar} shows the allowed region of $\beta -\Omega_{a}$ parameter space using both the Saskatoon experiment ($l=87$) and $\sigma_{8}$.  Figure~\ref{fig:237sigpar} shows the allowed region of $\beta -\Omega_{a}$ parameter space using both the Saskatoon experiment ($l=237$) and $\sigma_{8}$.  Notice that although large scale structure in the form of $\sigma_{8}$ allows larger values of $\beta$ than the Saskatoon data, in general $\beta$ should be at most fifty to be consistent with $\sigma_{8}$.

\begin{figure}
\begin{center}
\begin{picture}(0,0)%
\includegraphics{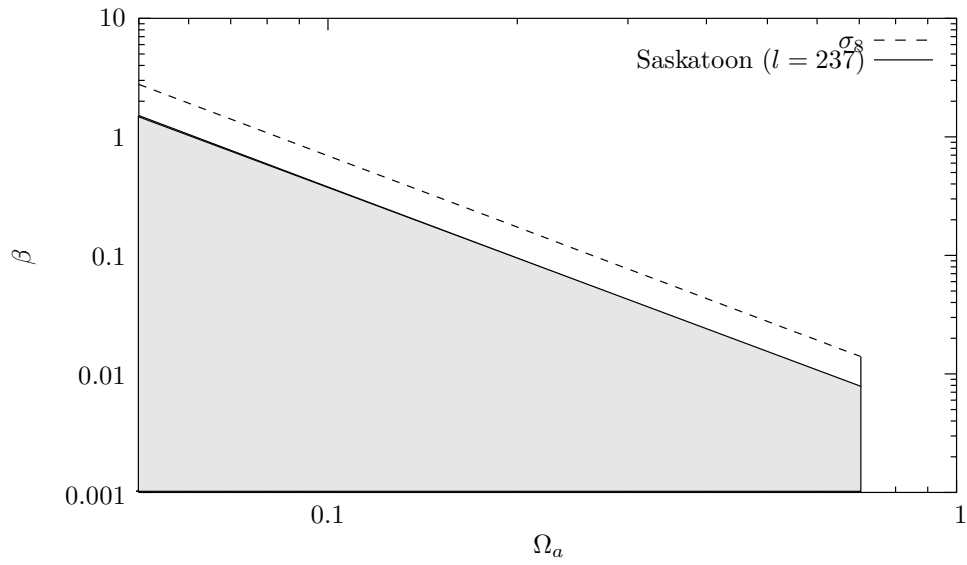}%
\end{picture}%
\setlength{\unitlength}{0.00083300in}%
\begingroup\makeatletter\ifx\SetFigFont\undefined%
\gdef\SetFigFont#1#2#3#4#5{%
  \reset@font\fontsize{#1}{#2pt}%
  \fontfamily{#3}\fontseries{#4}\fontshape{#5}%
  \selectfont}%
\fi\endgroup%
\begin{picture}(5949,3404)(316,-2967)
\put(1017,-2627){\makebox(0,0)[rb]{\smash{\SetFigFont{10}{12.0}{\familydefault}{\mddefault}{\updefault}0.001}}}
\put(1017,-1882){\makebox(0,0)[rb]{\smash{\SetFigFont{10}{12.0}{\familydefault}{\mddefault}{\updefault}0.01}}}
\put(1017,-1137){\makebox(0,0)[rb]{\smash{\SetFigFont{10}{12.0}{\familydefault}{\mddefault}{\updefault}0.1}}}
\put(1017,-392){\makebox(0,0)[rb]{\smash{\SetFigFont{10}{12.0}{\familydefault}{\mddefault}{\updefault}1}}}
\put(1017,353){\makebox(0,0)[rb]{\smash{\SetFigFont{10}{12.0}{\familydefault}{\mddefault}{\updefault}10}}}
\put(2279,-2751){\makebox(0,0)[b]{\smash{\SetFigFont{10}{12.0}{\familydefault}{\mddefault}{\updefault}0.1}}}
\put(6227,-2751){\makebox(0,0)[b]{\smash{\SetFigFont{10}{12.0}{\familydefault}{\mddefault}{\updefault}1}}}
\put(3659,-2937){\makebox(0,0)[b]{\smash{\SetFigFont{10}{12.0}{\familydefault}{\mddefault}{\updefault}$\Omega_{a}$}}}
\put(425,-1137){\makebox(0,0)[b]{\smash{\SetFigFont{10}{12.0}{\familydefault}{\mddefault}{\updefault}\begin{rotate}{90}$\beta$\end{rotate}}}}
\put(5634,216){\makebox(0,0)[rb]{\smash{\SetFigFont{10}{12.0}{\familydefault}{\mddefault}{\updefault}$\sigma_{8}$}}}
\put(5634, 92){\makebox(0,0)[rb]{\smash{\SetFigFont{10}{12.0}{\familydefault}{\mddefault}{\updefault}Saskatoon ($l=237$)}}}
\end{picture}
\caption[$\beta-\Omega_{a}$ Parameter Space for Saskatoon Data Point $l=237$ and $\sigma_{8}$ in the case of $\Omega_{\nu}=0.2$ with 2 equally massive neutrinos.]{Acceptable regions of the axion parameter space using the Saskatoon experiment and $\sigma_{8}$.  The solid line is the upper boundary of the allowed parameter space for the Saskatoon experiment at $l=237$ and the dashed line is the upper boundary of the parameter space for $\sigma_{8}$.  In this plot $\Omega_{b}=0.096$, $\Omega_{\nu}=0.2$, $N_{\nu}=2$ and $h=0.5$.} \label{fig:237nupar}
\end{center}
\end{figure}

As can be seen from these figures, CDM models with a mixture of isocurvature perturbations cannot satisfy the Saskatoon data and $\sigma_{8}$ simultaneously.  Satisfying the Saskatoon data requires a relatively small value of $\beta$, which makes $\sigma_{8}$ too large.  Satisfying the values of $\sigma_{8}$  makes $\beta$ larger, which lowers the power spectrum too much to satisfy the Saskatoon constraints.  Interestingly, however, the two constraints can be satisfied simultaneously if we introduce a hot dark matter component ($\Omega_{\nu}=0.2$ with 2 equally massive neutrinos), as is evident in Figure~\ref{fig:237nupar}.  There are two reasons for this.  First, adding HDM increases the height of the Doppler peak because of the ISW effect.  This increases the allowed region of $\beta-\Omega_{a}$ parameter space.  Second, the free-streaming of neutrinos reduces the small scale power, lowering $\sigma_{8}$ and the allowed region of $\beta-\Omega_{a}$ parameter space.

\section{Discussion}
The Saskatoon and $\sigma_{8}$ restrictions on $\beta$ have interesting consequences for axion isocurvature perturbations and inflation.  The apparent rise in the angular power spectrum at $l \sim 200$, as indicated by the Saskatoon data, is a strong indication that the isocurvature perturbations do not dominate the isentropic ones.   Although high precision satellite experiments may be able to detect the presence of isocurvature fluctuations, as yet there are not enough data to determine the exact nature of any isocurvature contribution.  We may, however, speculate about the consequences for axions and inflation, depending on whether or not isocurvature perturbations are detected.  Each one of these possibilities has interesting consequences for both axion and inflation models.  Let us examine each in turn.

\subsection{A Universe With No Isocurvature Perturbations}
Before proceeding, recall that for $f_{PQ} \stackrel{<}{\sim}1.6\times 10^{18}$GeV, $\beta$ is given by Equation~\ref{eq:powerl}:
\begin{eqnarray}
\beta &\approx &3 \times (10^{4}-10^{5})\left(\frac{m_{Pl}}{f_{PQ}}\right)^{0.82}\left[\frac{\Omega_{a}}{(\Omega_{0}h)^{2}}\right] \left[\left(\frac{f_{PQ}}{f_{a}}\right)^{2}f(\bar{\Theta})\right](-n_{T}) \nonumber ,
\end{eqnarray}
while for $f_{PQ}\stackrel{>}{\sim}1.6\times 10^{18}$GeV, $\beta$ is given by Equation~\ref{eq:powerg}:
\begin{eqnarray}
\beta &\approx &(0.6-1.0)\times 10^{5}\sqrt{\frac{m_{Pl}}{f_{PQ}}}\left[\frac{\Omega_{a}}{(\Omega_{0}h)^{2}}\right] \left[\left(\frac{f_{PQ}}{f_{a}}\right)^{2}f(\bar{\Theta})\right](-n_{T}) \nonumber
\end{eqnarray}
In the case that no isocurvature perturbations are observed, $\beta$ is either very small or zero.  The trivial solution then would be $\Omega_{a} = 0$.  Let us assume, however, that axions are discovered.  In that case, it seems reasonable to assume that axions constitute a sizable fraction of the dark matter, i.e., that $\Omega_{a}$ is ${\cal O}(1)$.  Throughout the rest of the discussion, we shall assume that this is indeed the case.  Note that even if we assume that $\Omega_{a}$ is ${\cal O}(1)$, if $f_{PQ}$, $f_{a}$ and $n_{T}$ are all unknown, it is not possible to say anything definitive about either inflation or axions, other than $\frac{H_{I}}{f_{a}\bar{\Theta}} \ll 1$.

\begin{figure}
\begin{center}
\begin{picture}(0,0)%
\includegraphics{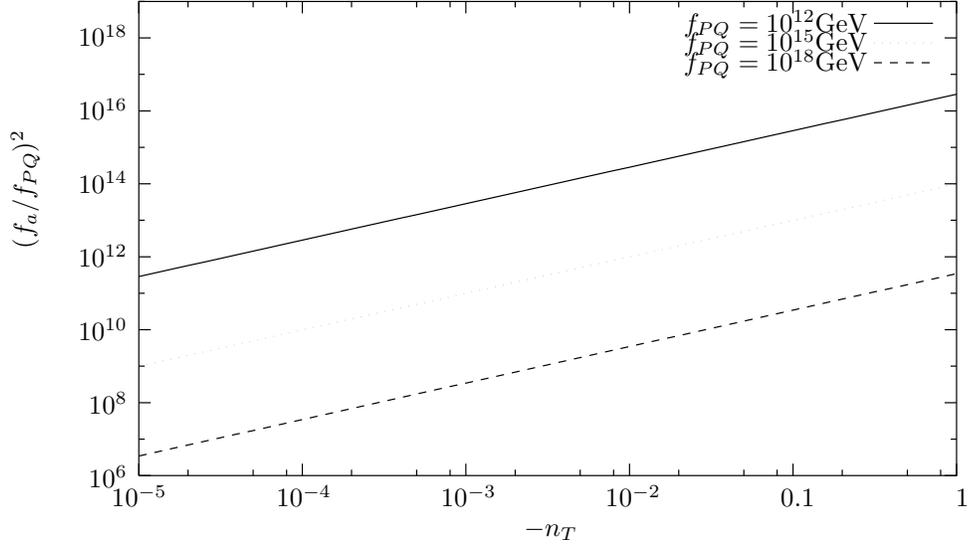}%
\end{picture}%
\setlength{\unitlength}{0.00083300in}%
\begingroup\makeatletter\ifx\SetFigFont\undefined%
\gdef\SetFigFont#1#2#3#4#5{%
  \reset@font\fontsize{#1}{#2pt}%
  \fontfamily{#3}\fontseries{#4}\fontshape{#5}%
  \selectfont}%
\fi\endgroup%
\begin{picture}(5961,3394)(304,-2967)
\put(1017,-2627){\makebox(0,0)[rb]{\smash{\SetFigFont{10}{12.0}{\familydefault}{\mddefault}{\updefault}$10^{6}$}}}
\put(1017,-2169){\makebox(0,0)[rb]{\smash{\SetFigFont{10}{12.0}{\familydefault}{\mddefault}{\updefault}$10^{8}$}}}
\put(1017,-1710){\makebox(0,0)[rb]{\smash{\SetFigFont{10}{12.0}{\familydefault}{\mddefault}{\updefault}$10^{10}$}}}
\put(1017,-1252){\makebox(0,0)[rb]{\smash{\SetFigFont{10}{12.0}{\familydefault}{\mddefault}{\updefault}$10^{12}$}}}
\put(1017,-793){\makebox(0,0)[rb]{\smash{\SetFigFont{10}{12.0}{\familydefault}{\mddefault}{\updefault}$10^{14}$}}}
\put(1017,-335){\makebox(0,0)[rb]{\smash{\SetFigFont{10}{12.0}{\familydefault}{\mddefault}{\updefault}$10^{16}$}}}
\put(1017,124){\makebox(0,0)[rb]{\smash{\SetFigFont{10}{12.0}{\familydefault}{\mddefault}{\updefault}$10^{18}$}}}
\put(1091,-2751){\makebox(0,0)[b]{\smash{\SetFigFont{10}{12.0}{\familydefault}{\mddefault}{\updefault}$10^{-5}$}}}
\put(2118,-2751){\makebox(0,0)[b]{\smash{\SetFigFont{10}{12.0}{\familydefault}{\mddefault}{\updefault}$10^{-4}$}}}
\put(3145,-2751){\makebox(0,0)[b]{\smash{\SetFigFont{10}{12.0}{\familydefault}{\mddefault}{\updefault}$10^{-3}$}}}
\put(4173,-2751){\makebox(0,0)[b]{\smash{\SetFigFont{10}{12.0}{\familydefault}{\mddefault}{\updefault}$10^{-2}$}}}
\put(5200,-2751){\makebox(0,0)[b]{\smash{\SetFigFont{10}{12.0}{\familydefault}{\mddefault}{\updefault}0.1}}}
\put(6227,-2751){\makebox(0,0)[b]{\smash{\SetFigFont{10}{12.0}{\familydefault}{\mddefault}{\updefault}1}}}
\put(425,-1137){\makebox(0,0)[b]{\smash{\SetFigFont{10}{12.0}{\familydefault}{\mddefault}{\updefault}\begin{rotate}{90}$(f_{a}/f_{PQ})^{2}$\end{rotate}}}}
\put(3659,-2937){\makebox(0,0)[b]{\smash{\SetFigFont{10}{12.0}{\familydefault}{\mddefault}{\updefault}$-n_{T}$}}}
\put(5634, 92){\makebox(0,0)[rb]{\smash{\SetFigFont{10}{12.0}{\familydefault}{\mddefault}{\updefault}$f_{PQ}=10^{15}$GeV}}}
\put(5634,-32){\makebox(0,0)[rb]{\smash{\SetFigFont{10}{12.0}{\familydefault}{\mddefault}{\updefault}$f_{PQ}=10^{18}$GeV}}}
\put(5634,216){\makebox(0,0)[rb]{\smash{\SetFigFont{10}{12.0}{\familydefault}{\mddefault}{\updefault}$f_{PQ}=10^{12}$GeV}}}
\end{picture}
\caption[No isocurvature perturbations:  $(f_{PQ}/f_{a})^{2}$ vs. $-n_{T}$]{Possible values of $(f_{PQ}/f_{a})^{2}$ as a function of $-n_{T}$ for (from top to bottom) $f_{PQ} = 10^{12}$GeV, $10^{15}$GeV and $10^{18}$GeV, in the case that no isocurvature perturbations are detected ($\beta = 10^{-5} \ll 1$).  Here we have assumed that $\Omega_{a}=0.95$.} \label{fig:fant}
\end{center}
\end{figure}

Suppose, however, that the tensor spectral index $n_{T}$ is measurable.  Typical inflationary models seem to make several rough predictions for $n_{T}$ (see Table~\ref{tab:modelnT}):  $-n_{T} \ll 1$, as in the new and ``unnatural" natural inflation scenarios,  $-n_{T}$ is ${\cal O}(0.01)$, as in the chaotic inflation model, or $-n_{T}>0.3$, as in extended inflation.   First, consider the case that $-n_{T} \ll 1$ and no isocurvature perturbations are observed.  This would correspond to $\beta \ll 1$.  Since measuring $n_{T}$ will be extremely challenging, the errors themselves will more than likely make it impossible to differentiate between $-n_{T}=10^{-4}$ and $-n_{T}=0$, for example.   With this uncertainty in $-n_{T}$, it would probably be impossible to say anything about $f_{a}/f_{PQ}$, even though $\beta \ll 1$.  If, on the other hand, $-n_{T}$ is ${\cal O}(0.01)$ or larger, then one can see from the formulae for $\beta$ that either $\Omega_{a} \ll 1$ or $(f_{PQ}/f_{a})^{2} \ll 1$.  The latter case suggests that inflation occurred well above the energy scale $f_{PQ}$, and that the PQ field probably had some kind of chaotic distribution, as in the chaotic inflation scenario.  Figure~\ref{fig:fant} shows $(f_{a}/f_{PQ})^{2}$ as a function of $n_{T}$ for $\beta =10^{-5} \ll 1$. 

Now let us go one step further and assume that either or both of the Kyoto \cite{Matsuki1,Matsuki2} or LLNL \cite{vanBibber} experiments find axions with a mass $m_{a}$.  Then $f_{PQ}/N$ can be determined, since the axion mass $m_{a}$ is related to the PQ symmetry breaking scale by
\begin{equation}
  m_{a} \simeq 0.62\mbox{eV}\frac{10^{7}\mbox{GeV}}{f_{PQ}/N}
\end{equation}
This might have interesting consequences for inflation or axions.  As before, assume that $\Omega_{a}$ is ${\cal O}(1)$ and $-n_{T} \sim 0.01$.  If $f_{PQ}$ is known, it is possible to place a lower bound on $f_{a}$, or equivalently, the value of the inflaton field $\varphi_{50}$ fifty $e$-foldings before the end of inflation.  Consider an example:  If $\Omega_{a} \approx 1$ and $-n_{T} = 0.01$, then $f_{a} = \varphi_{50} \stackrel{>}{\sim} 10^{4} f_{PQ}$.  In this sense, even the absence of isocurvature perturbations would provide important information about inflation.  Note, however, that there is also a different way to interpret this result.  Although interesting, such a scenario also seems unlikely, in that it requires the PQ symmetry breaking phase transition to span at least four orders of magnitude.  It might be more sensible in this case that $\Omega_{a} \sim 0$.

\subsection{A Universe With Isocurvature Perturbations}
A more intriguing possibility is that axion isocurvature fluctuations do in fact make a detectable contribution to the power spectra.\footnote{In the discussion that follows, we will assume that there is no other source of isocurvature fluctuations.}  The CMB anisotropies could then be used to constrain the allowed values of $\beta$ and $\Omega_{a}$.  In addition, from the equations for $\beta$, we can see that this precludes an exactly scale-invariant gravitational wave spectrum.\footnote{Even if the power spectrum of gravitational waves were not observed, the observation of isocurvature perturbations would be a strong signal for a tilted power spectrum of gravitational waves.}  Such a detection would more than likely also mean that $\Omega_{a}$ is indeed ${\cal O}(1)$ and that axions make a significant contribution to the matter density.  But without knowing $f_{PQ}$, $f_{a}$ and $n_{T}$, it would probably not be possible to say much more than this.

As before, assume that $-n_{T}$ is measured.  Now things begin to get interesting.  First, suppose that $-n_{T} \sim 0.01$, as predicted in the chaotic inflation model.  As we saw in Section~\ref{sec-CMBparam}, Saskatoon constrains $\beta$ to be less than a few, and high-precision satellite experiments could conceivably limit $\beta$ in the same way.  Furthermore, suppose that $\Omega_{a} \sim 1$ and $\beta \sim 1$.  This would mean that $(f_{a}/f_{PQ})^{2} \sim 10^{4}$, which is reminiscent of Linde's chaotic inflation with a time-varying $f_{a}$.  Moreover, if axions were discovered with a mass $m_{a}$, this would fix $f_{PQ}$ and it would be possible to estimate $f_{a} = \varphi_{50}$ fifty $e$-foldings before the end of inflation.

\begin{figure}
\begin{center}
\begin{picture}(0,0)%
\includegraphics{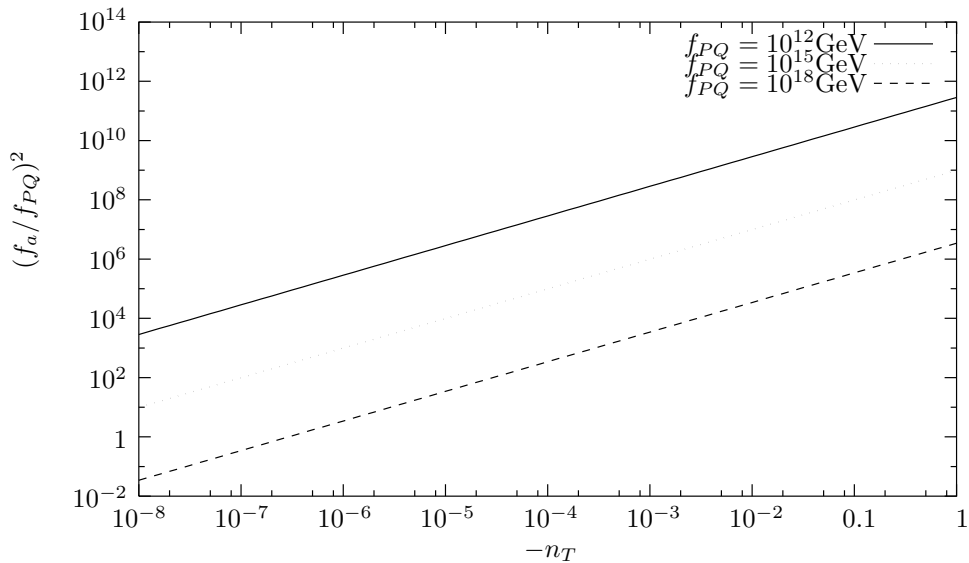}%
\end{picture}%
\setlength{\unitlength}{0.00083300in}%
\begingroup\makeatletter\ifx\SetFigFont\undefined%
\gdef\SetFigFont#1#2#3#4#5{%
  \reset@font\fontsize{#1}{#2pt}%
  \fontfamily{#3}\fontseries{#4}\fontshape{#5}%
  \selectfont}%
\fi\endgroup%
\begin{picture}(5961,3404)(304,-2967)
\put(1017,-2627){\makebox(0,0)[rb]{\smash{\SetFigFont{10}{12.0}{\familydefault}{\mddefault}{\updefault}$10^{-2}$}}}
\put(1017,-2254){\makebox(0,0)[rb]{\smash{\SetFigFont{10}{12.0}{\familydefault}{\mddefault}{\updefault}1}}}
\put(1017,-1882){\makebox(0,0)[rb]{\smash{\SetFigFont{10}{12.0}{\familydefault}{\mddefault}{\updefault}$10^{2}$}}}
\put(1017,-1509){\makebox(0,0)[rb]{\smash{\SetFigFont{10}{12.0}{\familydefault}{\mddefault}{\updefault}$10^{4}$}}}
\put(1017,-1137){\makebox(0,0)[rb]{\smash{\SetFigFont{10}{12.0}{\familydefault}{\mddefault}{\updefault}$10^{6}$}}}
\put(1017,-764){\makebox(0,0)[rb]{\smash{\SetFigFont{10}{12.0}{\familydefault}{\mddefault}{\updefault}$10^{8}$}}}
\put(1017,-392){\makebox(0,0)[rb]{\smash{\SetFigFont{10}{12.0}{\familydefault}{\mddefault}{\updefault}$10^{10}$}}}
\put(1017,-19){\makebox(0,0)[rb]{\smash{\SetFigFont{10}{12.0}{\familydefault}{\mddefault}{\updefault}$10^{12}$}}}
\put(1017,353){\makebox(0,0)[rb]{\smash{\SetFigFont{10}{12.0}{\familydefault}{\mddefault}{\updefault}$10^{14}$}}}
\put(1091,-2751){\makebox(0,0)[b]{\smash{\SetFigFont{10}{12.0}{\familydefault}{\mddefault}{\updefault}$10^{-8}$}}}
\put(1733,-2751){\makebox(0,0)[b]{\smash{\SetFigFont{10}{12.0}{\familydefault}{\mddefault}{\updefault}$10^{-7}$}}}
\put(2375,-2751){\makebox(0,0)[b]{\smash{\SetFigFont{10}{12.0}{\familydefault}{\mddefault}{\updefault}$10^{-6}$}}}
\put(3017,-2751){\makebox(0,0)[b]{\smash{\SetFigFont{10}{12.0}{\familydefault}{\mddefault}{\updefault}$10^{-5}$}}}
\put(3659,-2751){\makebox(0,0)[b]{\smash{\SetFigFont{10}{12.0}{\familydefault}{\mddefault}{\updefault}$10^{-4}$}}}
\put(4301,-2751){\makebox(0,0)[b]{\smash{\SetFigFont{10}{12.0}{\familydefault}{\mddefault}{\updefault}$10^{-3}$}}}
\put(4943,-2751){\makebox(0,0)[b]{\smash{\SetFigFont{10}{12.0}{\familydefault}{\mddefault}{\updefault}$10^{-2}$}}}
\put(5585,-2751){\makebox(0,0)[b]{\smash{\SetFigFont{10}{12.0}{\familydefault}{\mddefault}{\updefault}0.1}}}
\put(6227,-2751){\makebox(0,0)[b]{\smash{\SetFigFont{10}{12.0}{\familydefault}{\mddefault}{\updefault}1}}}
\put(425,-1137){\makebox(0,0)[b]{\smash{\SetFigFont{10}{12.0}{\familydefault}{\mddefault}{\updefault}\begin{rotate}{90}$(f_{a}/f_{PQ})^{2}$\end{rotate}}}}
\put(3659,-2937){\makebox(0,0)[b]{\smash{\SetFigFont{10}{12.0}{\familydefault}{\mddefault}{\updefault}$-n_{T}$}}}
\put(5634,-32){\makebox(0,0)[rb]{\smash{\SetFigFont{10}{12.0}{\familydefault}{\mddefault}{\updefault}$f_{PQ}=10^{18}$GeV}}}
\put(5634, 92){\makebox(0,0)[rb]{\smash{\SetFigFont{10}{12.0}{\familydefault}{\mddefault}{\updefault}$f_{PQ}=10^{15}$GeV}}}
\put(5634,216){\makebox(0,0)[rb]{\smash{\SetFigFont{10}{12.0}{\familydefault}{\mddefault}{\updefault}$f_{PQ}=10^{12}$GeV}}}
\end{picture}
\caption[Isocurvature perturbations:  $(f_{PQ}/f_{a})^{2}$ vs. $-n_{T}$]{Possible values of $(f_{PQ}/f_{a})^{2}$ as a function of $-n_{T}$ for $f_{PQ} = 10^{12}$GeV, $10^{15}$GeV and $10^{18}$GeV, in the case that isocurvature perturbations are detected ($\beta = 1$).  Here we have assumed that $\Omega_{a}=0.95$.} \label{fig:fant1}
\end{center}
\end{figure}

On the other hand, suppose that $-n_{T}$ were much less than one.  Once again, it would probably not be possible to tell if $-n_{T}$ were $10^{-4}$ or zero.  In this case, it will be difficult to say anything about inflation or $f_{a}/f_{PQ}$.  This is because $-n_{T} \stackrel{<}{\sim} 10^{-6}$ when $f_{a}/f_{PQ}$ switches from being greater than one (which corresponds to Linde's chaotic inflation scenario) to being less than or equal to one (which would correspond to either Pi's model or PQ symmetry breaking before inflation begins).   Figure~\ref{fig:fant1} shows some of the values of $(f_{PQ}/f_{a})^{2}$ as a function of $n_{T}$ for several values of $f_{PQ}$ in the case that $\beta = 1$.  Unfortunately, even if isocurvature perturbations are detected, it will not be possible to say anything definitive about inflation if $-n_{T} \ll 1$, despite the fact that different models predict different values of $f_{a}/f_{PQ}$.
    
\section{Conclusions}
As we have seen, the quadrupole anisotropy measured by COBE, $\Delta T \sim 18 \mu$K, strongly constrains inflationary models in a universe that contains axions.  First, consider the case when PQ symmetry breaking occurs before inflation.  For the canonical PQ symmetry breaking parameters $\bar{\Theta} \sim 1$ and $f_{PQ} \sim 10^{12}$GeV, the Hubble constant must be rather small, $H_{I} \stackrel{<}{\sim} 10^{8}$GeV.  Even if we let $f_{PQ}$ range up to $m_{Pl}$, $H_{I}$ must still be less than about $10^{11}$GeV to satisfy the COBE quadrupole.  In this case, neither chaotic inflation nor exponential inflation produces isocurvature fluctuations of an acceptable magnitude.  The ``unnatural" regime of natural inflation can produce both acceptable and interesting isocurvature fluctuations if $m_{Pl}/3 \stackrel{<}{\sim} f \stackrel{<}{\sim} 5 \times 10^{18}$GeV.  New inflation can also produce acceptable isocurvature perturbations, but they are probably unobservable unless $f_{PQ}$ and $\sigma$ are near the Planck scale.

If, however, $f_{a}$ is still evolving to its SSB value of $f_{PQ}/\sqrt{2}$ during inflation, the chaotic, exponential and new models of inflation fare a little better.  In the case of chaotic inflation, making $f_{a}$ a dynamical variable eases the restrictions on $H_{I}$.  In particular, if $f_{a} \stackrel{>}{\sim}10^{18}$GeV, it is possible to have a Hubble constant as large as $10^{14}$GeV, as predicted in chaotic and exponential inflation models.  A drawback of this scenario, however, is that for the Hubble constant to be this large, and for the isocurvature perturbations to be compatible with COBE, $f_{PQ}$ must be about $10^{12}$GeV.  This seems somewhat unnatural though, since it requires the PQ symmetry breaking phase transition to span some six orders of magnitude.  A model like that of Pi, in which the PQ field is responsible for inflation, does quite well in producing acceptable and interesting isocurvature fluctuations if $\sigma \sim 10^{16}$GeV, right near the GUT scale.  

Recent work has indicated that the microwave background may be able to measure the matter content of the universe, including $\Omega$, $\Omega_{b}$ and $\Omega_{\nu}.$  Motivated by these results, we have considered the possibility that the microwave background may be able to give indications about the cold dark matter content as well.  Here we considered the effects of a mixed cold dark matter model, namely axions and neutralinos, which is quite sensible from a particle physics point of view, particularly in the context of string theory.  If the primordial fluctuations are purely isentropic, the two types of cold dark matter would be indistinguishable.  But if there are isocurvature axion perturbations that are comparable to the isentropic perturbations, axions may leave a distinct signature on the microwave background.  This signature depends not only on the amplitude $\beta$ of the perturbations, but also on the value of $\Omega_{a}.$  Unfortunately, these two parameters are degenerate and it seems unlikely that CMB experiments or large scale structure measurements will be able to extract the independent values of $\beta$ and $\Omega_{a}$.

An attractive benefit of adding an isocurvature component is the hope of reconciling the COBE quadrupole with $\sigma_{8}$ \cite{Kawasaki}.  However, it does not appear likely that simply adding some amount of isocurvature power, as is the case for axion isocurvature perturbations, will be able to solve the problems of cold dark matter.  Solving these problems appears to require adding some hot dark matter, a cosmological constant or a tilted power spectrum.\footnote{Although the amount of tilt required to bring CDM into agreement with large scale structure  seems to predict too little anisotropy on small angular scales.}

The strength of the isocurvature contribution is rather strongly constrained by the Saskatoon data, but more weakly constrained by $\sigma_{8}$.  According to the Saskatoon data, unless $\Omega_{a}$ is small compared to $\Omega_{x}$, $\beta$ should be $\stackrel{<}{\sim}0.1$ for $\l=87$ and $\stackrel{<}{\sim}10^{-3}-10^{-2}$ for $\l=237$.  Presently it is too early to determine the exact nature of any isocurvature contribution.  In the next few years, however, the MAP and PLANCK satellite experiments should either be able to identify or strongly constrain any isocurvature contribution.  This will have some interesting consequences for axions and inflationary models.  First suppose that isocurvature perturbations are detected.  This in itself would suggest that $\Omega_{a}$ is ${\cal O}(1)$.  Then if it were possible to measure $n_{T}$, and depending upon its value, it might be possible to say something about various inflationary scenarios.  For example, if $-n_{T} \sim 0.01$ and $\beta \sim 1$, this would imply that $f_{a} \sim 10^{2} f_{PQ}$.  Knowing $m_{a}$ or $f_{PQ}$, it would then be possible to estimate $f_{a}$ and get an idea of $\varphi_{50}$, the value of the inflaton field fifty $e$-foldings before the end of inflation.  If, however, $-n_{T} \ll 1$, it will probably be impossible to determine $f_{a}/f_{PQ}$ and really say anything about $\varphi_{50}$ or the model of inflation.  If $-n_{T} \ll 1$, it is also unlikely that isocurvature fluctuations can be used to differentiate between inflationary models.  If no isocurvature perturbations are observed, it is not possible to say very much, although a measurement of $n_{T} \sim 0.01$ would imply that either $\Omega_{a} \ll 1$ or that $f_{PQ}/f_{a} \ll 1$.

I would very much like to thank Roberto Peccei, Ned Wright and Graciela Gelmini for enlightening conversations.


\begin{thebibliography}{199}
\bibitem{PeeblesPPC}P. J. E. Peebles, \begin{em}Principles of Physical Cosmology\end{em} (Princeton University Press, 1992).
\bibitem{KT}E. Kolb and M. S. Turner, \begin{em}The Early Universe\end{em} (Addison Wesley, 1990).
\bibitem{cmbr}D. J. Fixsen et al., astro-ph/9605054 (1996).
\bibitem{Penzias}A. A. Penzias and R. W. Wilson, \begin{em}Astrophys. J.\end{em} {\bf 142}, 419 (1965).
\bibitem{Turnerbaryon}M. S. Turner, astro-ph/9610158.
\bibitem{Guth}A. H. Guth, Phys. Rev. {\bf D23}, 347 (1981).
\bibitem{Lindechaotic}A. D. Linde, \begin{em}Phys. Lett.\end{em} {\bf 129B}, 177 (1983).
\bibitem{Turnererice}M. S. Turner, astro-ph/9704062.
\bibitem{Davies}T. Bunch and P. C. W. Davies, \begin{em}Proc. Roy. Soc. London\end{em} {\bf A360}, 117 (1978).

\bibitem{tHooft1}G. `t Hooft, \begin{em}Phys. Rev. Lett.\end{em} {\bf 37}, 8 (1976).
\bibitem{tHooft2}G. `t Hooft, \begin{em}Phys. Rev. Lett.\end{em} {\bf D14}, 3432 (1976).
\bibitem{PQ1}R. D. Peccei and H. R. Quinn, \begin{em}Phys. Rev. Lett.\end{em} {\bf 38}, 1440 (1977).
\bibitem{PQ2}R. D. Peccei and H. R. Quinn, \begin{em}Phys. Rev.\end{em} {\bf D16}, 1791 (1977).
\bibitem{Weinberg}S. Weinberg, \begin{em}Phys. Rev. Lett.\end{em} {\bf 40}, 223 (1978).
\bibitem{Wilczek}F. Wilczek, \begin{em}Phys. Rev. Lett.\end{em} {\bf 40}, 279 (1978).

\bibitem{Drees}M. Drees and M. M. Nojiri, \begin{em}Phys. Rev.\end{em} {\bf D48}, 3483 (1993).
\bibitem{Griest} K. Griest, M. Kamionkowski, and M. S. Turner, \begin{em}Phys. Rev.\end{em} {\bf D41}, 3565 (1990).
\bibitem{JKG}G. Jungman, M. Kamionkowski and K. Griest, \begin{em}Phys. Rep.\end{em} {\bf 267}, 195 (1996).
\bibitem{search}J. Ellis, T. Falk, K. Olive, and M. Schmitt, (1996), hep-ph 9610410.

\bibitem{ShafiMDM}Q. Shafi and F. W. Stecker, \begin{em}Phys. Rev. Lett.\end{em} {\bf 53}, 1292 (1984).
\bibitem{DavisMDM}M. Davis, F. Summers and D. Schlegel, \begin{em}Nature\end{em} {\bf 359}, 393 (1992).
\bibitem{TaylorMDM}A. N. Taylor and M. Rowan-Robinson, \begin{em}Nature\end{em} {\bf 359}, 396 (1992).
\bibitem{vanDalenMDM} A. van Dalen and R. K. Schaefer, \begin{em}Astrophys. J.\end{em} {\bf 398}, 33 (1992).
\bibitem{KlypinMDM}A. Klypin, J. Holtzman, J. Primack and E. Reg\"{o}s, \begin{em}Astrophys. J.\end{em} {\bf 416}, 1 (1993).
\bibitem{tiltedCDM}F. Adams, et al.  \begin{em}Phys. Rev.\end{em} {\bf D47}, 426 (1993).
\bibitem{Peebleslambda}P. J. E. Peebles, \begin{em}Astrophys. J.\end{em} {\bf 258}, 415, (1984).
\bibitem{Turnerlambda1}M. S. Turner, G. Steigman and L. Krauss, \begin{em}Phys. Rev. Lett.\end{em} {\bf 52}, 2090 (1984).
\bibitem{Efstathioulambda}G. Efstathiou et al., \begin{em}Nature\end{em}, {\bf 348}, 705 (1990).
\bibitem{Turnerlambda2}M. S. Turner, \begin{em}Phys. Scr.\end{em} {\bf T36}, 167 (1991).

\bibitem{MaBertschinger}C. P. Ma and E. Bertschinger, \begin{em}Astrophys. J.\end{em} {\bf 455}, 7 (1995).
\bibitem{Chandra}S. Chandrasekhar, \begin{em}Radiation Transfer\end{em} (Dover, 1960).
\bibitem{Kosowsky}A. Kosowsky, \begin{em}Ann. Phys.\end{em} {\bf 246}, 49 (1996).
\bibitem{Hu}W. Hu, astro-ph/9508126 and references therein (1995).
\bibitem{BE1}J. R. Bond and G. Efstathiou, \begin{em}Astrophys. J.\end{em} {\bf 285}, L47 (1984).
\bibitem{BE2}J. R. Bond and G. Efstathiou, \begin{em}MNRAS\end{em} {\bf 226}, 655 (1987).

\bibitem{SZ}U. Seljak and M. Zaldarriaga, \begin{em}Astrophys. J.\end{em} {\bf 469}, 437 (1996).

\bibitem{Jungman}G. Jungman, et al., \begin{em}Phys. Rev.\end{em} {\bf D54}, 1332 (1996).
\bibitem{BET}J. R. Bond, G. Efstathiou and M. Tegmark, astro-ph/9702100 (1997).
\bibitem{HuSpergel}W. Hu, D. N. Spergel and M. White, \begin{em}Phys. Rev.\end{em} {\bf D55}, 3288 (1997).
\bibitem{CrittendenTurok}R. G. Crittenden and N. G. Turok, Phys. Rev. Lett. {\bf 75}, 2642 (1995).
\bibitem{Turokpeak}N. G. Turok, Phys. Rev. {\bf D54}, 3686 (1996).

\bibitem{PeeblesLSS}P. J. E. Peebles, \begin{em}Large Scale Structure in the Universe\end{em} (Princeton University Press, 1980).
\bibitem{Steinhardt}P. Steinhardt, in \begin{em}Particle and Nuclear Astrophysics and Cosmology in the Next Millenium\end{em}, ed. E. W. Kolb and R. D. Peccei (World Scientific, 1994).

\bibitem{Bertschinger}E. Bertchinger, in \begin{em}Cosmology and Large Scale Structure\end{em}, ed. R. Schaffer et al. (Elsevier Science, Netherlands, 1996).
\bibitem{Bardeen}J. M. Bardeen, P. J. Steinhardt and M. S. Turner, \begin{em}Phys. Rev.\end{em} {\bf D28}, 679 (1983).
\bibitem{Turnertilt}M. S. Turner, \begin{em}Phys. Rev.\end{em} {\bf D48}, 3502 (1993).
\bibitem{Seljakpol}U. Seljak, D. Spergel and M. Zaldarriaga, astro-ph/9702157 (1997).
\bibitem{SZpol}U. Seljak and M. Zaldarriaga, \begin{em}Phys. Rev. Lett.\end{em} {\bf 78}, 2054 (1997).
\bibitem{Kamionpol1}M. Kamionkowski, A. Kosowsky and A. Stebbins, \begin{em}Phys. Rev. Lett.\end{em} {\bf 78}, 2058 (1997)
\bibitem{Kamionpol2}M. Kamionkowski, A. Kosowsky and A. Stebbins, astro-ph/9611125 (1996).
\bibitem{axTurner}M. S. Turner, \begin{em}Phys. Rev.\end{em} {\bf D33}, 889 (1985).
\bibitem{Altarelli}G. Altarelli, et al., in \begin{em}Proceedings of the Rencontres de Hanoi\end{em}, CERN preprint CERN-PPE/94-71 (1994).
\bibitem{Abe}K. Abe, et al., \begin{em}Phys. Rev.\end{em} {\bf D51}, 962 (1995).
\bibitem{Bennett}C. L. Bennett et al., \begin{em}Astrophys. J.\end{em} {\bf 464}, L1 (1996).
\bibitem{Gorski}M. G\'{o}rski et al., \begin{em}Astrophys. J.\end{em} {\bf 464}, L11 (1996).
\bibitem{Belinsky}V. Belinsky, L. Grishchuk, I. Khalatnikov and Ya. B. Zel'dovich, \begin{em}Phys. Lett.\end{em} {\bf 155B}, 232 (1985).
\bibitem{Jensen}L. Jensen (unpublished).
\bibitem{Lindeaxion}A. D. Linde, \begin{em}Phys. Lett.\end{em} {\bf B259}, 38 (1991).
\bibitem{Abbott}L. Abbott and M. Wise, \begin{em}Nucl. Phys.\end{em} {\bf B244}, 541 (1984).
\bibitem{Lucchin}F. Lucchin and S. Matarrese, \begin{em}Phys. Rev.\end{em} {\bf D32}, 1316 (1985).
\bibitem{Fabbri}R. Fabbri, F. Lucchin and S. Matarrese, \begin{em}Phys. Lett.\end{em} {\bf 166B}, 49 (1986).
\bibitem{extended}D. La and P. J. Steinhardt, \begin{em}Phys. Rev. Lett.\end{em} {\bf 62}, 376 (1989).
\bibitem{Lindenew}A. D. Linde, \begin{em}Phys. Lett.\end{em} {\bf 108B}, 389 (1982).
\bibitem{Albrecht}A. Albrecht and P. J. Steinhardt, \begin{em}Phys. Rev. Lett.\end{em} {\bf 48}, 1220 (1982).
\bibitem{Pi}S.-Y. Pi, \begin{em}Phys. Rev. Lett.\end{em} {\bf 52}, 1725 (1984).
\bibitem{natural}K. Freese, J. Frieman and A. Olinto, \begin{em}Phys. Rev. Lett.\end{em} {\bf 65}, 3233 (1990).

\bibitem{EB}G. Efstathiou and J. R. Bond, \begin{em}MNRAS\end{em} {\bf 218}, 103 (1986).
\bibitem{Stompor}R. Stompor, A. J. Banday and K. M. G\'{o}rski, \begin{em}Astrophys. J.\end{em} {\bf 463}, 8 (1996).
\bibitem{Kawasaki}M. Kawasaki, N. Sugiyama and T. Yanagida, \begin{em}Phys. Rev.\end{em} {\bf D54}, 2442 (1996).
\bibitem{Page}C. B. Netterfield et al., \begin{em}Astrophys. J.\end{em} {\bf 474}, 47 (1997).
\bibitem{PD}J. A. Peacock and S. J. Dodds, \begin{em}MNRAS\end{em} {\bf 267}, 1020 (1994).
\bibitem{Bunn}E. F. Bunn and M. White, Astrophys. J. {\bf 480}, 6 (1997).
\bibitem{BondMyers1}J. R. Bond and S. Myers, in \begin{em}Trends in Astroparticle Physics\end{em}, ed. D. Cline and R. Peccei, Singapore, World Scientific, p. 262 (1991).
\bibitem{Carlberg}R. Carlberg et al., \begin{em}J. R. Astron. Soc. Canada\end{em} {\bf 88}, 39 (1994).
\bibitem{BondMyers2}J. R. Bond and S. Myers, \begin{em}Astrophys. J. Supplement\end{em}. {\bf 103}, 63 (1996).
\bibitem{WEF}S. D. M. White, G. Efstathiou and C. Frenk, \begin{em}MNRAS\end{em} {\bf 262}, 1023 (1993).
\bibitem{Viana}P. T. P. Viana and A. Liddle, \begin{em}MNRAS\end{em} {\bf 281}, 323 (1996).
\bibitem{Dekel}A. Dekel, \begin{em}ARA\&A\end{em} {\bf 32}, 371 (1994).
\bibitem{Strauss}M. A. Strauss and J. Willick, \begin{em}Phys. Rep.\end{em} {\bf 261}, 271 (1995).
\bibitem{Loveday}J. Loveday et al., \begin{em}Astrophys. J.\end{em} {\bf 400}, L43 (1992).
\bibitem{Matsuki1}S. Matsuki and K. Yamamoto, \begin{em}Phys. Lett.\end{em}  {\bf B263}, 523 (1991).
\bibitem{Matsuki2}S. Matsuki, I. Ogawa and K. Yamamoto, \begin{em}Phys. Lett.\end{em}  {\bf B336}, 573 (1994).
\bibitem{vanBibber}K. van Bibber et al., \begin{em}Int. J. Mod. Phys.\end{em} {\bf D3} Suppl., 33 (1994).


\end{thebibliography}
\end{document}